\documentclass[a4paper,11pt]{article}
\pdfoutput=1 % if your are submitting a pdflatex (i.e. if you have
\usepackage{jheppub} 

\usepackage[T1]{fontenc} % if needed

\newcommand\numubar{\bar\nu_\mu}

\newcommand\nuebar{{\bar\nu_e}}

\title{The mass-hierarchy and CP-violation 
discovery reach of the
LBNO long-baseline neutrino experiment}

\author[o]{S.K.~Agarwalla,}
\author[ao]{L.\,Agostino,}
\author[ae]{M. Aittola,}
\author[h]{A.~Alekou,}
\author[an]{B.\,Andrieu,}
\author[w]{D.\,Angus,}
\author[h]{F. Antoniou, }
\author[b]{A.\,Ariga,}
\author[b]{T.\,Ariga,}
\author[u]{R.\,Asfandiyarov,}

% ...
\affiliation[b]{University of Bern, Albert Einstein Center for Fundamental Physics, Laboratory for High Energy Physics (LHEP), Bern, Switzerland}
% ...
\affiliation[h]{CERN, Geneva, Switzerland}
% ...
\affiliation[o]{Institute of Physics, Sachivalaya Marg, Sainik School Post, Bhubaneswar 751005, India}
% ...
\affiliation[u]{University of Geneva, Section de Physique, DPNC, Geneva, Switzerland}
\affiliation[w]{Institute for Particle Physics Phenomenology, Department of Physics, Durham University, United Kingdom}
% ...
\affiliation[ae]{Oulu Southern Institute and Department of Physics, University of Oulu, Finland}
% ...
\affiliation[an]{UPMC, Universit\'e Paris Diderot, CNRS/IN2P3, Laboratoire de Physique Nucl\'eaire et de Hautes Energies (LPNHE), Paris, France}
\affiliation[ao]{APC, AstroParticule et Cosmologie, Universit\'e Paris Diderot, CNRS/IN2P3, CEA/Irfu, Observatoire de Paris, Sorbonne Paris Cit\'e, 10, rue Alice Domon et L\'eonie Duquet, 75205 Paris Cedex 13, France}
% ...

\usepackage{subfig}

\abstract{
The next generation neutrino observatory proposed by the LBNO collaboration will 
address fundamental questions in particle and astroparticle physics.
The experiment consists of a far detector, in its first stage a 20 kt LAr double phase TPC and a magnetised iron calorimeter, 
situated at 2300 km from CERN and a near detector based on a high-pressure argon gas TPC. 
The long baseline provides a unique opportunity to study neutrino flavour oscillations over their 
1st and 2nd oscillation maxima exploring the $L/E$ behaviour,
and distinguishing effects arising from $\delta_{CP}$ and matter. 

In this paper we have reevaluated the physics potential of this setup for determining the 
mass hierarchy (MH) and discovering CP-violation (CPV), 
using a conventional neutrino beam from the CERN SPS with a power of 750 kW. 
We use conservative assumptions on the knowledge of oscillation parameter priors and systematic uncertainties.
The impact of each systematic error and the precision of oscillation prior is shown. We demonstrate that the first stage of LBNO can determine unambiguously the MH to $>5\sigma$~C.L. over the whole phase space. We show that the statistical treatment of the experiment is of very high importance, resulting in the conclusion that 
LBNO has $\sim 100\%$ probability to determine the MH in at most 4-5 years of running.
Since the knowledge of MH is indispensable to extract $\delta_{CP}$ from the data, the first LBNO phase 
can convincingly give evidence for CPV on the $3\sigma$~C.L. using today's knowledge on oscillation 
parameters and realistic assumptions on the systematic uncertainties.
}

\begin{document}
{\small\maketitle}
\flushbottom

\section{Introduction}
The main goals of the proposed LBNO~\cite{Stahl:2012exa}
next-generation long-baseline neutrino and antineutrino oscillation experiment 
are to discover CP-violation in the leptonic sector (CPV or $\delta_{CP}\neq0$ and $\pi$) and 
determine the neutrino mass hierarchy (MH or $sign(\Delta m_{31}^2)=\pm 1$). 
Discovery is defined according to usual practice in experimental high-energy
physics as the ability to exclude the wrong hypothesis
with at least a $5\sigma$~confidence level (C.L.), while a $3\sigma$~C.L. would
correspond to an evidence for the tested hypothesis.
Since propagation through Earth impacts neutrinos and antineutrinos differently,
 neutrino oscillations in matter can mimic a CP-asymmetry induced 
 by $\delta_{CP}$ and also affect the
determination of the $\delta_{CP}$ value. 
Hence, to decouple genuine CP-phase from matter induced effects,
the strategy of LBNO is to exploit the $L/E$ dependence of 
the ${\nu_\mu}\rightarrow\nu_e$   and ${\bar\nu_\mu}\rightarrow\bar\nu_e$
appearance probabilities with a  
wide-band beam at a baseline of 2300~km. Separate information
on neutrinos and antineutrinos is obtained by 
changing the horn focusing polarity of the beam. 
The disappearance channels ($\nu_\mu\rightarrow\nu_\mu$ and 
$\bar\nu_\mu\rightarrow\bar\nu_\mu$) 
will constrain the atmospheric parameters and the muon charge identification will 
independently determine the $\nu_\mu$/$\bar\nu_\mu$ fluxes at the far distance.
The $\nu_\mu\rightarrow\nu_\tau$
and $\bar\nu_\mu\rightarrow\bar\nu_\tau$ appearance channels
will also be accessible with an unprecedented precision.
Unlike the attempts of infering MH with atmospheric neutrinos,
the accelerator-based approach of LBNO addresses both fundamental problems
of CPV and MH in clean and straightforward conditions, profiting from the ability
to reverse the focusing horns polarity and from the well-controlled  fluxes,  which
characterise accelerator-based neutrino beams.

In this paper, we present an updated study of the sensitivity 
of LBNO, and discuss the impact of 
systematic errors and of the {\it a priori} knowledge of oscillation parameters. 
Following a realistic and incremental approach~\cite{Agarwalla:2011hh},
the initial phase of LBNO foresees an underground $\sim$20~kton
fiducial mass double-phase liquid Argon TPC complemented by a magnetised muon detector
and coupled to a conventional neutrino beam from the CERN SPS, monitored by 
a magnetised near detector system. We show that this first realistic phase already provides 
conclusive and well-motivated physics opportunities.
We employ a Monte-Carlo technique simulating a very large number
of toy experiments to estimate the confidence level of the MH and CPV measurements.
With the capability of reversing the horn focusing polarity, and
even under pessimistic assumptions on systematic errors, a few years of running at
the CERN SPS suffice for
LBNO alone to produce a direct and
guaranteed discovery of MH ($>5\sigma$~C.L.)  over the full phase space
of oscillation parameters, 
and a unique sensitivity to CPV  through the exploration of the 
first and second oscillation maxima.
Neglecting any systematic error, LBNO in its first phase, has the power to reach
a CPV discovery level $> 5\sigma$'s C.L., the actual significance depending on how far from
zero and $\pi$ the true value
of $\delta_{CP}$ is.
The actually attainable CPV reach is sensitive to the prior knowledge of the oscillation parameters
and to the achievable systematic errors on fluxes, cross-sections and detector-related effects. 
With conservative assumptions on
the systematic errors and after
$\sim$12~years of running at the CERN SPS,
a significance for CPV above $> 3\sigma$'s C.L. will be reached for $\sim25(40)$\% of the $\delta_{CP}$ 
values, under the expectation that $\sin^22\theta_{13}$ will be known from reactor experiments with a precision
of $\pm10(2.5)$\%. Several sources of systematic effects need to be addressed, in order to reduce the overall error
balance and reach a discovery level. In particular, improvements in
the present knowledge of the neutrino interaction differential cross-sections could increase 
the expected CPV discovery reach. Alternatively, a second phase of LBNO with
an increased exposure with far more detector mass and beam power, aimed at
reducing the statistical error around the 2nd oscillation maximum, 
would allow to reach a $> 5\sigma$ CPV discovery level  over
a wide range of $\delta_{CP}$ values, even under the present conservative assumptions
on systematic errors, thanks to the increased dependence on $\delta_{CP}$ at the 2nd maximum.

\section{Phenomenology for LBNO}

The discovery that neutrinos change flavour while propagating in space -- the phenomenon of \emph{neutrino oscillations} -- has historically
been triggered by deep underground astrophysics experiments with neutrinos, first observing the Sun and, later on, 
the neutrinos generated in the interaction of cosmic rays with the Earth's atmosphere -- the atmospheric neutrinos. At the same time the detection of a handful of neutrinos from the supernova SN1987A gave a fundamental input on astrophysical models.
The firmly established flavour oscillations 
imply that neutrinos have small though non-vanishing and non-degenerate rest masses, and the existence
of a physically observable mixing in the leptonic sector.

New physics is a key ingredient to resolve questions that the Standard Model (SM) cannot answer. 
In this context, neutrino masses and oscillations are, to this day, the only experimentally established 
Beyond the Standard Model (BSM) physics.
In the framework of three neutrino family scenario, the weak eigenstates $\nu_\alpha$ ($\alpha$ = e,$\mu$,$\tau$)
are given as linear combinations of the mass eigenstates $\nu_i$ (of definite mass $m_i$, i=1,2,3) via the 
Pontecorvo-Maki-Nakagawa-Sakata (PMNS)~\cite{Maki:1962mu,Pontecorvo:1967fh} matrix $U$ as $\nu_\alpha = \sum_{i} U_{\alpha i}\nu_i$.
The $3\times 3$ unitary matrix $U$ is generally
parameterized by the three mixing angles $\theta_{12}$, $\theta_{13}$, $\theta_{23}$, and
the  phase $\delta_{CP}$ (where we have neglected Majorana phases):
\begin{eqnarray}
U &=&
\left(
\begin{array}{ccc}
U_{e1}   & U_{e2}   & U_{e3} \\
U_{\mu1} & U_{\mu2} & U_{\mu3} \\
U_{\tau1} & U_{\tau2} & U_{\tau3} \\
\end{array}
\right)  = 
\left(
\begin{array}{ccc}
1 & 0   & 0 \\
0 & c_{23} & s_{23} \\
0 & -s_{23} & c_{23} \\
\end{array}
\right)
\left(
\begin{array}{ccc}
c_{13} & 0   & s_{13}e^{-i\delta} \\
0 & 1 & 0 \\
-s_{13}e^{i\delta} & 0 & c_{13} \\
\end{array}
\right)
\left(
\begin{array}{ccc}
c_{12} & s_{12}  & 0 \\
-s_{12} & c_{12} & 0 \\
0 & 0 & 1 \\
\end{array}
\right)
\end{eqnarray}
where \(c_{ij}\) and \(s_{ij}\) represent $\cos(\theta_{ij})$ and
$\sin(\theta_{ij})$, respectively.
The parameter $\delta_{CP}$ is the  phase that controls the $CP$ asymmetry.

The present level of understanding of the PMNS matrix already represents an incredible experimental
achievement, which will culminate in the determination of the  phase $\delta_{CP}$. 
Global analyses~\cite{Machado:2011ar,Fogli:2012ua,Tortola:2012te,GonzalezGarcia:2012sz} of 
all neutrino oscillation data, including the recent highest precision measurements,
indicate that the ``minimal'' three-neutrino PMNS framework is sufficient to 
completely describe
the observed oscillation phenomenology (apart from some ``anomalies'' in terrestrial
short baseline experiments).
The  question of the CP-violation in the leptonic sector (CPV) remains
an unresolved and urgent problem of particle physics.
All data are self-consistent and are
compatible with any
value for $\delta_{CP}$ in $0\leq\delta_{CP}\leq 2\pi$
within the $2\sigma$ confidence range.

The neutrino mass hierarchy (MH) or 
the sign of $\Delta m^2_{31}\equiv m^2_3 - m^2_1$ is also not yet known: whether 
$m_3$ is the heaviest mass eigenvalue (normal hierarchy, $\Delta m^2_{31} > 0$)
or
$m_3$ is the lightest one (inverted hierarchy, $\Delta m^2_{31} < 0$)
remains to be experimentally determined, and presently there
is no evidence indicating a preference for either value.
Such an experimental determination is a crucial ingredient to 
resolve the problem of CPV. It is also relevant
for the understanding of 
the origin of neutrino masses, which is expected to relate to
BSM physics. In particular, an inverted hierarchy would be a strong 
hint that some unexpected physics is underlying the masses and
flavour problem, and an important ingredient for leptogenesis scenarios. 
The mass hierarchy is important to interpret
cosmological observations probing the hot
dark matter fraction. Likewise,  future data from
supernova bursts will be more easily interpreted with the MH known.
Finally, the existence of an inverted mass hierarchy would be an useful 
input for neutrino-less double beta decay searches, which aim at
testing the Majorana nature of neutrinos~(see e.g. \cite{Piquemal:2010zz}).

Both CPV and MH problems can be addressed with accelerator-based
long-baseline neutrino oscillation
experiments via the
electron appearance channels $\nu_\mu\rightarrow \nu_e$ and $\bar\nu_\mu\rightarrow \bar\nu_e$.
Including higher order terms and the effect of coherent forward scattering pointed out by Wolfenstein
in case of neutrino oscillations in matter~\cite{Wolfenstein:1977ue},
the $\nu_\mu\rightarrow \nu_e$ oscillation probability can be approximated as~\cite{Arafune:1997hd}:
\begin{eqnarray}
P(\nu_{\mu} \rightarrow \nu_{\rm e})
&\simeq &
 4  c_{13}^2 s_{13}^2 s_{23}^2
 \left\{
  1 + \frac{a}{\Delta m^2_{31}} \cdot 2 (1 - 2 s_{13}^2)
 \right\}\sin^2 \frac{\Delta m^2_{31} L}{4 E}
 \nonumber \\
&+&
 c_{13}^2 s_{13} s_{23}
 \left\{
  - \frac{aL}{ E} s_{13} s_{23} (1 - 2 s_{13}^2)
  +
 \frac{\Delta m^2_{21}L}{ E} s_{12}
    (-s_{13} s_{23} s_{12} + c_{\delta} c_{23} c_{12})
 \right\} \sin \frac{\Delta m^2_{31} L}{2 E}
 \nonumber \\
&-&
 4 \frac{\Delta m^2_{21} L}{2 E} 
 s_{\delta} c_{13}^2 s_{13} c_{23} s_{23} c_{12}
 s_{12} \sin^2 \frac{\Delta m^2_{31} L}{4 E}
 \label{eq:oscprob}
\end{eqnarray}
where $c_{ij} = \cos\theta_{ij}$,   $s_{ij} = \sin\theta_{ij}$, 
$c_\delta = \cos\delta_{CP}$, $s_\delta = \sin\delta_{CP}$, and $a=2\sqrt{2}G_F n_e E$,
with $n_e$ representing the electron density of the traversed medium.
The corresponding probability for $\numubar \to \nuebar$ transition is obtained by replacing $\delta_{CP} \rightarrow -\delta_{CP}$
and $a \rightarrow -a$.
The CP-violating effects of $\delta_{CP}$ are modulated by those of all three 
mixing angles and their interplay, resulting in complicated dependencies 
and leading to an
{\it a priori}\, eight-fold parameter degeneracy~\cite{Barger:2001yr}. In addition,
coherent forward scattering in matter affects oscillations, and also produces an asymmetry 
between neutrinos and anti-neutrinos.

\begin{figure}[hptb]
\begin{center}
\includegraphics[width=0.49\textwidth]{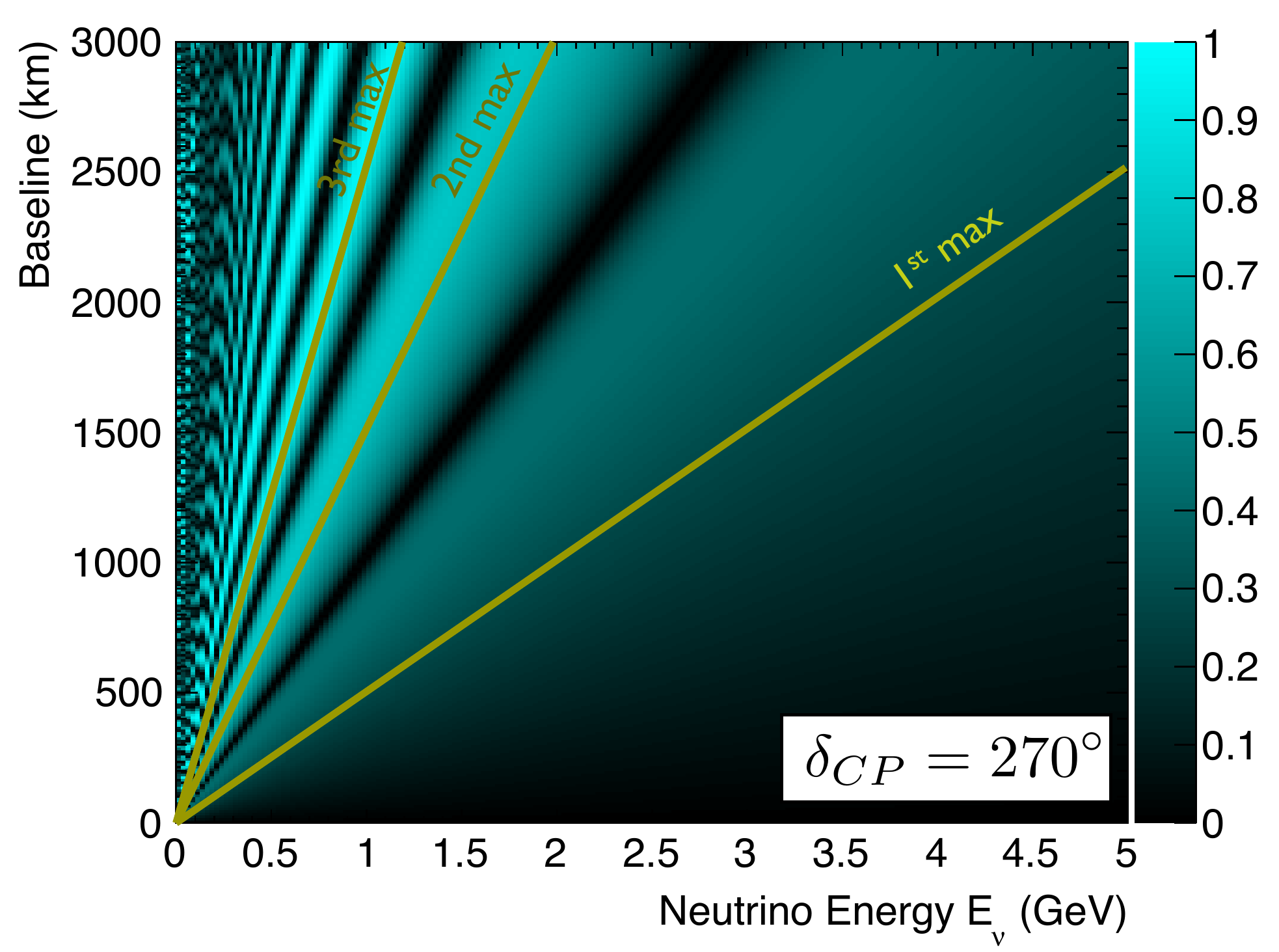} 
\includegraphics[width=0.49\textwidth]{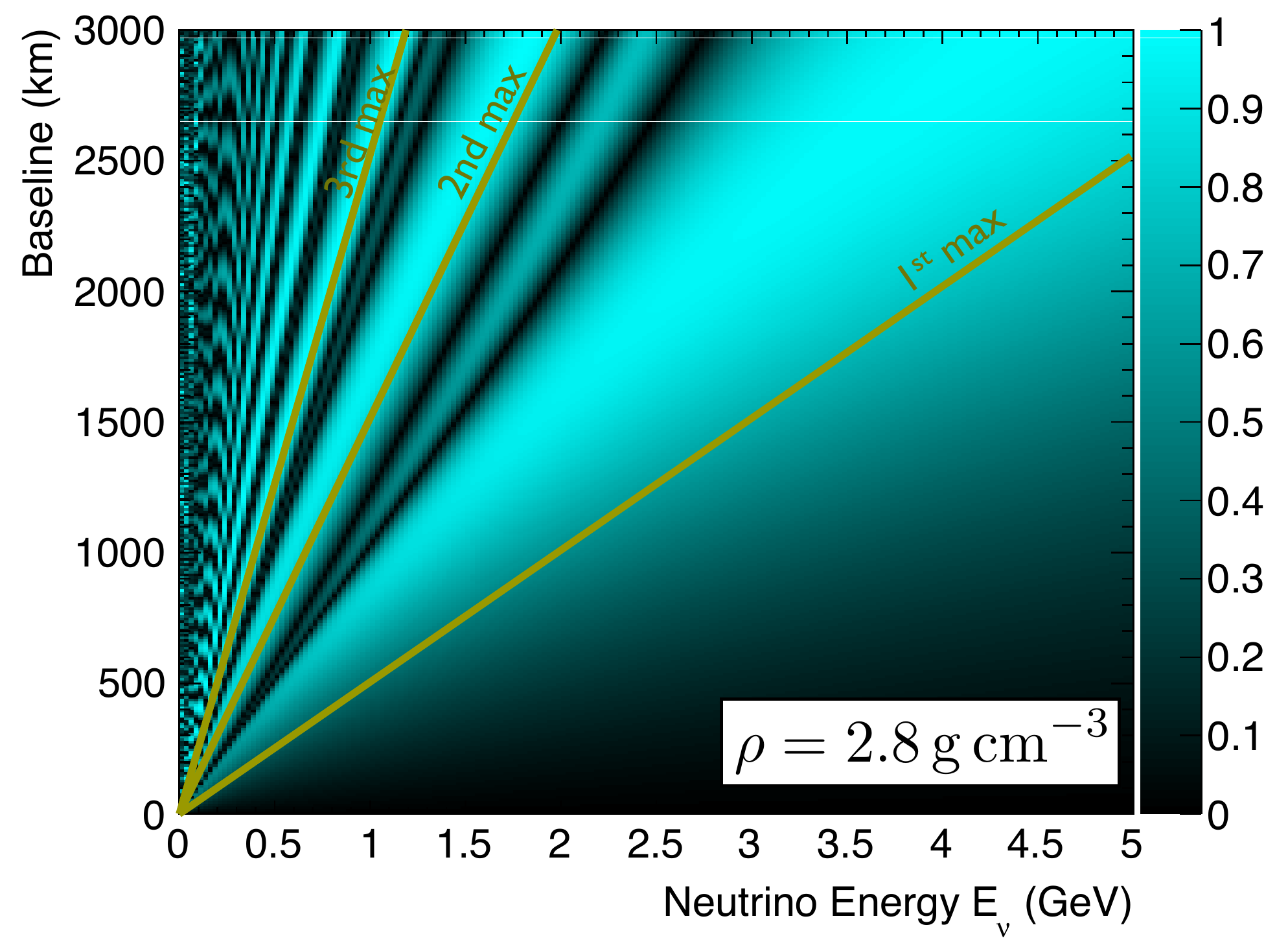} 
\caption{\small The two asymmetries, $abs({\mathcal A}_{CP}^{vac}(\delta_{CP}))$ and $abs({\mathcal A}_{CP}(\rho))$ 
in the ($E_\nu$,$L$) plane. The 1st, 2nd and 3rd oscillation maxima are represented by yellow lines
at constant $L/E_\nu$'s.}
\label{fig:oscasym}
\end{center}
\end{figure}

Several  ideas have emerged worldwide in order to advance the field, and
have converged into rather well defined
projects such as LBNO~\cite{Stahl:2012exa}, LBNE~\cite{Adams:2013qkq} and Hyper-Kamiokande~\cite{Abe:2011ts}. 
A general consensus, reflected in the above mentioned setups,
is that new generation very large scale and deep underground neutrino detectors will be needed 
to satisfactorily address open questions such as CPV and MH. 
In this context, it is handy to define two asymmetries between the probability of oscillations of
neutrinos and antineutrinos, 
one related to the CP effect computed in vacuum ${\mathcal A}_{CP}^{vac}(\delta_{CP})$:
\begin{equation}
{\mathcal A}_{CP}^{vac}(\delta_{CP})  \equiv  \left(\frac{P^{vac}(\nu)-P^{vac}(\bar\nu)}{P^{vac}(\nu)+P^{vac}(\bar\nu)}\right) 
\end{equation}
and the other to the matter effects ${\mathcal A}_{CP}(\rho)$ computed in matter for a fixed value
of $\delta_{CP}$:
\begin{equation}
{\mathcal A}_{CP}(\rho)  \equiv  \left(\frac{P^{mat}(\nu)-P^{mat}(\bar\nu)}{P^{mat}(\nu)+P^{mat}(\bar\nu)}\right)
\end{equation}
where $\rho$ represents the traversed Earth matter density (in the constant density approximation).
These two variables, plotted in the two dimensional plane of the neutrino energy $E_\nu$
versus the baseline $L$, are shown in Figure~\ref{fig:oscasym}.
In these graphs, the black regions correspond to combinations of neutrino energy and baseline at which
the oscillation phenomena 
is \emph{insensitive}\, to the effect, while the light (blueish) regions correspond to those where the
effect is largest. 

These features lead to the following phenomenological observations, 
which were taken into account in the definition of the LBNO experimental setup:
\begin{itemize}
\item The CP asymmetries increase at the next orders oscillation maxima.
This is understood by the
fact that ${\mathcal A}_{CP}^{vac}(\delta_{CP})$ has an envelope determined by~\cite{Arafune:1997hd}:
\begin{eqnarray}
%%\left. \frac{P(\nu)-P(\bar\nu)}{P(\nu)+P(\bar\nu)}\right|_{a=0}
% \sin^2 \frac{\delta m^2_{31} L}{4 E}
\frac{2 s_{\delta}  c_{12}
 s_{12}}{   s_{13} }\cot\theta_{23}
 \frac{\Delta m^2_{21} L}{2 E}
% \sin^2 \frac{\delta m^2_{31} L}{4 E}
\end{eqnarray}
which grows as $L/E$. The 2nd maximum is hence more sensitive to CPV than the 1st maximum.
Hence, access to the 2nd maxima extends the sensitivity to CPV, in particular 
when the measurement at the first maximum becomes systematic dominated. 
\item The matter asymmetry covers a broad region just below, and dominates around, the 1st maximum.
\item The energy dependence of the probability can resolve
several parameter degeneracies, and allows in particular disentangling the CP-driven and
the matter-driven effects, if the baseline is large enough.
\item Conversely, if the mass hierarchy is unknown, or if the matter effects are treated as a source
of systematic error, degeneracies reduce the ability to significantly detect CPV.
\item Assuming an energy threshold of about 1~GeV, which is a realistic value
taking into account on-axis conventional neutrino beam fluxes from high energy
protons, and the vanishing
neutrino cross-sections at low energies (in particular for antineutrinos),
the measurement of the 2nd maximum requires a baseline greater than $1500$~km:
\begin{eqnarray}
E_\nu^{2nd\,max}\gtrsim 1\,\rm{GeV} \Longrightarrow L\gtrsim 1500~\rm{km}
\end{eqnarray}
\end{itemize}

In order to attack the challenging problem of CPV and the related MH determination,
LBNO adopts a combination of methods, by 
precisely measuring the disappearance and appearance energy 
spectrum shapes (in particular, peak position and height for 1st and 2nd oscillation maximum and minimum) 
with high resolution, and by comparing  neutrino- and antineutrino-induced oscillations.
The liquid Argon and magnetized iron  detectors will in fact provide complementary 
studies of all three active transitions $\nu_\mu\rightarrow \nu_\mu$, 
$\nu_\mu\rightarrow \nu_e$ and $\nu_\mu\rightarrow \nu_\tau$ charged current events
over the optimized range of neutrino energies.
They can also probe the active-sterile transition with neutral current events.
A precise investigation of the oscillation probabilities as a function of energy 
and a comparison of neutrino and antineutrino behaviours will
verify if they follow the expectations from 3-generation neutrino mixing. 

As will be exposed in this paper, LBNO is optimised to best perform these measurements
and yields a definitive resolution of MH and a significant exploration of CPV. 
The goal of the near detectors
will be to precisely predict the unoscillated neutrino fluxes, using well-developed
and tested techniques in present long-baseline experiments, such as T2K~\cite{Abe:2012av}.
Hadro-production and neutrino cross-section campaigns will cover the relevant
region for LBNO.
The 2300~km baseline is adequate to have an excellent separation of the asymmetry due to 
the matter effects (i.e. the mass hierarchy measurement) 
and the CP asymmetry due to the $\delta_{CP}$  phase, and thus to break the parameter degeneracies. 
Therefore, the existence of matter and CP-violation induced effects will be examined 
without over-relying on theoretical modelling and assumptions, and the standard neutrino paradigm 
will be tested explicitly. Hence, the LBNO approach to extract
MH and $\delta_{CP}$ value is clean and straightforward.

\section{The LBNO experimental setup - an incremental approach}
\begin{figure}[!htb]
\begin{center}
\mbox{
\includegraphics[angle=0,width=0.5\textwidth]{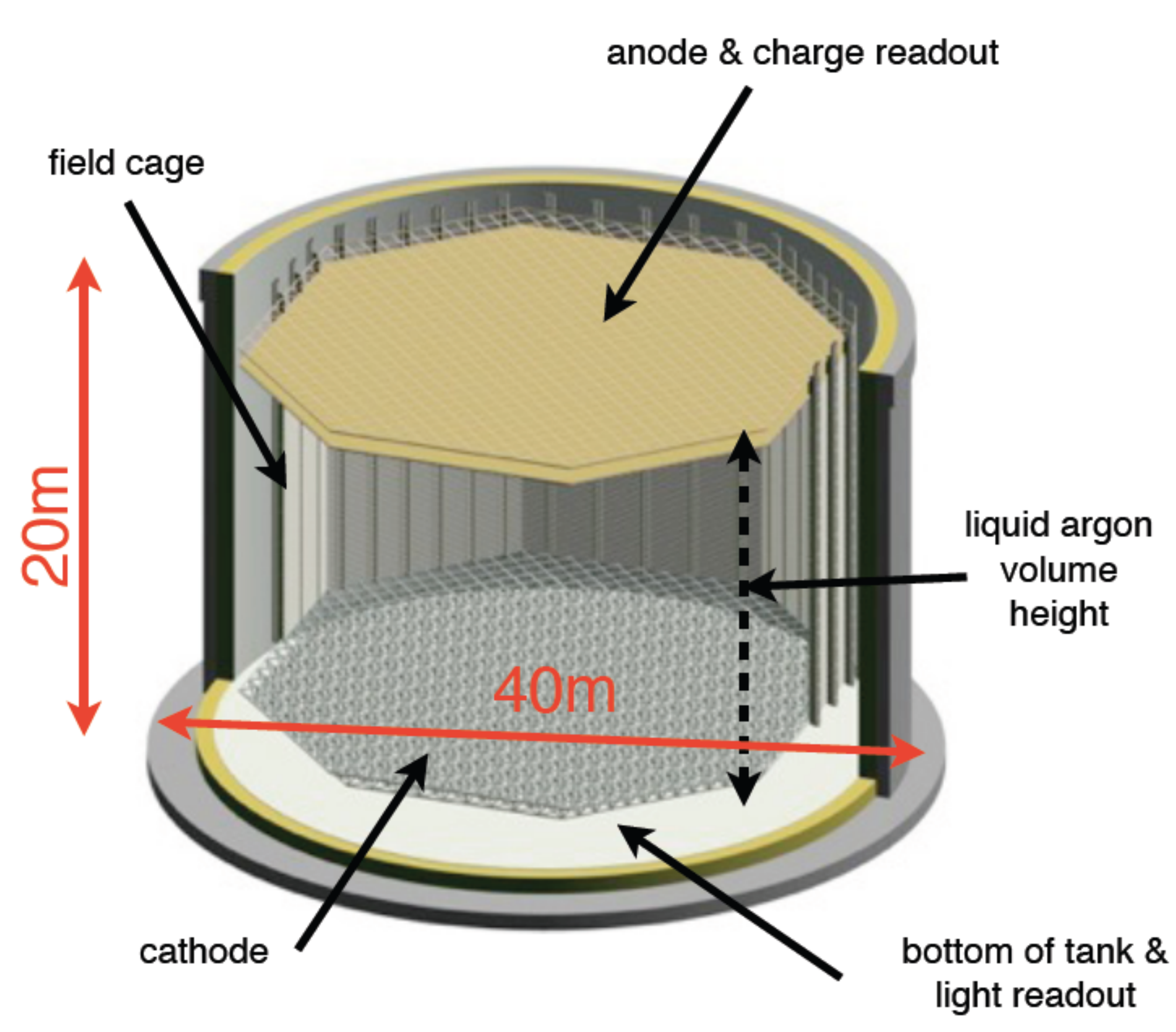}
\includegraphics[angle=0,width=0.5\textwidth]{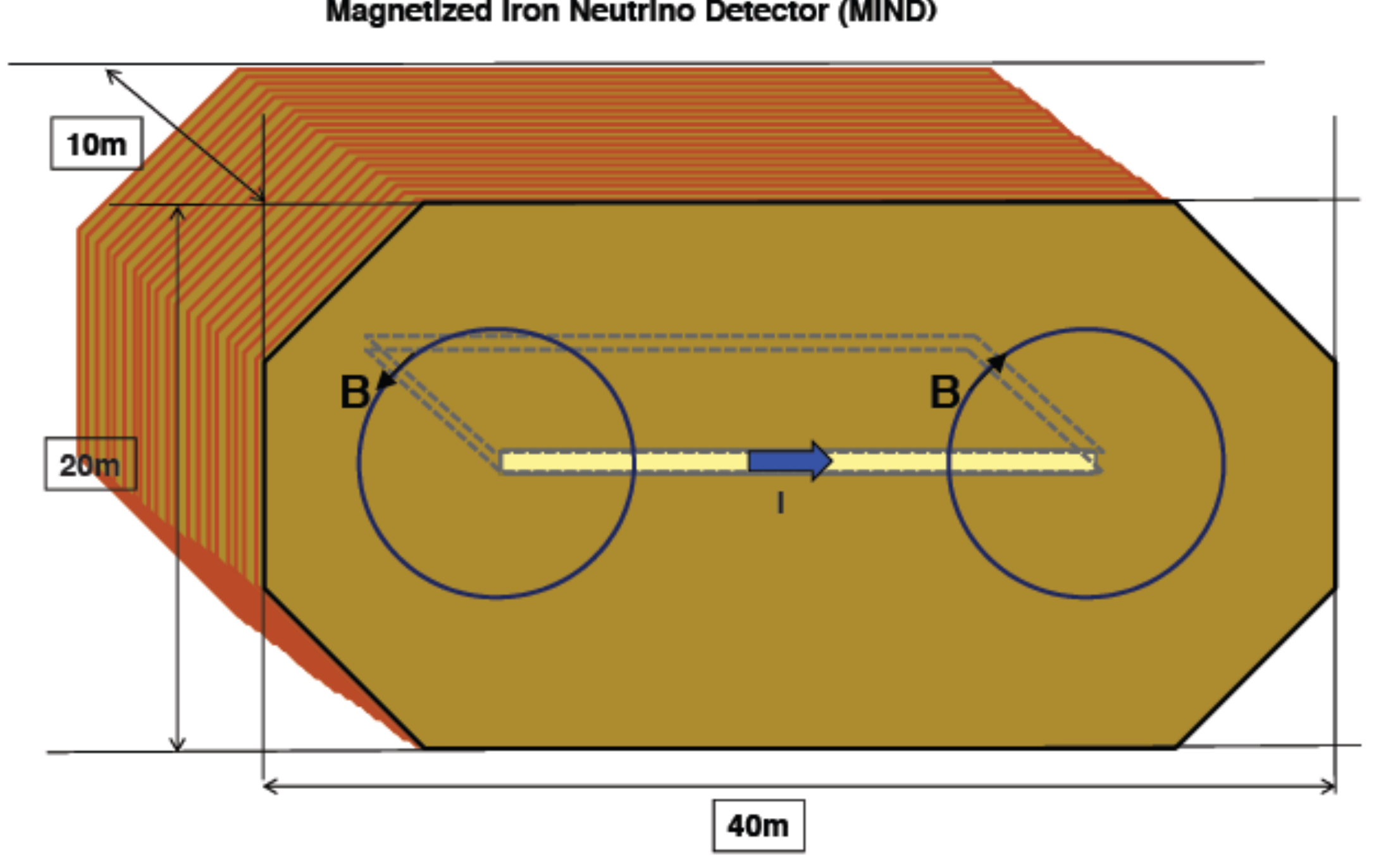}
}
\caption{\small{Schematic view of the 20 kt LAr detector (left panel) and the 35 kt MIND detector (right panel)} }
\label{fig:LAt}
\end{center}
\end{figure}

As today's state-of-art is set by the Super-Kamiokande detector (SK)
with its 22.5~kton fiducial mass, new detectors should be more than an order of magnitude larger 
than SK or use technologies which can outperform the Water Cherenkov technique, such
as the liquid Argon (LAr) Time Projection Chamber~\cite{Rubbia:1977zz}.
Taking into account the latest knowledge of oscillation parameters, the construction feasibility of such 
a large underground laboratory, the detector itself and the involved costs, the LBNO Expression of 
Interest~\cite{Stahl:2012exa} proposes \emph{an incremental approach} 
at the Pyh\"asalmi mine. The incremental approach is motivated by physics,
technical and financial aspects.
From the point of view of oscillation physics, the
priority for the underground far detectors is the initial
20~kton double phase LAr LEM-TPC (GLACIER~\cite{Rubbia:2004tz,Rubbia:2010}) combined 
with a magnetized muon detector (MIND~\cite{Abe:2007bi,Cervera:2010rz}) in one of two large
underground caverns (see Ref.~\cite{Stahl:2012exa} for details).
Schematic views of a 20 kton LAr detector and a 35 kt MIND detector are
shown in Figure~\ref{fig:LAt}. In the current engineering concept, the
20~kton LAr detector has a total LAr mass of 32.5~kton, and an instrumented active mass of 
22.8~kton~\cite{Stahl:2012exa}. In the simulations performed for this paper, the field cage
of the 20~kton detector is approximated with a  cylinder of radius 33~m and height 20~m, corresponding to 
an instrumented volume of  17100~m$^3$ and an active mass of 23.9~kton.

LBNO builds upon the results of several years of design studies funded by the European Commission (EC).
LAGUNA  was  organized as a 3-years long  project, to
  carry out underground sites investigations and develop a concept for a facility
  able to host the new underground neutrino observatory.
It was primarily motivated by the fact that, although
Europe currently has four national underground laboratories 
(at Boulby (UK), Canfranc (Spain), Gran Sasso (Italy), and Modane (France)),
 none of them is large enough to host a next-generation  observatory.
LAGUNA selected seven potential underground sites 
(Boulby (UK), Canfranc (Spain), Fr\'ejus (France), Pyh\"asalmi (Finland),
Sieroczowice (Poland), Slanic (Romania), Umbria (Italy))
to study, and compared them
in order to identify the scientifically and 
technically most appropriate and cost-effective strategy towards a 
large scale European neutrino observatory.
  One of the key conclusion of LAGUNA is that all of the seven considered underground sites are
\emph{in principle} technically feasible,
 and able to host the desired types of detectors. 
 With the reduced impact of all the other factors, site selection should be based on 
 physics arguments. 
A second phase called LAGUNA-LBNO has been
funded by the EC and initiated in October 2011.
It  further evaluated the findings of LAGUNA, and in particular, 
it  assessed the underground construction of the large detectors, their commissioning,
and the long-term operation of the facility. 
 LAGUNA-LBNO is in addition specifically considering
  long-baseline  beams
 from CERN~\cite{Rubbia:2010fm}. 
From the seven pre-selected LAGUNA sites, the two deepest,
 Fr\'ejus (overburden of 4800~m.w.e.)
and Pyh\"asalmi (overburden of 4000~m.w.e.), were found particularly attractive 
and retained further
attention. 
Careful simulations and a detailed analysis of the key findings of
LAGUNA have motivated the choice of the CERN-Pyh\"asalmi baseline
(2300~km) \textit{as the first priority}. Two main physics criteria were
considered in order to optimize the choice of the site, informed by the
first indications of large $\theta_{13}$~\cite{Abe:2011sj}, which have since then being
strongly confirmed by reactors~\cite{An:2012eh} and T2K~\cite{Abe:2013xua,Abe:2013hdq}.

\section{The LBNO physics programme}
For long-baseline physics in the post-$\theta_{13}$ discovery era, 
the LAr and magnetized iron  detectors provide complementary 
studies of the three active transitions $\nu_\mu\rightarrow \nu_\mu$, 
$\nu_\mu\rightarrow \nu_e$ and $\nu_\mu\rightarrow \nu_\tau$ charged current events
over a range of neutrino energies optimized via tuning of the conventional beam focusing,
and can also probe the active-sterile transition by measuring neutral current events.
A precise investigation of the oscillation probabilities as a function of energy 
and a comparison of neutrino and antineutrino behaviours will
verify if they follow the expectations from 3-generation neutrino mixing. 
The 2300~km baseline is adequate to have an excellent separation of the asymmetry due to 
the matter effect (i.e. the mass hierarchy measurement) 
and the CP asymmetry due to the $\delta_{CP}$  phase.
This is optimised to break the parameter degeneracies and
provide a definitive resolution of MH and a significant exploration of CPV in the neutrino sector.
These measurements are performed
without over-relying on theoretical modelling and assumptions and the standard neutrino paradigm. 
This is different from extracting
the MH and $\delta_{CP}$ value from global fits of all available data.

The chosen location in one of the deepest mine in Europe ($\sim 4000$~m.w.e.) will also provide a unique
opportunity to observe very rare phenomena with a LAr detector, independent of the CERN beam events.
Proton decay can be explored in many different -- often background free -- decay channels.
After 10 years of exposure, 
the sensitivity on the proton lifetime will reach
$\tau_p\geq 2\times 10^{34}$\,years at 90\% C.L. in the  $p\rightarrow K\bar{\nu}$ 
channel. 
In addition, other exclusive decay channels will be investigated, 
such as $p\rightarrow e^{+}\pi^{0}$  and $p\rightarrow \mu^{+}\pi^{0}$. 
Measuring many different channels helps to distinguish between models.
Furthermore, 5600 atmospheric neutrino events per year will be measured.
Atmospheric neutrinos, detected with good energy- and angular resolution and flavor identification  
are a new tool to perform oscillation physics complementary to the CERN beam, and could
be a new method to obtain a radiography of the Earth's interior via matter effects. 
The neutrino burst from a galactic supernova (SN) explosion would be observed with high statistics
in the electron neutrino channel,
providing invaluable information on the inner mechanism of the SN explosion and on neutrino
oscillations, not accessible to other setups.
For a supernova explosion at 10 kpc, $\sim$~10,000 neutrino interactions will be recorded in the 
active LAr volume.
Unknown sources of astrophysical neutrinos, like for instance those that could arise 
from annihilation processes of WIMP particles in astrophysical objects could
also be observed.

\section{New CERN beam layout}

The beam under design is a conventional third generation neutrino beam facility based 
on the CNGS~\cite{CNGS} technology. Initially, the facility will use protons
from an upgraded CERN SPS accelerator, reaching 750~kW of nominal beam power.  
This high-intensity operation goes beyond the record intensity of 565~kW ever achieved in the SPS~\cite{SPSint}, and 60\% above the operational beam power for CNGS. The main limitations to achieve such intensities come from beam losses in both PS and SPS, and due to limited RF power at SPS. In Table~\ref{tabl:spspower} the expectations for the SPS potential in delivering intense beams for a future neutrino program are described, coming as stretched goal within the foreseen LHC injector upgrades (LIU) project~\cite{LIUproj}. 
Table~\ref{tabl:CN2PYparam} summarises the key parameters for the beam. The quoted yearly intensities correspond to 200 days of running mode with 80\% efficiency for the accelerators, and $60\div85$\% of beam sharing with other users for the case of SPS.

\begin{table}[b]
\centering
\caption{\label{tabl:spspower} Present, All-time Record, and Possible Future SPS Parameters for Neutrino Type Beams}
\begin{tabular}{ l  c  c  c } 
\hline
 & {\bf CNGS} &  {\bf RECORD} &  {\bf LBNO} \\ \hline
$E_\text{SPS}$ [GeV]	& 400 & 400& 400 \\ 
Bunch spacing [ns]	& 5& 5 & 5 \\ 
 $I_\text{bunch}$ [$\times 10^{10}$]	& 1.05	& 1.3 & 	1.7 \\ 
$ N_\text{bunches}$ 	& 4200& 	4200 & 4200 \\
$I_\text{SPS}$ [$\times10^{13}$] & 	4.4	& 5.3	& 7.0 \\ 
$I_\text{PS}$ [$\times10^{13}$]&	2.3&	3.0& 4.0 \\ 
PS cycle length [s]	& 1.2	& 1.2	& 1.2/2.4 \\ 
SPS cycle length [s]	&6.0	& 6.0	& 6.0/7.2 \\
$E_\text{PS}$  [GeV/c]	& 14	&14& 14 \\
Beam power [kW]	& 470	& 565& 747/622 \\ \hline
\end{tabular}
\end{table}

\begin{table}[h]
\centering
\caption{\label{tabl:CN2PYparam} Key parameters of the assumed LBNO proton beam from the SPS and HP-PS~\cite{ipac13_hpps}.}
\begin{tabular} { l  c  c } \hline
{\bf Parameter} & {\bf SPS beam } & {\bf HP-PS beam} \\ \hline
$E_\text{beam} [GeV] $ & 400 & $50\div75$ \\
$I_\text{beam} [ppp] $ & $7\times10^{13}$ & $2.5\div1.7\times10^{14}$ \\
Cycle length [s]	&  6 & 1 \\
$P_\text{beam} [MW] $	& 0.750 & 2  \\
POT$_\text{year}$ [$10^{21}$] & $0.10\div0.14$ & $3.46\div2.35$ \\  \hline
\end{tabular}

\end{table}

\begin{figure}[h]
\centering
 \includegraphics[width=0.70\linewidth]{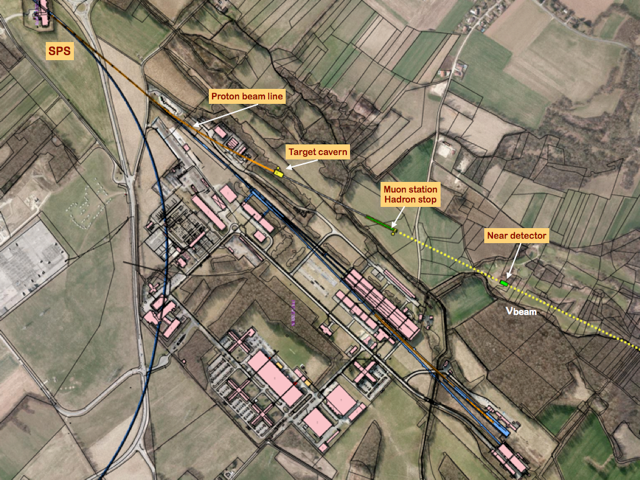}
\caption{\label{fig:nalayout} The baseline layout for the LBNO neutrino facility in the vicinity of the CERN North Area. }
\end{figure}

To profit from existing infrastructures for the target hall and near detector, in the baseline option, the LBNO
beam could be located near the SPS North Area as shown in Figure~\ref{fig:nalayout}. 
Under this hypothesis, for the first stage using the 400~GeV from SPS, the primary beam would be extracted from the TT2 channel and transported for about 400~m in the existing TT20 line. Then it would branch off to a new 480~m long transfer line required to match the direction and more importantly create the 10.4$^\circ$ downward slope required to point to the far detector.
For the fast extraction from SPS in the Long Straight Section-2 (LSS2), a novel scheme was developed to bypass the lack of space to install new kickers in the region whilst maintaining the elements required for the slow extraction to fixed-target experiments. The new scheme uses a non local extraction combining kickers in LSS6 and LSS2 sections. First tests with beam have shown encouraging results, further studies are planned after the restart of the CERN accelerators in 2015~\cite{spsextr}. 

A key constraint in the location and design of the secondary beam comes from the steep 18.1\% slope required due to the long-baseline. The combination of high-intensity and high beam energy put a constraint on the 
minimum distance between the target and the near detector location to ensure that the high-energy muons
are absorbed in the beam dump and the earth, preventing them from generating background in the near
detector. Considering the first phase with the primary beam at 400~GeV and assuming a rock density of 2.3~g/cm$^3$, preliminary Monte Carlo calculations suggest that a distance of at least 800~m is necessary if a passive
concrete shielding of about 100~m long is used in the beam dump. As a consequence,  the near detector cavern will be 144~m deeper than the target. Therefore, the option to branch off from the TT20 line at its upmost point very close to the surface is very attractive, as it allows the whole installation to be at smaller depths with significant cost savings. In this configuration the target cavern is located at -41~m, the hadron stop at -100~m and the near detector at -185~m, almost at the same level as the deepest point of LHC. 

Three upgrade scenarios are being considered for the neutrino beam. 
These primarily involve upgrade or alternative scenarios for the proton injector to the same 
target area beyond the initial operation with the present SPS:
(i) use of an upgraded high-energy PS or SPS, machines discussed in the HE-LHC option,
(ii) use of a new dedicated HP-PS synchrotron~\cite{ipac13_hpps},
(iii) use of a Neutrino Factory beam concept.
The prospects for a major upgrade of the LBNO neutrino beam are very attractive, 
offering a long term vision. The realization of
the HP-PS accelerator with MW power would expand the capability of the LBNO
facility and provide an interesting way to increase the exposure by a significant factor  
without prohibitively extending the running time. The chosen baseline of
2300~km is suitable to implement a Neutrino Factory, 
opening the path towards an ultimate exploration and an era 
of high-precision oscillation studies.

\section{Beam optimisation and expected raw event rates}
The neutrino beamline design is a central component of the LBNO optimisation, since
it will directly impact the long-baseline physics reach, affecting  for instance 
the mass hierarchy sensitivity,  the CP violation reach and the study of the 
other oscillation channels like the tau neutrino appearance.
Different beamline designs have been investigated and a preliminary
design has been established  in an effort to 
illustrate the compelling physics reach of the experiment. 

\begin{figure}[!h]
\begin{center}
\mbox{
\includegraphics[angle=0,width=0.7\textwidth]{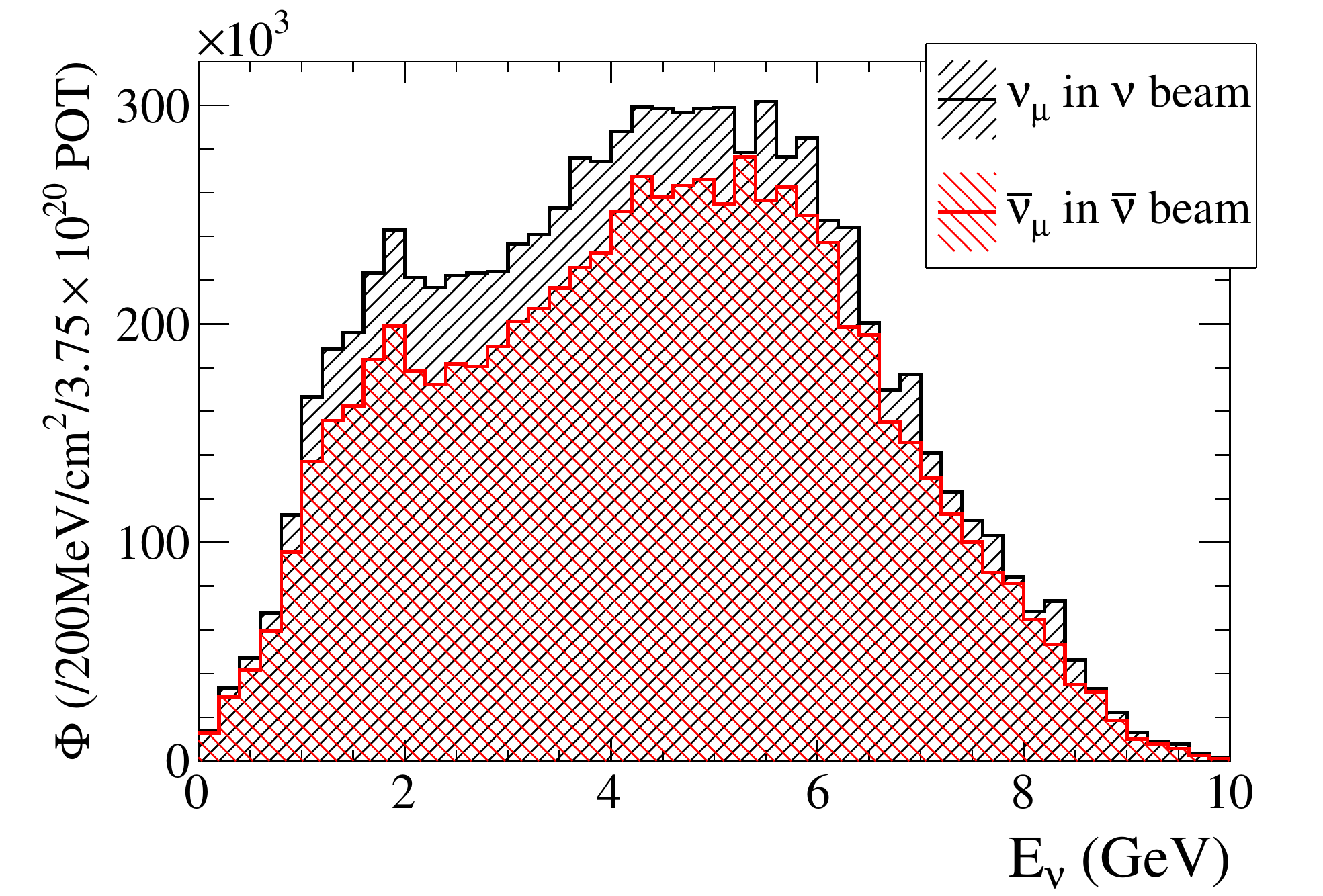}
}
\caption{\small{Neutrino and anti-neutrino flux for the CERN to Pyh\"asalmi beam.} }
\label{fig:flux}
\end{center}
\end{figure}

The neutrino beam line is tentatively composed of a target, two horns and a decay volume. 
The target is modelled as a 1 m long cylinder of graphite
with density $\rho=1.85$~g/cm$^3$ and 2 mm radius. 
The focusing system is based on a pair of parabolic horns
which we will denote as horn (upstream) and reflector (downstream) 
according to the current terminology. 
The decay tunnel is 300~m long and 3~m wide. 
A new beam optimisation is currently 
under way to investigate different possible beam optics, which should lead to enhanced 
rates and an optimised beam profile for the LBNO physics program. A further optimisation
of the decay tunnel could also increase the neutrino flux.
The unoscillated neutrino and anti-neutrino beam flux is shown in Figure~\ref{fig:flux}.

The expected charged (CC) and neutral current (NC)
interaction rates are computed using the optimisation of the focusing
optics for 400~GeV and are shown in Table~\ref{tab:lbl_event_rates}.
The rates under the assumption of the potential 2nd phase at 2~MW with the new
HP-PS 50~GeV are listed in Table~\ref{tab:lbl_event_ratesphasetwo} for completeness.
The oscillated rates are computed using the oscillation parameters from 
the global fit of Ref.~\cite{GonzalezGarcia:2012sz}.
    The NC interactions rate
    are for events with visible energy $> 0.5$ GeV. The rate is given for
    an exposure of 50 kt.yrs, so 2.5~years with the 20~kton baseline LAr detector. 
    
For comparison, the rates for LBNE
with the baseline of 1300~km are also shown and normalised to the same exposure. 
The parameters and flux of the LBNE beam correspond to those described in Ref.~\cite{Papadimitriou:2011fz}.
 
 Although the LBNE baseline of 1300~km is significantly shorter than the LBNO baseline of 2300~km,
 the expected rate of oscillated events in both setups are very similar, in both neutrino and antineutrino
 mode. This is explained by the fact that the longer baseline requires also higher energy neutrinos in order
 to keep the $L/E$ parameter of both setups around the atmospheric region. The resulting higher boost
 of the parent mesons, and the resulting higher neutrino energies, compensate for the increase distance,
 by making the beam more pencil-like and profiting from the linearly increasing total neutrino interaction
 cross-section.
    
\begin{table}[!htp]
 \centering
{\footnotesize
\begin{tabular}[c]{|l|c|c|c|c|c|ccc|} \hline
Beam & $\nu_{\mu}$ unosc. & $\nu_{\mu}$ osc. & $\nu_e$ beam & $\nu_{\mu}$  & $\nu_{\mu}\rightarrow\nu_{\tau}$ & \multicolumn{3}{c|}{$\nu_{\mu}\rightarrow\nu_e$ CC}  \\
       & CC & CC & CC &  NC  & CC & $\delta_{CP} = -\pi/2$, & $0$,&
$\pi/2$ \\ \hline \hline

LBNO: 2300 km NH& & & & & & & &  \\
 400 GeV, 750 kW& & & & & & & &  \\
 $1.5\times10^{20}$ POT/year& & & & & & & & \\
 
 50kt years $\nu$ & 3447 &907&22&1183&215&246&201&162\\
 50kt years $\bar{\nu}$ & 1284&330&5&543&98&20&27&29\\

 \hline

LBNO: 2300 km IH& & & & & & & &  \\
 400 GeV, 750 kW& & & & & & & &  \\
 $1.5\times10^{20}$ POT/year& & & & & & & & \\
 
 50kt years $\nu$ & 3447 &853&22&1183&239&71&43&32\\
 50kt years $\bar{\nu}$ & 1284&330&5&543&99&61&77&89\\
  
\hline 
\hline

LBNE Low energy beam& & & & & & & &  \\
 120 GeV, 700 kW, NH& & & & & & & &  \\
 $6\times10^{20}$ POT/year& & & & & & & & \\
 
 50kt years $\nu$ & 4882 &1765&44&1513&61&217&174&126\\
 50kt years $\bar{\nu}$ & 2506 &890&13&620&22&44&54&56\\
%% 50kt years $\bar{\nu}$ (GloBes xsec)  & 2004 &741&12&620&22&37&44&46\\

 \hline
LBNE Low energy beam& & & & & & & &  \\
 120 GeV, 700 kW, IH& & & & & & & &  \\
 $6\times10^{20}$ POT/year& & & & & & & & \\
 
 50kt years $\nu$ & 4882 &1713&44&1513&67&120&82&58\\
 50kt years $\bar{\nu}$ & 2506 &875&13&620&23&59&69&82\\
 
\hline
\end{tabular}
}
 \caption{Raw $\nu$ oscillation event rates at the far site with $E_{\nu} < 10$ GeV
 normalised to 50kt years, corresponding to 2.5~years
 of data-taking with the 20~kton baseline LAr detector. See text.}
\label{tab:lbl_event_rates}
\end{table}

\begin{table}[!htp]
 \centering
{\footnotesize
\begin{tabular}[c]{|l|c|c|c|c|c|ccc|} \hline
Beam & $\nu_{\mu}$ unosc. & $\nu_{\mu}$ osc. & $\nu_e$ beam & $\nu_{\mu}$  & $\nu_{\mu}\rightarrow\nu_{\tau}$ & \multicolumn{3}{c|}{$\nu_{\mu}\rightarrow\nu_e$ CC}  \\
       & CC & CC & CC &  NC  & CC & $\delta_{CP} = -\pi/2$, & $0$,&
$\pi/2$ \\ \hline \hline

LBNO: 2300 km NH & & & & & & & &  \\
 50 GeV, 2 MW& & & & & & & &  \\
 $3.0\times10^{21}$ POT/year& & & & & & & & \\
 
 50kt years $\nu$ & 8616 &2266&54&2955&539&615&502&406\\
 50kt years $\bar{\nu}$ & 3325&828&13&1360&249&44&65&73\\
  
\hline

LBNO: 2300 km IH& & & & & & & &  \\
 50 GeV, 2 MW& & & & & & & &  \\
 $3.0\times10^{21}$ POT/year& & & & & & & & \\
 
 50kt years $\nu$ & 8616 &2132&54&2955&596&177&109&79\\
 50kt years $\bar{\nu}$ & 3325&828&13&1360&249&154&192&224\\
\hline
\end{tabular}
}
 \caption{Same as Table~\ref{tab:lbl_event_rates} but under the assumption of the HP-PS 50~GeV accelerator 
 at 2~MW operation. These rates are shown for completeness, they are not used in the calculations
 shown in this paper.}
\label{tab:lbl_event_ratesphasetwo}
\end{table}

% some reminder on MH and CPV measurement method
\section{Mass hierarchy and CP violation measurements at LBNO}

%subsection on analysis method

\subsection{General principle}

Our primary goal is to determine the mass hierarchy and measure CP violation by observing $\nu_\mu$ to $\nu_e$ oscillations, through a precise measurement of the neutrino spectrum and the comparison of neutrino- and antineutrino-induced oscillations.
%%%% addition of introduction to the measurement, taken from the EoI
The 2300 km baseline is adequate to have an excellent separation of the asymmetry due to the matter effects (i.e. the mass hierarchy
measurement) and the CP asymmetry due to the $\delta_{CP}$  phase, and thus to break the parameter degeneracies.
The probabilities of $\nu_\mu\to\nu_e$ and $\bar\nu_\mu\to\bar\nu_e$  oscillations 
contain the spectral information which provides an unambiguous determination of the
oscillations parameters and allows discriminating between the two CP-conserving scenarios,
namely $\delta_{CP}=0$ and $\delta_{CP}=\pi$.

If the distance between source and detector is fixed, the oscillatory behaviour of
the neutrino flavour conversion can be easily translated to
that for the expected neutrino energy spectrum of the oscillated events.
If the neutrino energy spectrum of the oscillated events can be reconstructed with sufficiently 
good resolution in order to distinguish first and second maxima, 
the spectral information obtained is invaluable for the unambiguous determination of
the oscillation parameters.

\begin{figure}[tb]
\begin{center}
\includegraphics[width=0.49\textwidth]{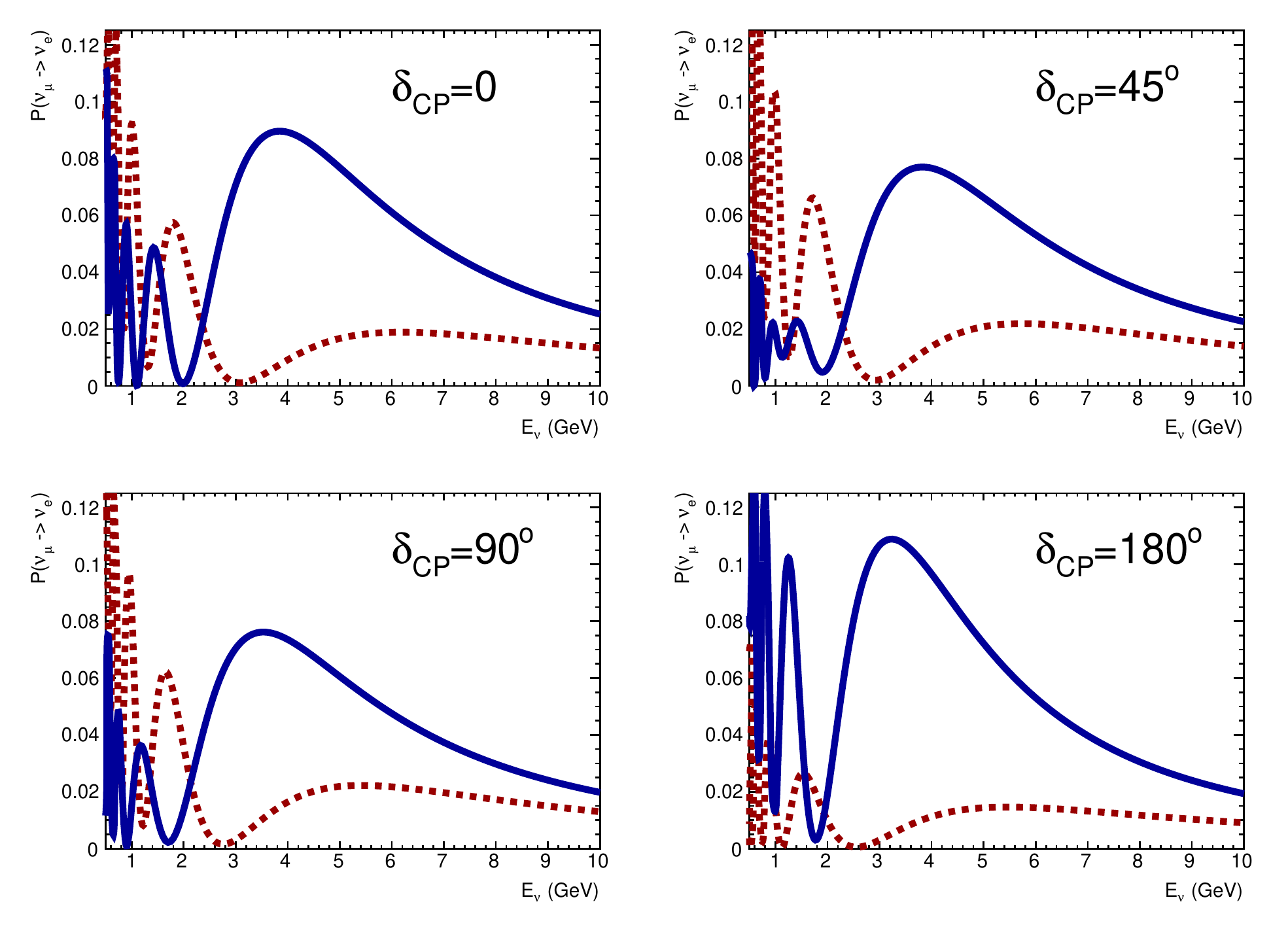} 
\includegraphics[width=0.49\textwidth]{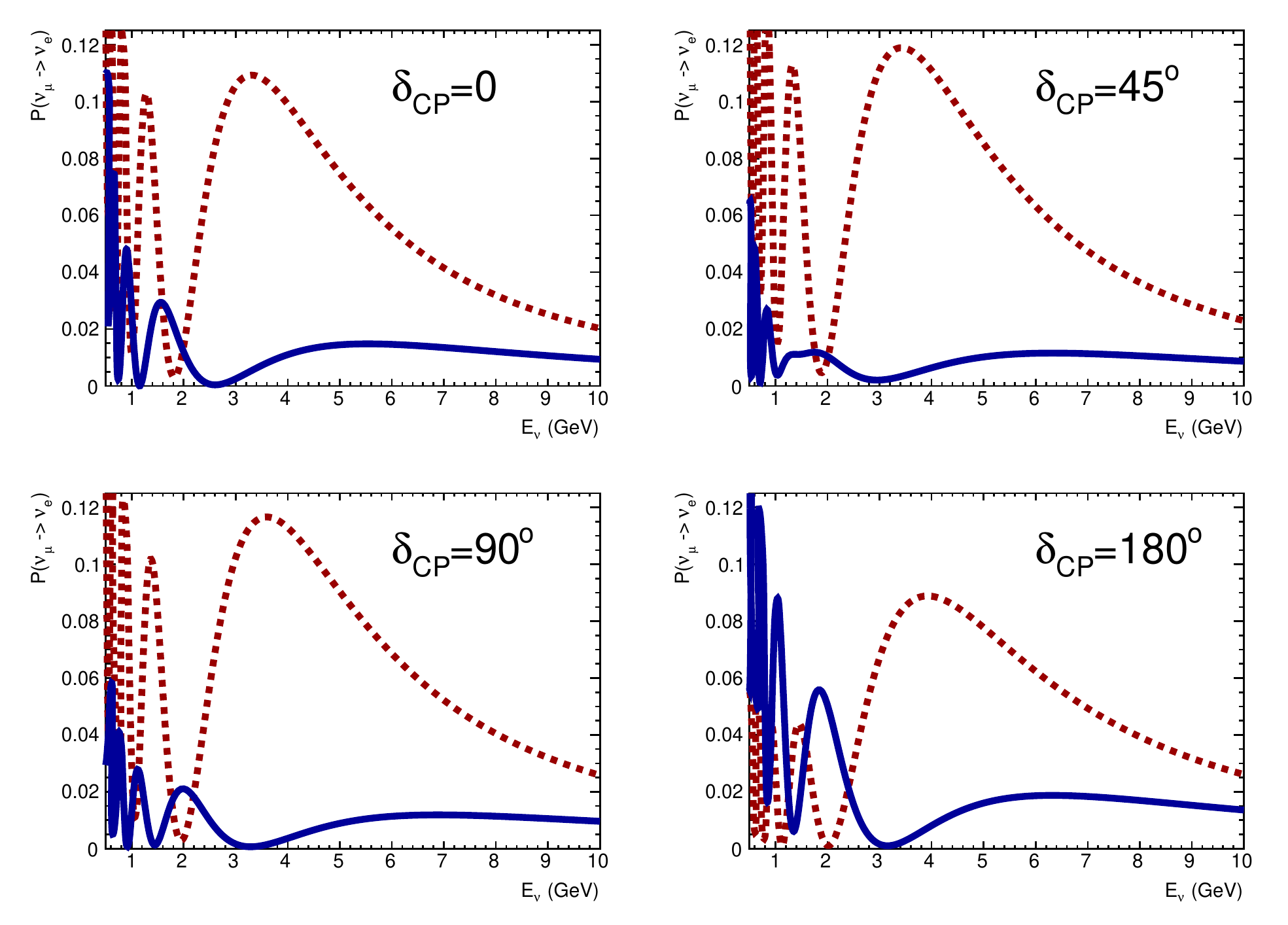} 
\caption{\small Oscillation probability of $\nu_{\mu} \to \nu_e$ (blue) and
$\bar\nu_{\mu} \to \bar\nu_e$ (red-dashed) for different values of $\delta_{cp}$ 
for (left) normal hierarchy $\Delta m^2_{31}>0$ (NH),
(right) inverted hierarchy $\Delta m^2_{31}<0$ (IH),
and $\sin^22\theta_{13}=0.09$. The spectral information provides an unambiguous
determination of the oscillation parameters and allows in principle to distinguish the two
CP-conserving scenarios, namely $\delta_{CP}=0$ and $\delta_{CP}=\pi$.}
\label{fig:osc2}
\end{center}
\end{figure}

The probabilities of $\nu_{\mu} \to \nu_e$ and
$\bar\nu_{\mu} \to \bar\nu_e$ oscillations for $\sin^22\theta_{13}=0.09$
and different values of $\delta_{CP}$ and normal hierarchy (NH) and inverted hierarchy (IH) are shown in 
Figure~\ref{fig:osc2}, as they are expected in LBNO at the 2300~km baseline. 
%%The same is shown in Figure~\ref{fig:osc2IH} in the case of an inverted mass hierarchy (IH).
%
The plots 
illustrate qualitatively
that the spectral information provides an unambiguous
determination of the oscillation parameters and allows discriminating between the two
CP-conserving scenarios.
The $\delta_{CP}$-phase and matter effects introduce
a well-defined energy dependence of the oscillation probability.
As a consequence,  the neutrino energy spectrum of the oscillated events need to be 
experimentally reconstructed with sufficiently 
good resolution in order to distinguish first and second maximum, and extract
unambiguous information on the oscillation parameters. The spectral measurement
will in addition verify the PMNS model, with possible unexpected differences between
neutrinos and antineutrinos than those predicted by $\delta_{CP}$, or other
non-standard deviations from the predicted $L/E$ dependence.

At the same time, the matter effects at 2300~km are large and the NH and IH scenarios
induce to an almost complete swap of behaviours between neutrinos and antineutrinos. This
is clearly exhibited in Figure~\ref{fig:osc2}. Hence, CP- and matter-induced
asymmetries are very different and distinguishable.

A sample of electron-like (e-like) events is thus a primary source of information. We consider the following background contributions to the signal e-like events:
\begin{itemize}
\item{Intrinsic $\nu_e$ contamination present in the beam (intrinsic $\nu_e$),}
\item{Electron events from $\nu_\tau$ charged current interaction with subsequent leptonic $\tau$ decay ($\nu_\tau \rightarrow e$ contamination),}
\item{Neutral current $\nu_\mu$ events with $\pi^0$ production (NC$\pi^0$),}
\item{Mis-identified muons from $\nu_\mu$ CC interactions (mis-id $\nu_\mu$).}
\end{itemize}
In addition, we use $\mu$-like events in the disappearance channel ($\nu_\mu \rightarrow \nu_\mu$ survival probability) to constrain the atmospheric oscillation parameters, $\Delta m^2_{31}$ and $\sin^2\theta_{23}$. 
A detailed description of neutrino event simulations and selection efficiency can be found in Ref.~\cite{Stahl:2012exa}. 

\subsection{Experimental observables}
\label{subsect:recovar}

\begin{figure}[!h]
\centering
\begin{tabular}{ccc}
\subfloat[All e-like]
{\includegraphics[width=0.31\textwidth]{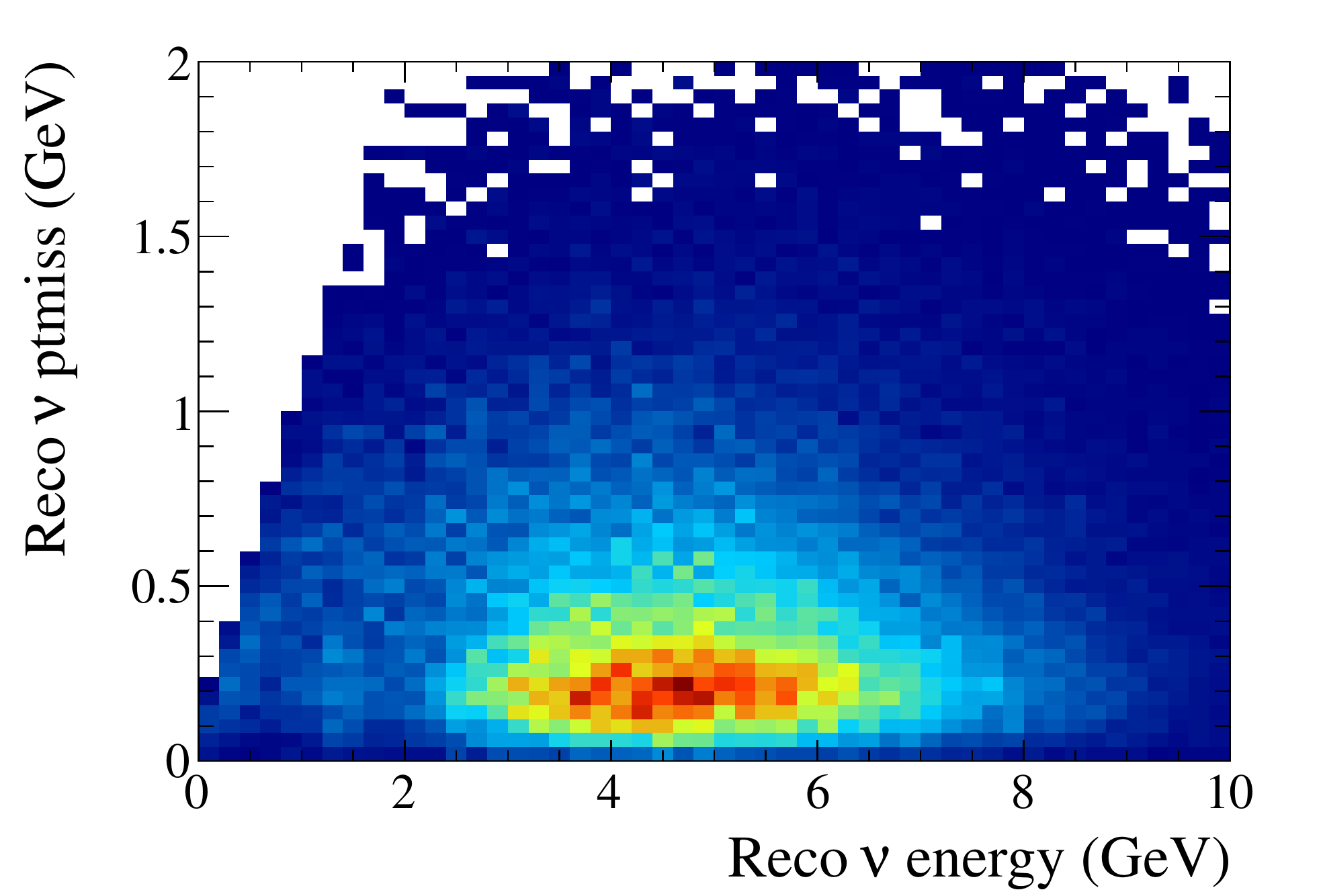}} &
\subfloat[Signal $\nu_e$]
{\includegraphics[width=0.31\textwidth]{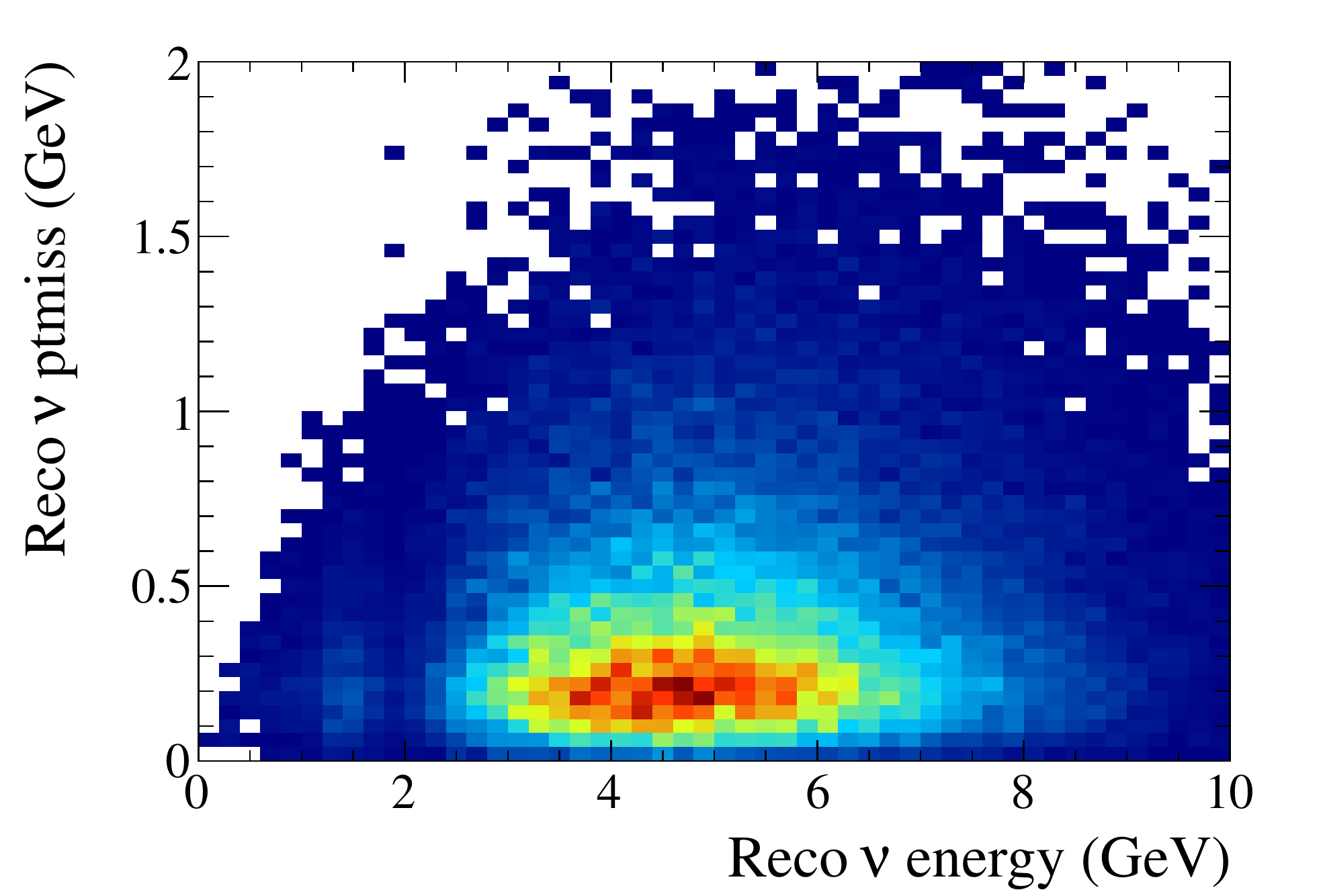}} &
\subfloat[Intrinsic $\nu_e$]
{\includegraphics[width=0.31\textwidth]{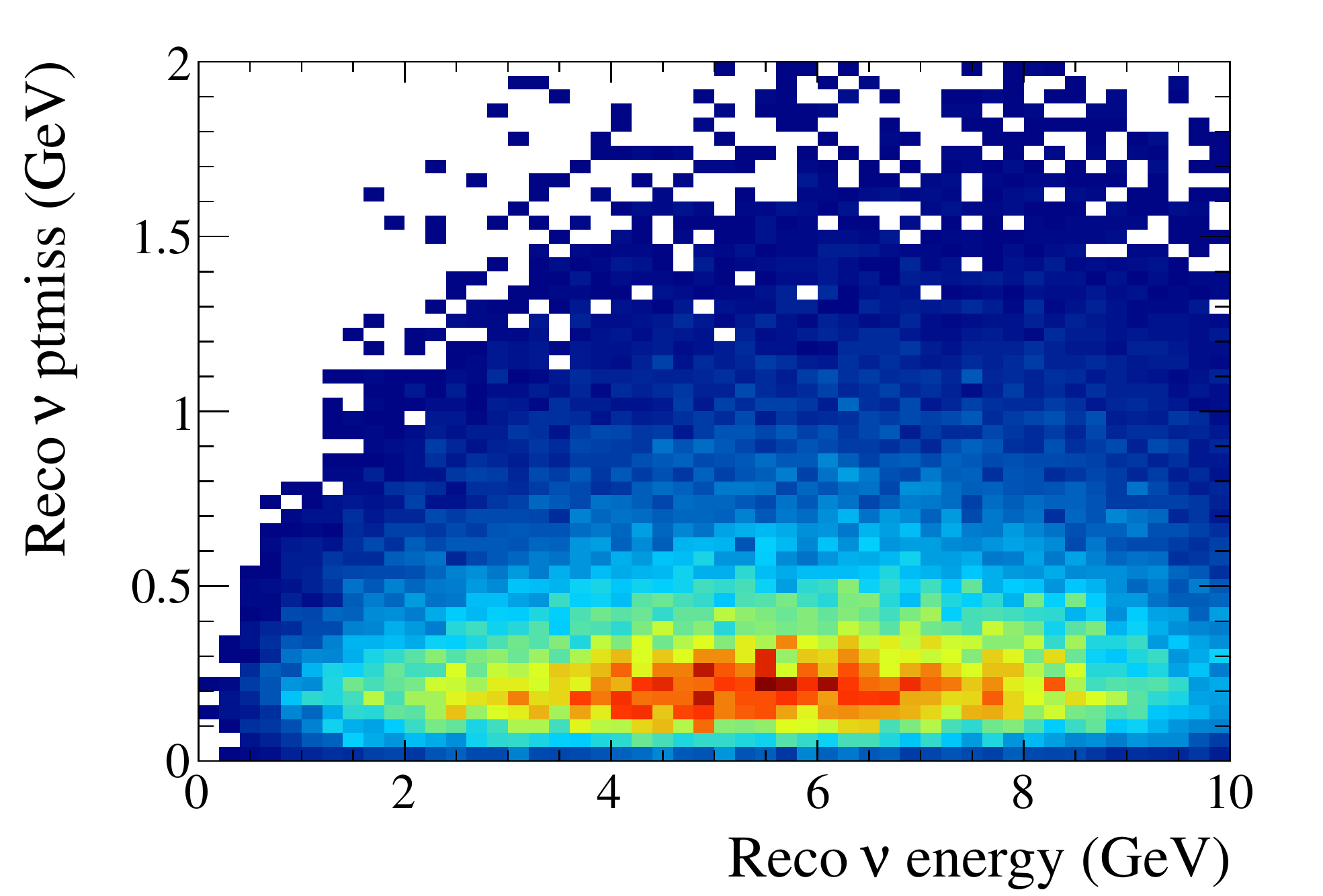}}\\
\subfloat[NC$\pi^0$]
{\includegraphics[width=0.31\textwidth]{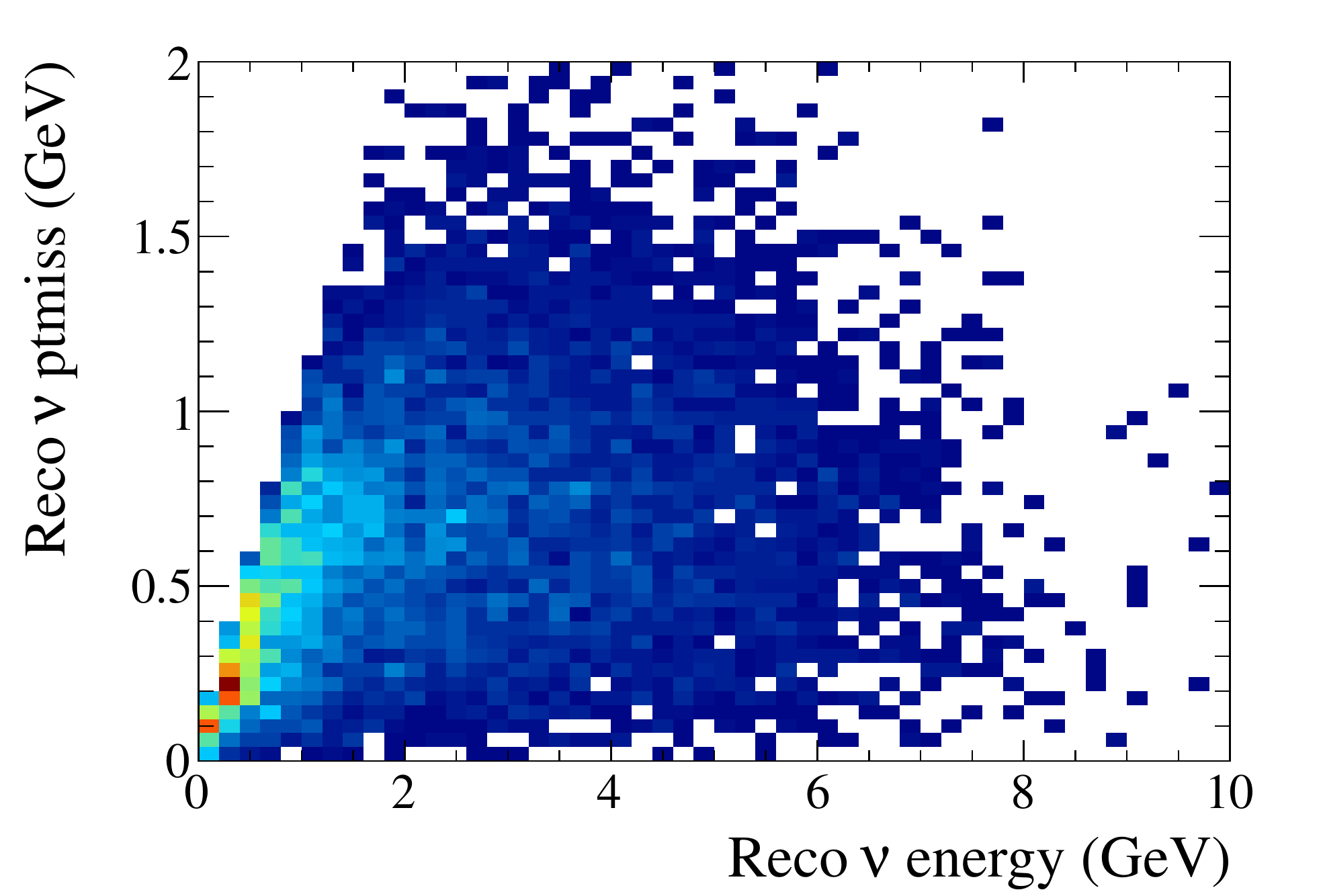}} &
\subfloat[$\nu_\tau \rightarrow e$ contamination]
{\includegraphics[width=0.31\textwidth]{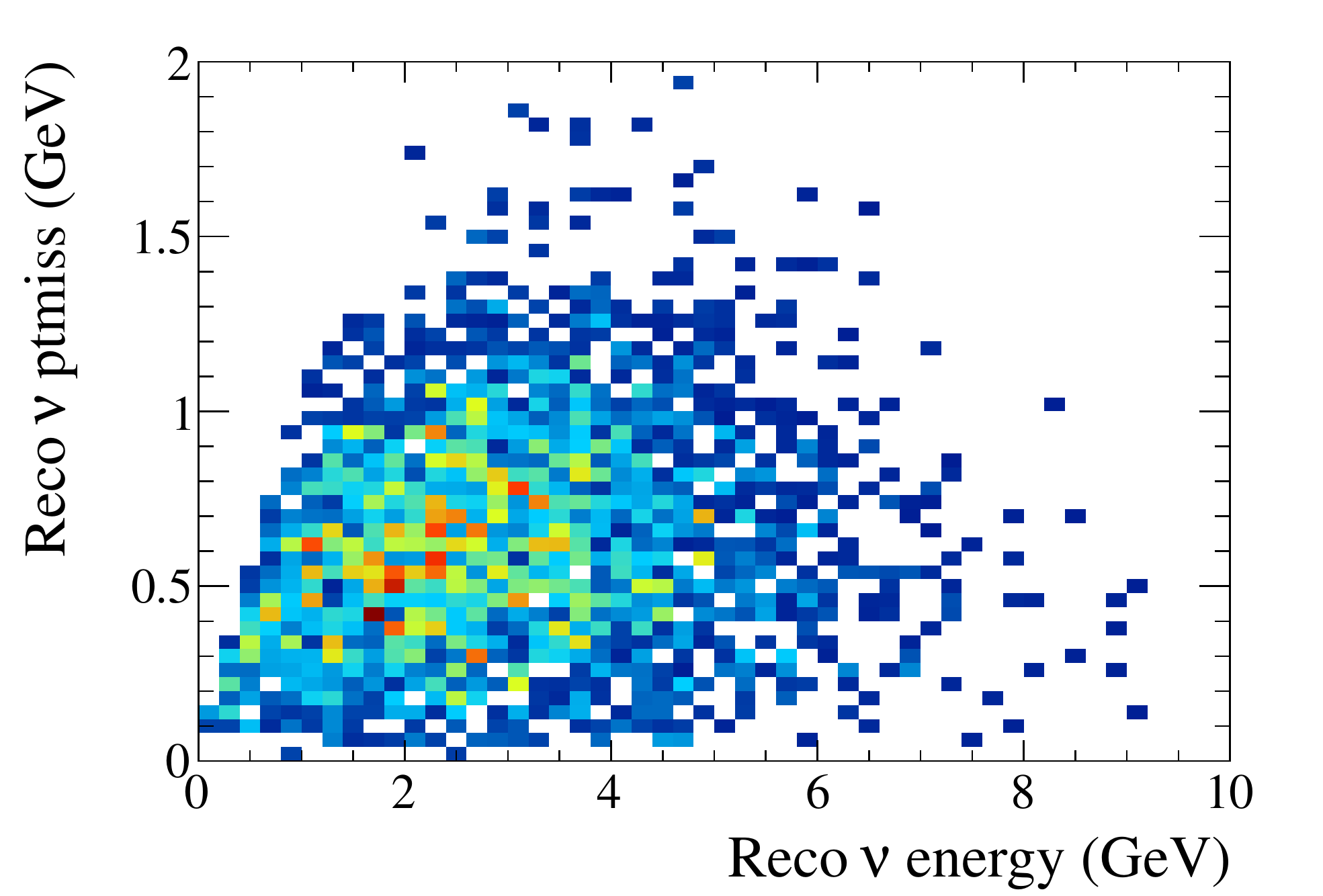}}  &
\subfloat[Mis-id $\nu_\mu$]
{\includegraphics[width=0.31\textwidth]{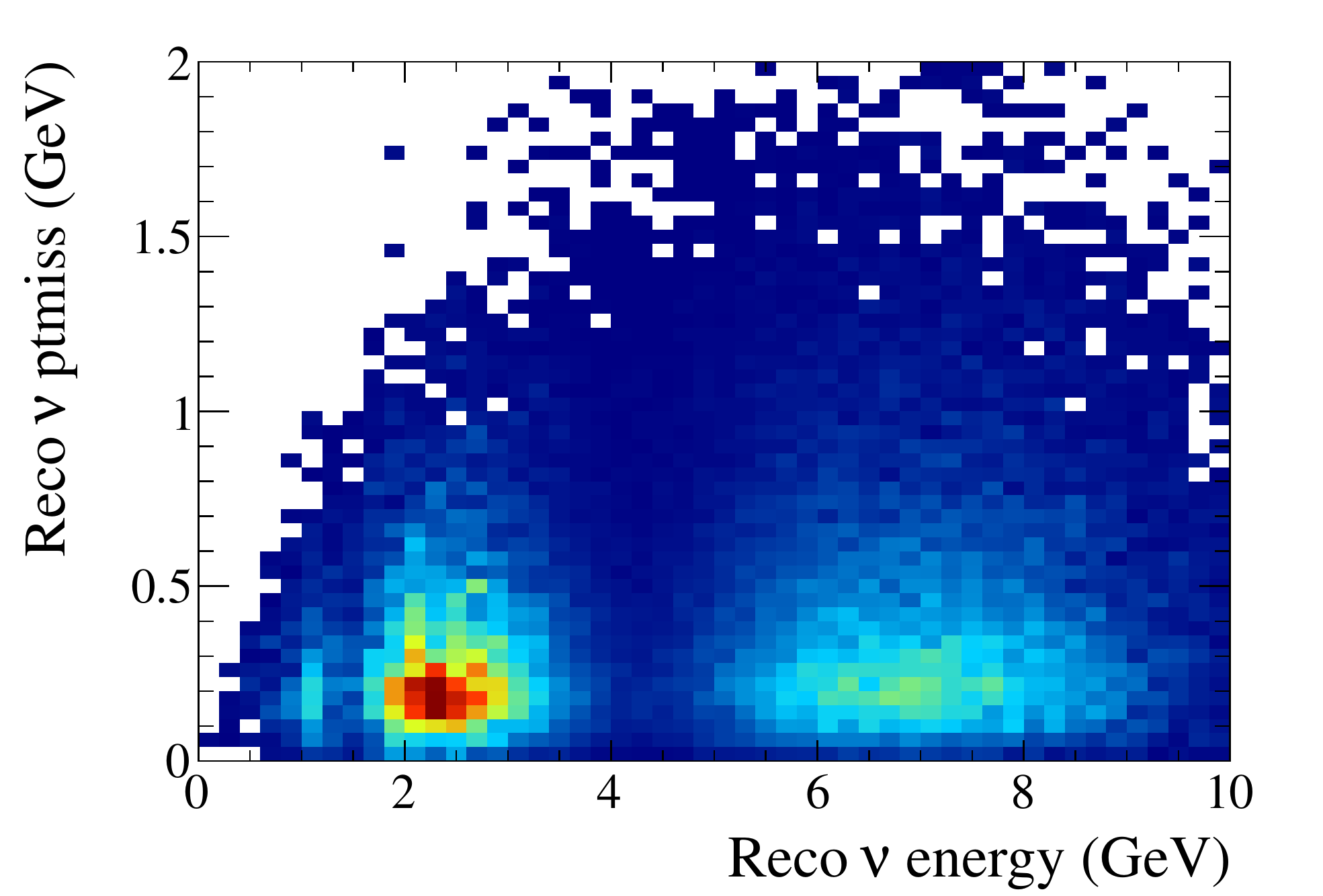}}
\end{tabular}
\caption{Example event distributions for various channels contributing to the e-like sample for $\delta_{CP} = 0$ and the case of the normal mass hierarchy.}
\label{fig:elikedist}
\end{figure}

We use reconstructed neutrino energy $E_{\nu}^{rec}$ and missing momentum in the transverse plane, defined by the incoming neutrino beam direction, $p_T^{miss}$ of each e-like event to construct bi-dimensional distributions useful to discriminate signal from background. Examples of such distributions are shown in Figure~\ref{fig:elikedist} for a value of $\delta_{CP} = 0$ and the case of normal mass hierarchy. As can be seen in the Figure, the shapes of signal and background contributions in the $E_{\nu}^{rec}-p_T^{miss}$ phase-space differ. 
%%%% added
In particular, NC $\pi^0$ interactions are characterized by low $E_{\nu}^{rec}$ values, while events originating from $\nu_\tau$ CC interactions tend to have larger $p_T^{miss}$ than the $\nu_e$ CC events because of the two neutrinos in the final state.
%%%%
This allows a better signal-background discrimination against neutral currents and tau charged current events, than if one were to use $E_\nu^{rec}$ information only. In the future, a cut-based or neural network analysis could be employed to further improve the purity of the e-like sample. In the case of the $\mu$-like events, only the reconstructed neutrino energy is used, since backgrounds
are less severe.

\subsection{Analysis method}
\label{subsect:anameth}

To evaluate the physics potential of the experiment it is mandatory do develop a sophisticated analysis package, 
which takes into account all the available experimental information and the sources of systematic uncertainties. 
The LBNO collaboration has developed such specific tools. 

We perform a fit of the oscillation parameters by minimizing the following $\chi^2$ with respect to the oscillation parameters $\mathbf{o}$ that are not fixed and systematic parameters $\mathbf{f}$  (Tables~\ref{tab:cuts} and~\ref{tab:cuts2}):
\begin{equation}
\chi^2 = \chi^2_{appear} + \chi^2_{disa} + \chi^2_{syst}.
\label{eq:totchi2}
\end{equation}

The $\chi^2_{appear}$ is the term corresponding to the electron appearance. It is given by: 
\begin{equation}
\begin{split}
\chi^2_{appear} = 2\sum_{+/-}\sum_{E_\nu^{rec},p_T^{miss}} & n_{e}(E_\nu^{rec},p_T^{miss}; \mathbf{o}_{test}, \mathbf{f}_{test}) - n_{e}(E_\nu^{rec},p_T^{miss}; \mathbf{o}_{true}, \mathbf{f}_{true}) \\
& + n_{e}(E_\nu^{rec},p_T^{miss}; \mathbf{o}_{true}, \mathbf{f}_{true})\ln{\frac{n_{e}(E_\nu^{rec},p_T^{miss}; \mathbf{o}_{true}, \mathbf{f}_{true})}{n_{e}(E_\nu^{rec},p_T^{miss}; \mathbf{o}_{test}, \mathbf{f}_{test})}},
\end{split}
\label{eq:chi2appear}
\end{equation}
where the subscript \textit{true} (\textit{test}) refers to the true (test) values of the $\mathbf{o}$ and $\mathbf{f}$ parameters. The \textit{true} parameters are those chosen by Nature, while \textit{test} refer to the parameter at which we compute the likelihood with respect to the \textit{true} value.
The first sum  in Eq.~\ref{eq:chi2appear} corresponds to adding the contributions from the neutrino and anti-neutrino beam running. The number of the e-like events in a given $E_\nu^{rec}-p_T^{miss}$ bin is determined according to:
\begin{equation}
\begin{split}
n_{e}(E_\nu^{rec},p_T^{miss}; \mathbf{o}, \mathbf{f}) &= f_{sig}n_{e-sig}(E_\nu^{rec},p_T^{miss};\mathbf{o}) \\ &+ f_{\nu_e}n_{\nu_e}(E_\nu^{rec},p_T^{miss};\mathbf{o}) + f_{\nu_\tau} n_{e,\nu_\tau}(E_\nu^{rec},p_T^{miss};\mathbf{o}) \\ &+ f_{NC}(n_{NC\pi^0}(E_\nu^{rec},p_T^{miss};\mathbf{o})+n_{mis-\nu_\mu}(E_\nu^{rec},p_T^{miss};\mathbf{o})),
\end{split}
\label{eq:nelike}
\end{equation}
where $n_{e-sig}$, $n_{\nu_e}$, $n_{e,\nu_\tau}$, $n_{NC\pi^0}$, and $n_{mis-\nu_\mu}$ are the number of events for signal, intrinsic beam $\nu_e$, electrons from tau decay, neutral current, and mis-identified $\nu_\mu$, respectively.

The information from the disappearance channel is contained in the $\chi^{2}_{disa}$ term of total $\chi^{2}$ in Eq.~\ref{eq:totchi2}, which is given by
\begin{equation}
\begin{split}
\chi^2_{disa} = 2\sum_{+/-}\sum_{E_\nu^{rec}} & n_{\mu}(E_\nu^{rec}; \mathbf{o}_{test}, \mathbf{f}_{test}) - n_{\mu}(E_\nu^{rec}; \mathbf{o}_{true}, \mathbf{f}_{true}) \\
& + n_{\mu}(E_\nu^{rec}; \mathbf{o}_{true}, \mathbf{f}_{true})\ln{\frac{n_{\mu}(E_\nu^{rec}; \mathbf{o}_{true}, \mathbf{f}_{true})}{n_{\mu}(E_\nu^{rec}; \mathbf{o}_{test}, \mathbf{f}_{test})}}.
\end{split}
\label{eq:chi2disa}
\end{equation}
The number of $\mu$-like events is the sum of signal ($n_{\mu-sig}$) and $\tau\rightarrow \mu$ background ($n_{\mu,\nu_\tau}$) contributions and is calculated as
\begin{equation}
n_{\mu}(E_\nu^{rec}; \mathbf{o}, \mathbf{f}) = f_{sig} n_{\mu-sig}(E_\nu^{rec};\mathbf{o}) + f_{\nu_\tau} n_{\mu,\nu_\tau}(E_\nu^{rec};\mathbf{o}).
\label{eq:nmulike}
\end{equation}

The oscillation and the systematic parameters are constrained through the $\chi^2_{syst}$ term:
\begin{equation}
\chi^2_{syst} = \sum_i \frac{(a_{0,i} - a_i)^2}{\sigma_{a_i}^2},
\end{equation}
where $a_{0,i}$ ($a_i$) is the prior (test) value of the \textit{i\textsuperscript{th}} parameter and $\sigma_{a_i}$ is the corresponding prior uncertainty. As will be shown in Section~\ref{sec:assumptions}, we use 6 priors for the neutrino oscillation parameters, and
4 for the normalisation uncertainties of signal and background.

In order to study the sensitivity of LBNO to CP violation, we define a test statistic $\Delta\chi^{2}$
\begin{equation}
\Delta\chi^{2} = \chi^{2}_{\delta_{CP}} - \chi^{2}_{best},
\label{eq:dchi2cpv_ts}
\end{equation}
where $\chi^{2}_{\delta_{CP}}$ is the minimized $\chi^2$ of Eq.~\ref{eq:totchi2} at a fixed value of $\delta_{CP}$ (true or test), while $\chi^{2}_{best}$ is the minimum $\chi^2$ obtained when $\delta_{CP}$ is allowed to vary over the full range of possible values. 

To evaluate sensitivity of the experiment to MH, we define the following test statistic $T$:
\begin{equation}
\label{eq:mhdeter}
T = \chi^2_{IH} - \chi^2_{NH}
\end{equation}
where $\chi^2_{IH}$ ($\chi^2_{NH}$) is obtained by minimizing the $\chi^{2}$ of Eq.~\ref{eq:totchi2},
marginalising with respect to systematic and oscillation parameters (including $\delta_{CP}$) around the negative (positive) $\Delta m^2_{31}$. %In the case of the IH, on average one expects $\chi^2_{IH} <  \chi^2_{NH}$ and therefore $T$ would favour the negative values. One the other hand, the opposite is true if the hierarchy is NH.  
The value of $T$ depends on $\delta_{CP}$.

The statistical method to determine MH and CPV is described in detail in the next section.

\subsection{Statistical approach to MH and CPV determination}
\label{sec:stat}
The sensitivity of an experiment to MH and CPV can be defined using the frequentist approach to the test of simple hypothesis, 
which we will review briefly.

The aim is to establish a criterion to assess that a ``null hypothesis'' $H_0$ is considered true 
and that an ``alternative hypothesis'' $H_1$ can be rejected. 
One chooses a test statistic $T$ that will be computed from the experimental data. 
This quantity is a stochastic variable with a certain probability density function ($PDF$) 
which should be different whether $H_0$ or $H_1$ is true. 
%One then decides on a critical value of $T$, $T_C$, which sets the confidence level at which $H_0$ 
%is accepted as the true hypothesis. 
One then defines a ``critical region'' such that, when $T$ has values within this region, $H_0$ is accepted as the 
true hypothesis. 
The probability to obtain a value of $T$ outside the critical 
region %defined by $T_C$ 
and consequently to reject $H_0$ even though it is true (``type I'' error or ``loss'') 
is usually denoted as $\alpha$. The confidence level ($CL$) with which one accepts $H_0$ as true is therefore
\begin{equation}
CL = 1 - \alpha.
\end{equation}
In addition, one must also consider the probability that $H_0$ is accepted as true even though $H_1$ is true
(``type II'' error or ``contamination''). 
This probability is usually denoted as $\beta$. The power of the test $p$, namely the probability of rejecting $H_0$ when it is false, is then 
\begin{equation}
p = 1 - \beta.
\end{equation}

To calculate the $CL$ and the power of the test, one has to know the respective $PDF$s of $T$ for $H_0$ and $H_1$. 
It has recently been pointed out~\cite{Qian:2012zn,Ge:2012wj} that for the case of MH and $T$ as in Eq.~\ref{eq:mhdeter}, the $PDF$ can be approximated by a Gaussian whose width is related to its mean $T_0$ as $\sigma=2\sqrt{T_0}$. Since in general the values of $T_0$ for NH and IH could be different, one has
\begin{equation}
\label{eq:pdfmh}
\begin{split}
& PDF(T|NH) = N(T_0^{NH},2\sqrt{T_0^{NH}}) \\
& PDF(T|IH) = N(-T_0^{IH},2\sqrt{T_0^{IH}}).
\end{split}
\end{equation}
For the long baseline oscillation experiments whose sensitivity comes primarily from the electron appearance
channel, $T_0^{NH}$ and $T_0^{IH}$ also have a strong dependency 
on the presently unknown CPV phase $\delta_{CP}$. We assume that also in the future MH will be determined
without precise knowledge on $\delta_{CP}$, hence study the problem as a function of the assumed
true value of $\delta_{CP}$.

\begin{figure}[tb]
\begin{center}
\includegraphics[width=0.49\textwidth]{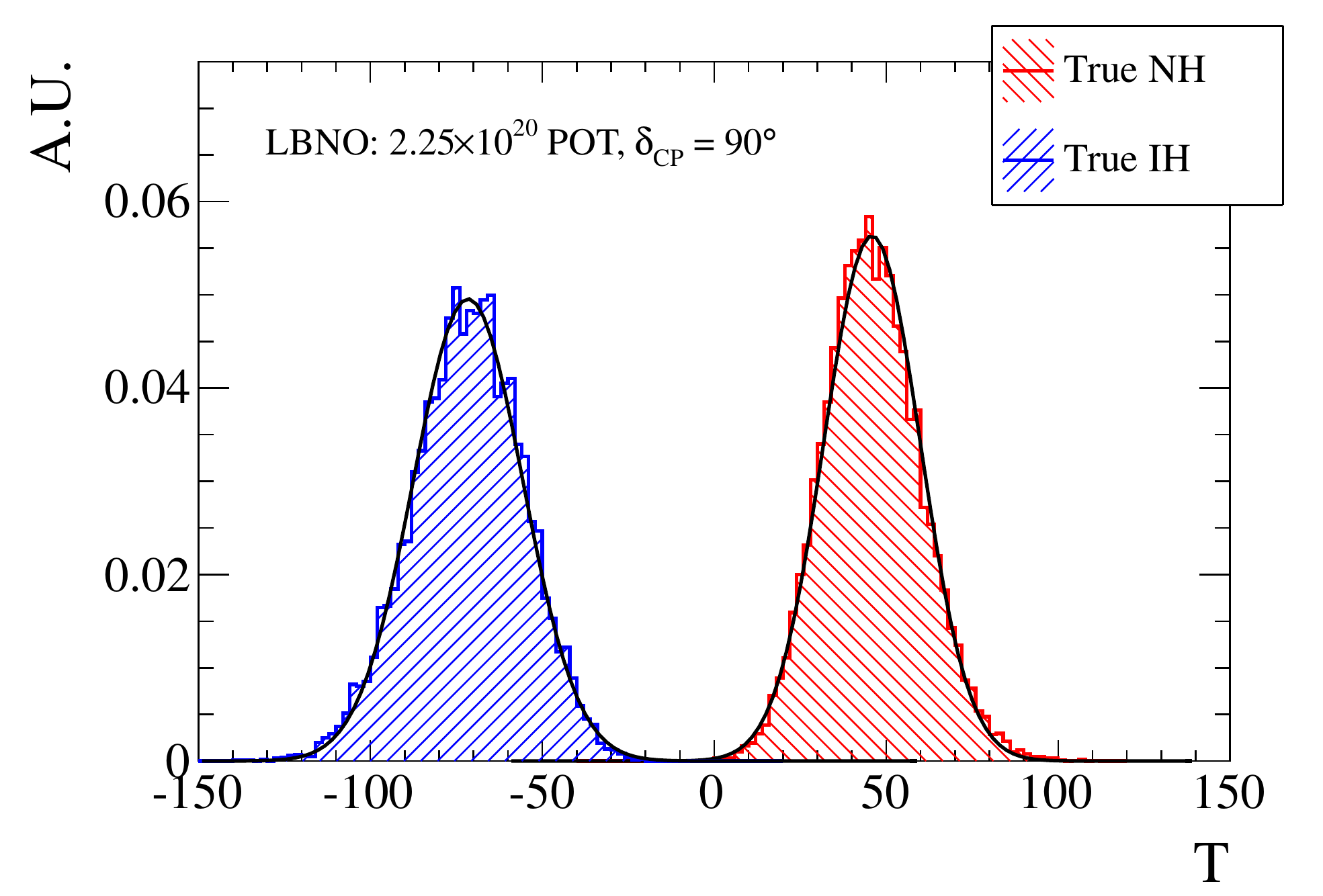} 
\includegraphics[width=0.49\textwidth]{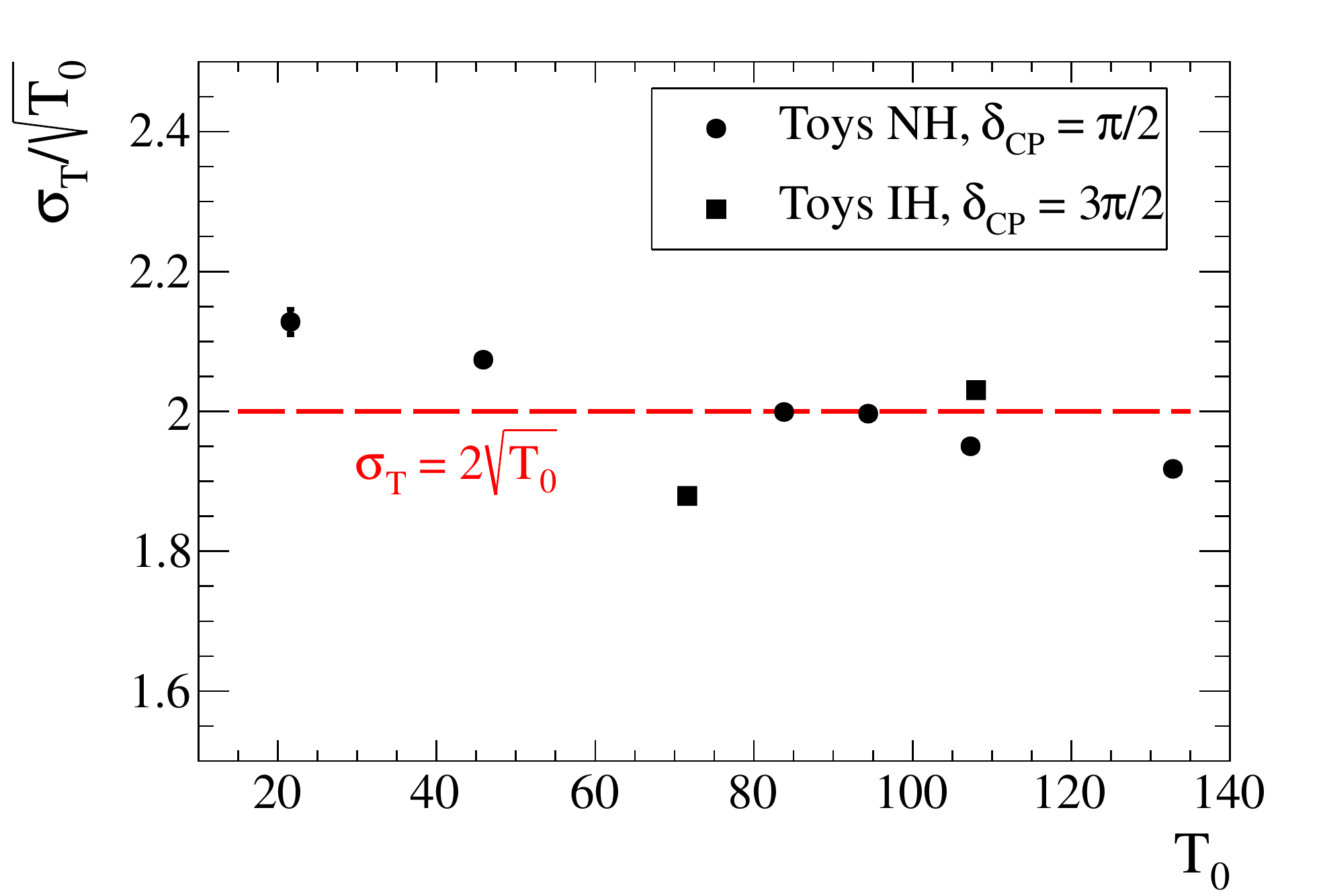} 
\caption{\small 
Distribution of the $T$ test statistic for $NH$ and $IH$ for an exposure of $2.25\times 10^{20}$ POT (or approx.
2 years of running) and $\delta_{CP} = \pi/2$ (left) and width of the distributions as a function of the average value $T_0$ (right).}
\label{fig:toymc}
\end{center}
\end{figure}

To verify that Eqs.~\ref{eq:pdfmh} are applicable for the LBNO case, 
we generated 20,000 toy data samples for different exposures, several values of  $\delta_{CP}$, and both hierarchy cases. Each data sample was then analyzed and $T$ was calculated according to Eq.~\ref{eq:mhdeter}. The distributions of T values for NH and IH obtained assuming $\delta_{CP} = \pi/2$ and a total exposure of $2.25\times 10^{20}$ POT ($50\%\nu:50\%\bar{\nu}$ running), or about two years of data taking, are shown in the left panel of Figure~\ref{fig:toymc}. Each distribution is fitted with a Gaussian function to determine the mean $T_0$ and the width $\sigma_{T}$. The right panel of Figure~\ref{fig:toymc} shows the ratio of $\sigma_T/\sqrt{T_0}$ as a function of $T_0$. As can be seen, the results from toy distributions deviate in some cases from the $\sigma_T=2\sqrt{T_0}$ rule. The effect of these deviations is, however, expected to be small and therefore we will subsequently assume that $PDF$ for $T$ is given by Eqs.~\ref{eq:pdfmh}.

To present the statistical treatment in the case of the MH determination, we begin by considering NH as our $H_0$. 
If one wants to achieve a given $CL$ for determining of the MH, 
we should consider values of $T$ larger than a given critical value as suggesting NH to be true. 
We should thus define a critical value $T_C^{\alpha}$, depending on the corresponding $\alpha$, such that
\begin{equation}
1-\alpha = \int_{T_C^\alpha}^\infty PDF(T|NH) \ dT 
\end{equation}
gives a desired CL. The power of the test is given by
\begin{equation}
p = 1-\beta = 1-\int_{-\infty}^{T_C^\alpha} PDF(T|IH) \ dT.
\end{equation}
The procedure is formally similar for IH as $H_0$.

As pointed out in~\cite{Blennow:2013oma}, if the $PDF$s of $T$ follow the distributions of equation~\ref{eq:pdfmh}, the critical value and the power become simple analytically functions of the chosen $\alpha$. For the case $H_0=NH$, we have
\begin{equation}
\label{eq:tcalpha}
T_C^\alpha = T_0^{NH}-\sqrt{8T_0^{NH}}\cdot \operatorname{Erfc}^{-1}(2\alpha)
\end{equation}
and
\begin{equation}
\label{eq:beta}
p = 1-\beta = 1- \frac{1}{2} \operatorname{Erfc}\left(\frac{T_0^{IH}+T_C^\alpha}{\sqrt{8T_0^{IH}}}\right)\quad\quad (\mathrm{for\,MH}).
\end{equation}
The calculation must be repeated for the case $H_0=IH$, with $T_0^{NH}$ and $T_0^{IH}$ interchanged.

The values of $T_0^{NH}$ and $T_0^{IH}$ depend on the unknown $\delta_{CP}$. 
As explained in~\cite{Blennow:2013oma}, one should therefore set $T_C^\alpha$ in the most pessimistic case, 
i.e. for the smallest absolute value of $T_0$ in either case. As shown later, 
this corresponds to choosing $T_0^{NH}$ ($T_0^{IH}$) for $\delta_{CP} = \pi/2$ ($\delta_{CP} = 3\pi/2$). 
We thus compute $T_{C}^\alpha$ as
\begin{equation}
\begin{split}
 & T_C^\alpha(T_0^{NH}(\delta_{CP}=\pi/2)) \mbox{~for~} H_0=NH,\\
 & T_C^\alpha(T_0^{IH}(\delta_{CP}=3\pi/2)) \mbox{~for~} H_0=IH.
\end{split}
\end{equation}
Once $T_C^\alpha$ is fixed, the statistical power will be a function of $\delta_{CP}$. For the test of $NH$, 
it will be largest for $\delta_{CP}=\pi/2$ and smallest for $\delta_{CP}=3\pi/2$ (where $T_0^{IH}$ is smallest), while the opposite is true for $IH$.

\begin{figure}[tb]
\begin{center}
\includegraphics[width=0.49\textwidth]{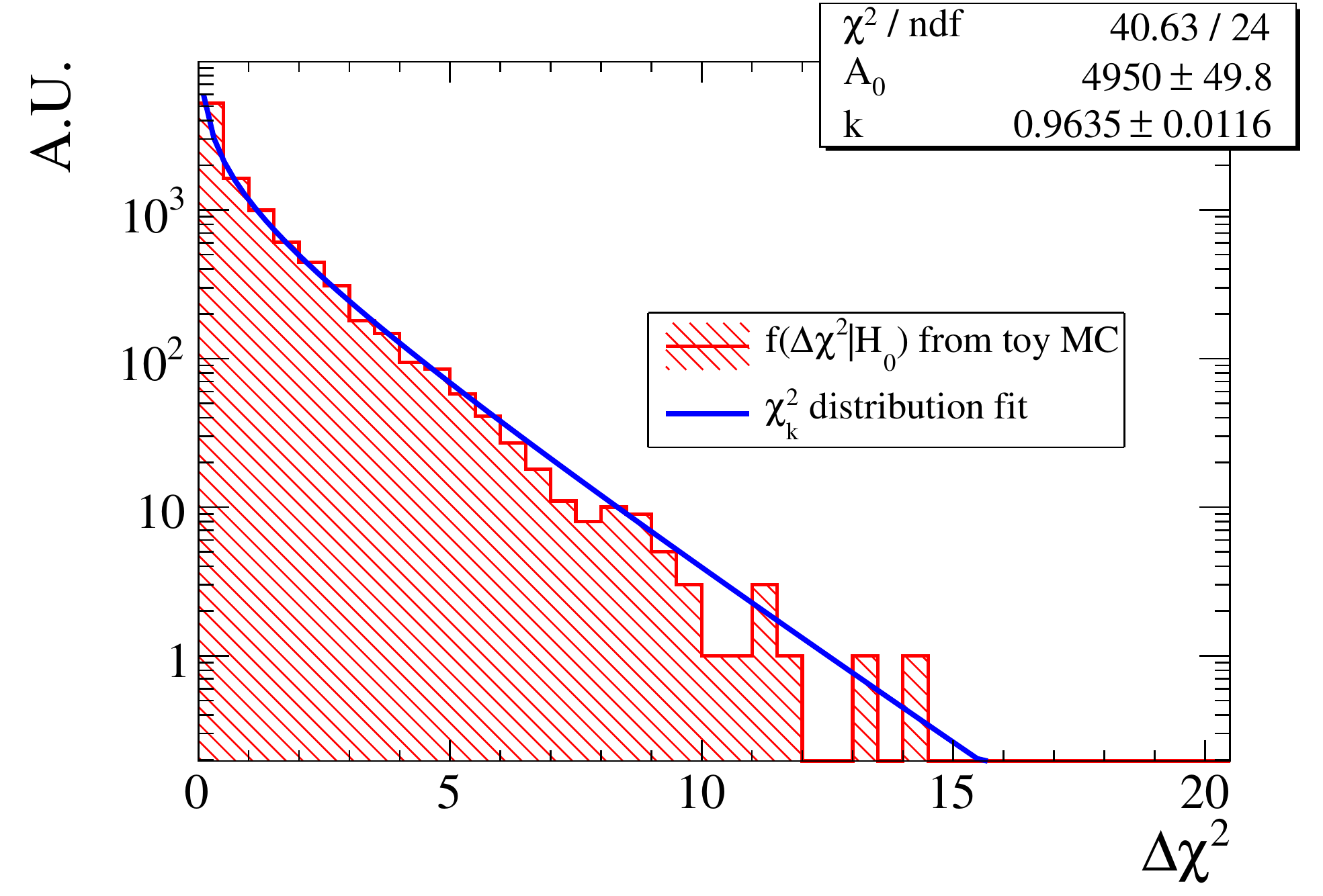} 
\includegraphics[width=0.49\textwidth]{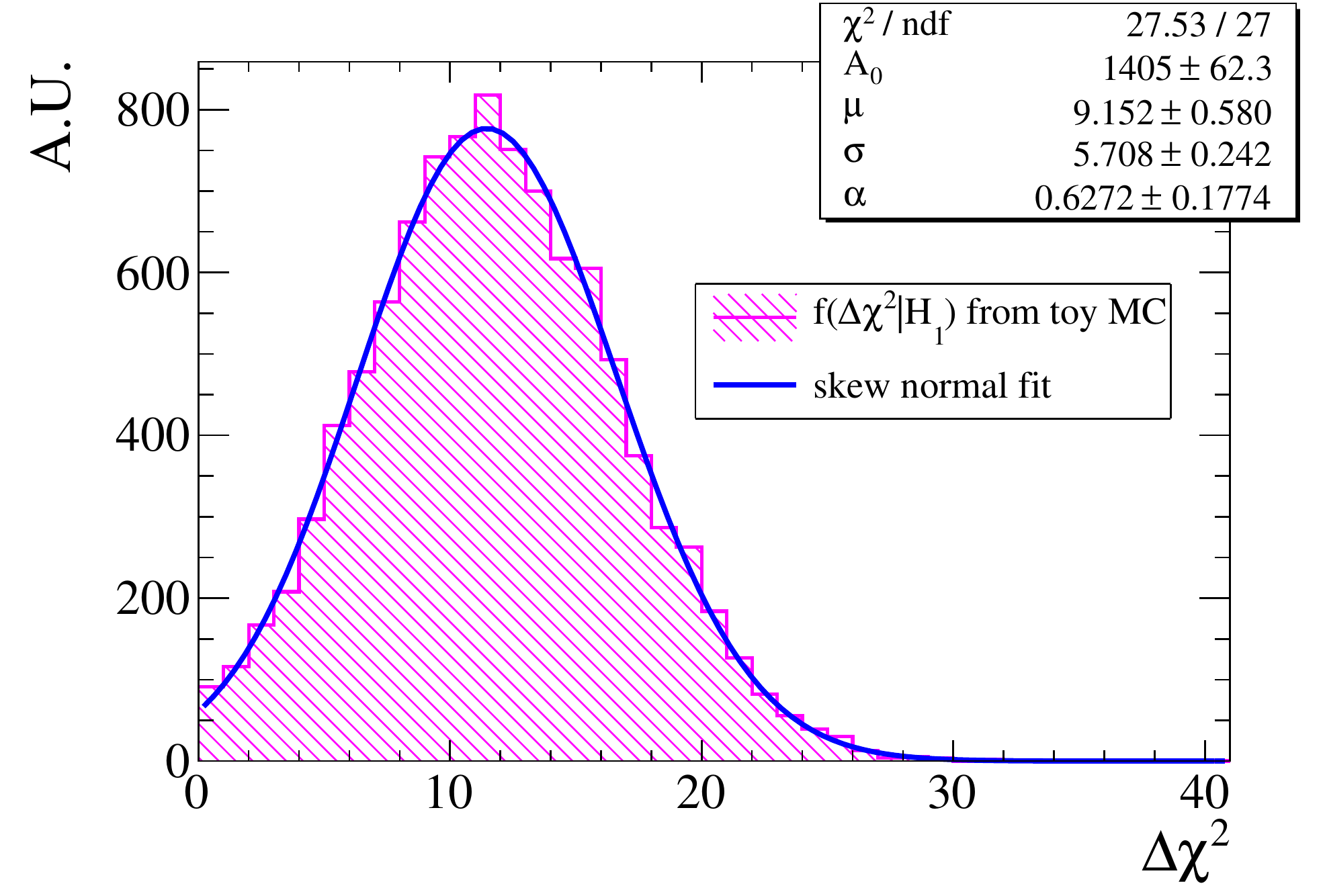} 
\caption{\small 
Distribution of the $\Delta\chi^2$ test statistic for $\delta_{CP}=0$ (left) and
$ \delta_{CP}=\pi/2$ (right) for an exposure of $15\times 10^{20}$ POT.
\label{fig:toymcCPV}}
\end{center}
\end{figure}

A similar approach was used to study the $\delta_{CP}$ phase determination. 
In this case, $H_0$ corresponds to $\delta_{CP}=0$ or $\pi$ and $H_1$ to ${0<\delta_{CP}<2\pi}$
(both hypotheses are composite). We are in the case of ``nested hypotheses''.
The distribution of the $\Delta\chi^{2}$ test statistic computed according to 
Eq.~\ref{eq:dchi2cpv_ts} follows a $\chi^2$ distribution with 1 d.o.f. when the null hypothesis is true,
as shown in Fig.~\ref{fig:toymcCPV}(left), and is independent of the exposure.
In the case of the alternative hypothesis, the distribution, shown in Fig.~\ref{fig:toymcCPV}(right) for the case of $\delta_{CP} = \pi/2$, is obtained using toy Monte Carlo simulations. To construct it, we generate toy data samples assuming the true values for $\delta_{CP}$ of 0 and $\pi$. We then compute the $\Delta\chi^{2}$ statistic as defined in Eq.~\ref{eq:dchi2cpv_ts} for a given test value of $\delta_{CP}$ for both cases, i.e,
\begin{equation}
\begin{split}
&\Delta \chi^2_0 = \chi^2(\delta_{CP}^{test}|\delta_{CP}^{true}=0)   - \chi^2_{best}(|\delta_{CP}^{true}=0) \\
&\Delta \chi^2_{\pi} = \chi^2(\delta_{CP}^{test}|\delta_{CP}^{true}=\pi) - \chi^2_{best}(|\delta_{CP}^{true}=\pi).
\end{split}
\end{equation}
Finally, we take the smallest of the two $\Delta\chi^2$s:
\begin{equation}
\label{eq:dchi2cpv}
\Delta \chi^2 = \min{(\Delta\chi^2_0, \Delta\chi^2_\pi)}.
\end{equation}
For computational purposes we approximate the resultant distribution analytically with a skew normal distribution. It should be noted that the average of this distribution defines the sensitivity for a given value of $\delta_{CP} \neq 0, \pi$ and this is the quantity that will be shown in the figures of section~\ref{sec:cpv}. 

Hence, for the CPV discovery, the null hypothesis will be accepted at a given $CL$ when
the value of $\Delta\chi^{2}$ is below a critical 
computed from the quantiles of the $\chi^2$(1 d.o.f.) distribution. 
The power of the test is the integral of the skew normal distribution above the critical value.

\subsection{Assumption on parameters and systematics}
\label{sec:assumptions}
\begin{figure}[!htb]
\begin{center}
\mbox{
\includegraphics[angle=0,width=0.55\textwidth]{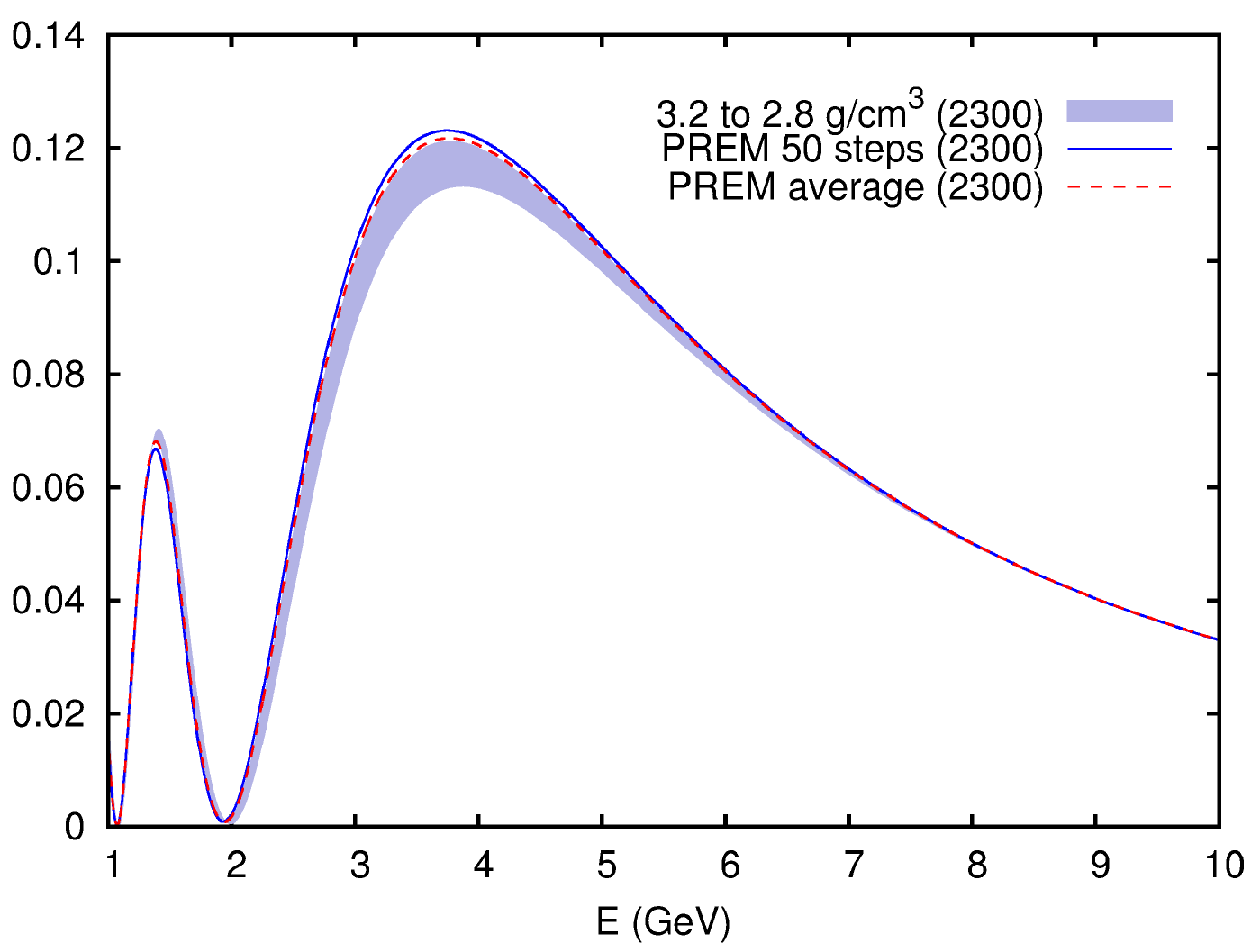}
}
\caption{\small{Oscillation $\nu_\mu\rightarrow\nu_e$ probability along
the CERN-Pyh\"asalmi baseline for (1) fixed average matter density varying
on the interval $3.2>\rho>2.8$~g/cm$^3$, (2) PREM 50 steps (3) PREM average.
In the band, the upper values correspond to a density of 3.2~g/cm$^3$.} }
\label{fig:matterdenseff}
\end{center}
\end{figure}

Assumptions on the oscillation parameters and uncertainties as well as on the beam line characteristics are shown in Table~\ref{tab:cuts}. 
Values take into account the results from ongoing experiments, in particular reactor experiments, and 
are based on the global analyses published in the literature.
Uncertainties are given at 1$\sigma$ and in percent. We have chosen to use the values available as of today, therefore our assumptions are conservative. In order to describe matter effects, we use a constant average density
approximation. We have compared the analytical oscillation probability obtained with the constant value to the one computed by integration of the oscillation amplitude in 50 steps through the Earth described by the Preliminary reference earth model (PREM)~\cite{Dziewonski:1981xy}.
As can be seen in Figure~\ref{fig:matterdenseff}, the assumed  value of 3.20~g/cm$^{3}$ describes best
the probability computed with the PREM. In the figure, the band corresponds to the oscillation probabilities
obtained by varying the density in the interval $3.2>\rho>2.8$~g/cm$^3$. The upper values of the band
are found for $\rho=3.20$~g/cm$^{3}$.

\begin{table}[ht]
\begin{center}
\begin{tabular}{cccc}
\hline
\hline
Name                                     &       Value                    &    error (1$\sigma$)   &  error (\%)   \\
\hline
\hline
L                  &  2300 km                                   & exact                      &     exact                  \\
\hline
$\Delta m^{2}_{21}$             & $7.6\times 10^{-5}$ $ eV^{2}$                                   & exact                        &      exact                \\
\hline
$ |\Delta m^{2}_{31} |\times 10^{-3}$ $eV^{2}$                & 2.420                                    & $ \pm  0.091    $                 &  $\pm     3.75$   \%                  \\  
\hline
$\sin^{2}  \theta_{12}$                                     & 0.31                                  & exact                         &       exact                 \\
\hline
$\sin^{2}2  \theta_{13}$                          & 0.10                                    &  $ \pm 0.01$                  &    $\pm    10  \%  $                \\
\hline
$\sin^{2}  \theta_{23}$                                & 0.440                                    &  $ \pm 0.044 $                  &     $\pm   10  \%  $                \\
\hline
Average density of traversed matter ($\rho$)              & 3.20 g/cm$^{3}$                                 &   $\pm 0.13$                          &        $ \pm 4  \%$                  \\
\hline
\hline                                                                                           

\hline
\end{tabular}
\end{center}
\caption{Assumptions on the values of the oscillation parameters and their uncertainties.}
\label{tab:cuts}
\end{table}

\begin{table}[!h]
\begin{center}
\begin{tabular}{ccc}
\hline
\hline 
Name  & Value        &       error (1$\sigma$)     \\
\hline
\hline 
Signal normalization ($f_{sig}$) & 1 &  $\pm 5\%$ \\
\hline
Beam electron contamination normalization ($f_{\nu_e}$) & 1 & $\pm 5\%$  \\
\hline
Tau normalization ($f_{\nu_\tau}$)   & 1            & $\pm 20\% - \pm 50\%$ \\  
\hline
$\nu$ NC and $\nu_{\mu}$ CC background ($f_{NC}$) & 1 & $\pm 10\%$  \\
\hline
\hline                                                                                           
\hline
\end{tabular}
\caption{Assumptions on event normalization uncertainties (bin-to-bin correlated errors).}
\label{tab:cuts2}
\end{center}
\end{table}

The assumptions on systematic errors on signal and background normalization are shown in Table~\ref{tab:cuts2}.
The systematic error on the tau normalization is set to 50\% for the mass hierarchy determination and to  20\% for the $\delta_{CP}$ sensitivity studies.  
This reduction is due to the fact that the experiment will be able to constrain $\nu_\tau$ cross section with the data accumulated during first few years of running performing  specific tau neutrino appearance channel measurements to constrain the production rate.

These errors are assumed to be fully correlated among the energy bins.

%%%%%%%%%%%%%%%%%%%%%%%%%%%%%%%%%%%%%%%%%%%%
%% MH measurement
%%%%%%%%%%%%%%%%%%%%%%%%%%%%%%%%%%%%%%%%%%%%
\section{Mass Hierarchy determination}
\label{sec:mass}

\begin{figure}[!ht]
\begin{center}
\mbox{
\includegraphics[angle=0,width=0.5\textwidth]{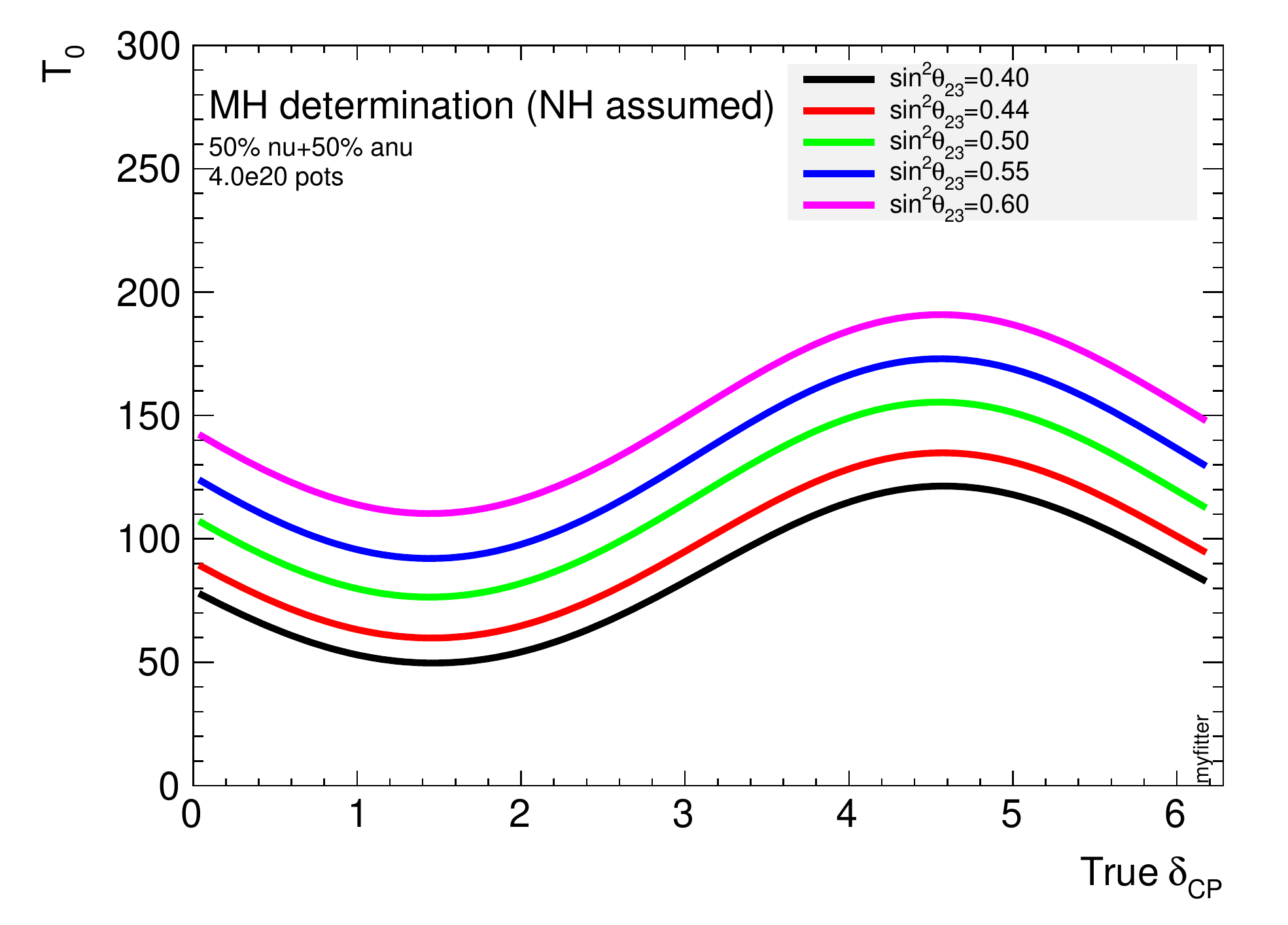}
\includegraphics[angle=0,width=0.5\textwidth]{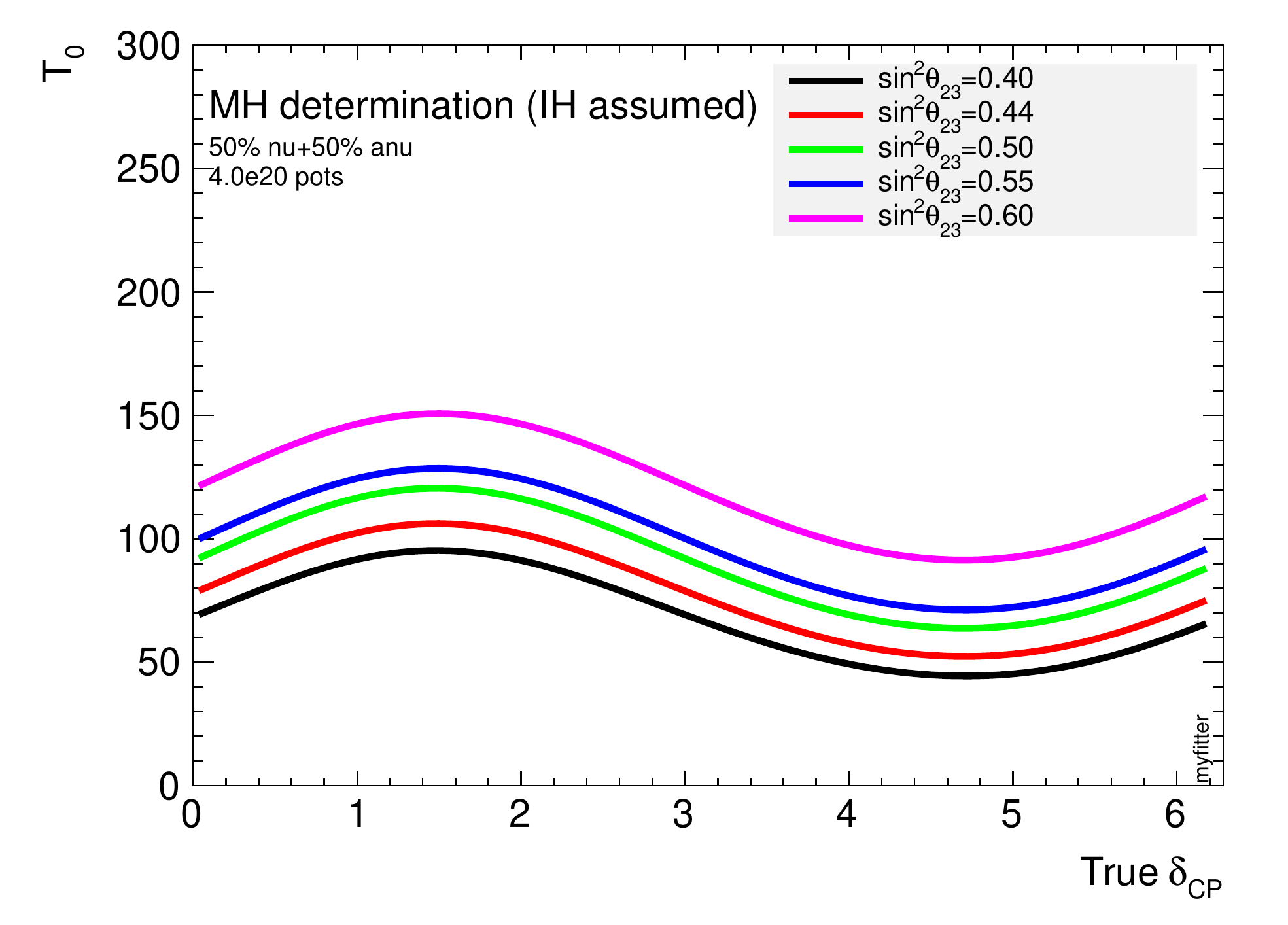}}
\caption{\small{Mean value of the mass hierarchy test statistic as a function of true $\delta_{CP}$ for
a total exposure of $4\times 10^{20}$~pots
(or about 4 years of running at the SPS) and LBNO 20~kton detector. 
Left: Normal Hierarchy assumed. Right: Inverted Hierarchy assumed. } }
\label{fig:mass}
\end{center}
\end{figure}

Mass hierarchy sensitivity studies have been developed using a 50\% $\nu$ 50\% $\bar{\nu}$ sharing, which is optimised for this measurement, and $4\times10^{20}$ POT, corresponding to about 4 years of nominal data taking with the SPS at 750~kW.   
The mean value $T_0$ of the test statistic $T$ defined in Eq.~\ref{eq:mhdeter} for mass hierarchy determination is shown in Figure~\ref{fig:mass} as a function of $\delta_{CP}$, for the assumed Normal and Inverted true mass hierarchy. 
%%%% added 
The parameter with the largest impact on the matter oscillation probabilities, and thus on the mass hierarchy determination, is the $\theta_{23}$ mixing angle.
%%%%
Figure~\ref{fig:mass}  shows also the effect of varying $\theta_{23}$ by a large amount in the two octants. This study proves that the present uncertainty on $\theta_{23}$ does not compromise the sensitivity %at $5 \sigma$ 
of our experiment.
%%%% added
The other sources of uncertainty considered above can only have a smaller impact.%%%%

\begin{figure}[!ht]
\begin{center}
\mbox{
\includegraphics[angle=0,width=0.5\textwidth]{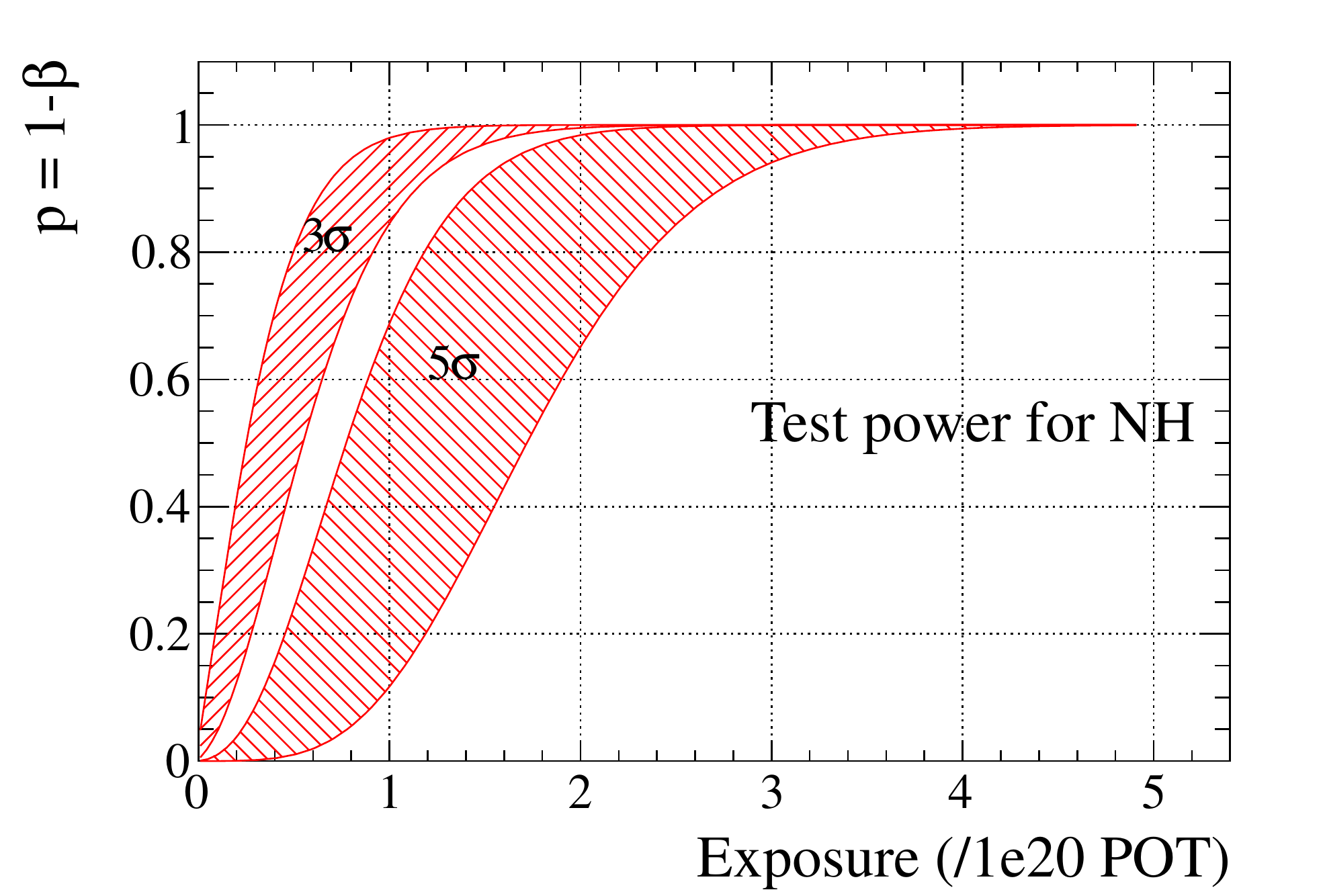}
\includegraphics[angle=0,width=0.5\textwidth]{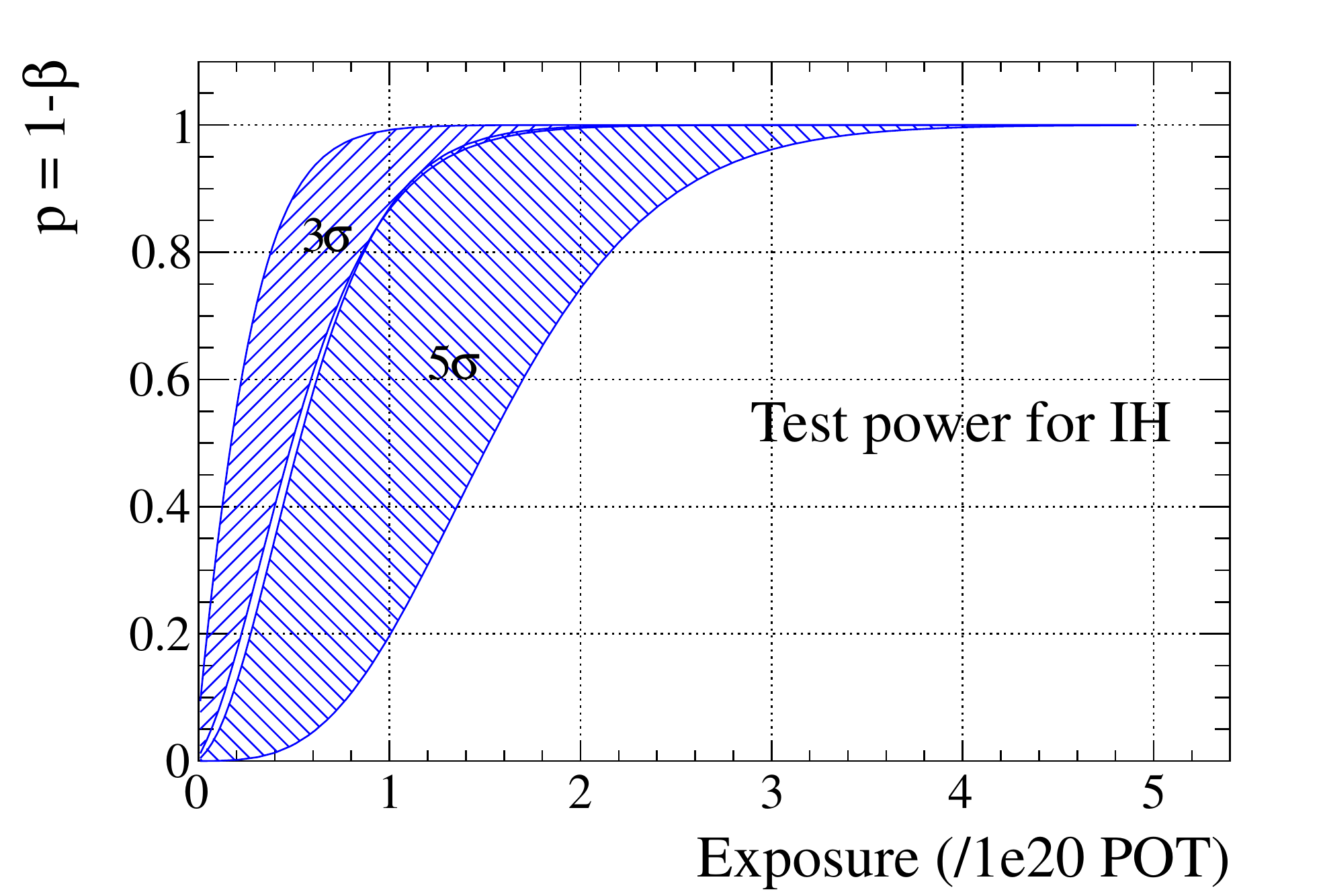}
}
\caption{\small{Statistical power as a function of exposure for the test of NH (left) and IH (right) for $3\sigma$ and $5\sigma$ CL. The nominal central values for oscillation parameters have been assumed and
the shaded bands correspond to the variation of $\delta_{CP}$. } }
\label{fig:mhpower}
\end{center}
\end{figure}

Following the statistical method discussed in section~\ref{sec:stat}, the power of LBNO for $NH$ and the $IH$ 
determination at a confidence level of $3\sigma$ or $5\sigma$ 
is shown in Figure~\ref{fig:mhpower} as a function of exposure.
The shaded area corresponds to the variation of $\delta_{CP}$ and the extreme values are reached 
for $\delta_{CP}=\pi/2$ or $3\pi/2$, as has been explained above. 
One can see that LBNO has a probability of essentially 100\% to discover the MH in either case for any value of $\delta_{CP}$. 
An exposure of slightly more that $2\times10^{20}$pot will guarantee that 
a $3\sigma$ CL is obtained, while a $5\sigma$ CL will be reached with less than $4\times10^{20}$pot, 
corresponding to about 4 years of SPS running.

%%%%%%%%%%%%%%%%%%%%%%%%%%%%%%%%%%%%%%%%%%%%
%% CP violation measurement
%%%%%%%%%%%%%%%%%%%%%%%%%%%%%%%%%%%%%%%%%%%%
%\section{CP violation measurement and the impact of oscillation parameters and their uncertainties}
\section{Study of the sensitivity to CP violation}
\label{sec:cpv}

\subsection{Beam focusing mode optimisation}

\begin{figure}[tb]
\begin{center}
\mbox{
\includegraphics[angle=0,width=0.85\textwidth]{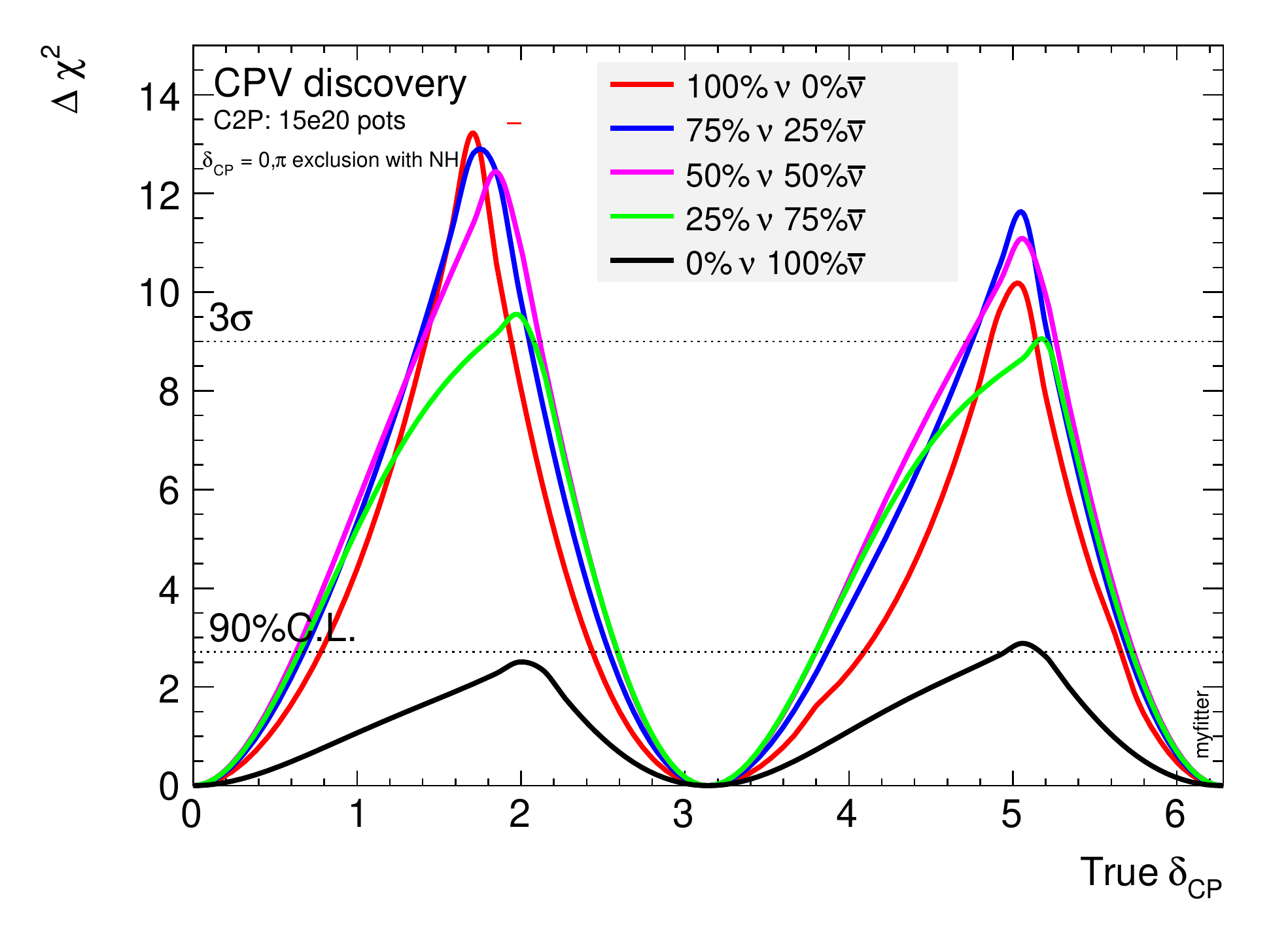}
}
\mbox{
\includegraphics[angle=0,width=0.81\textwidth]{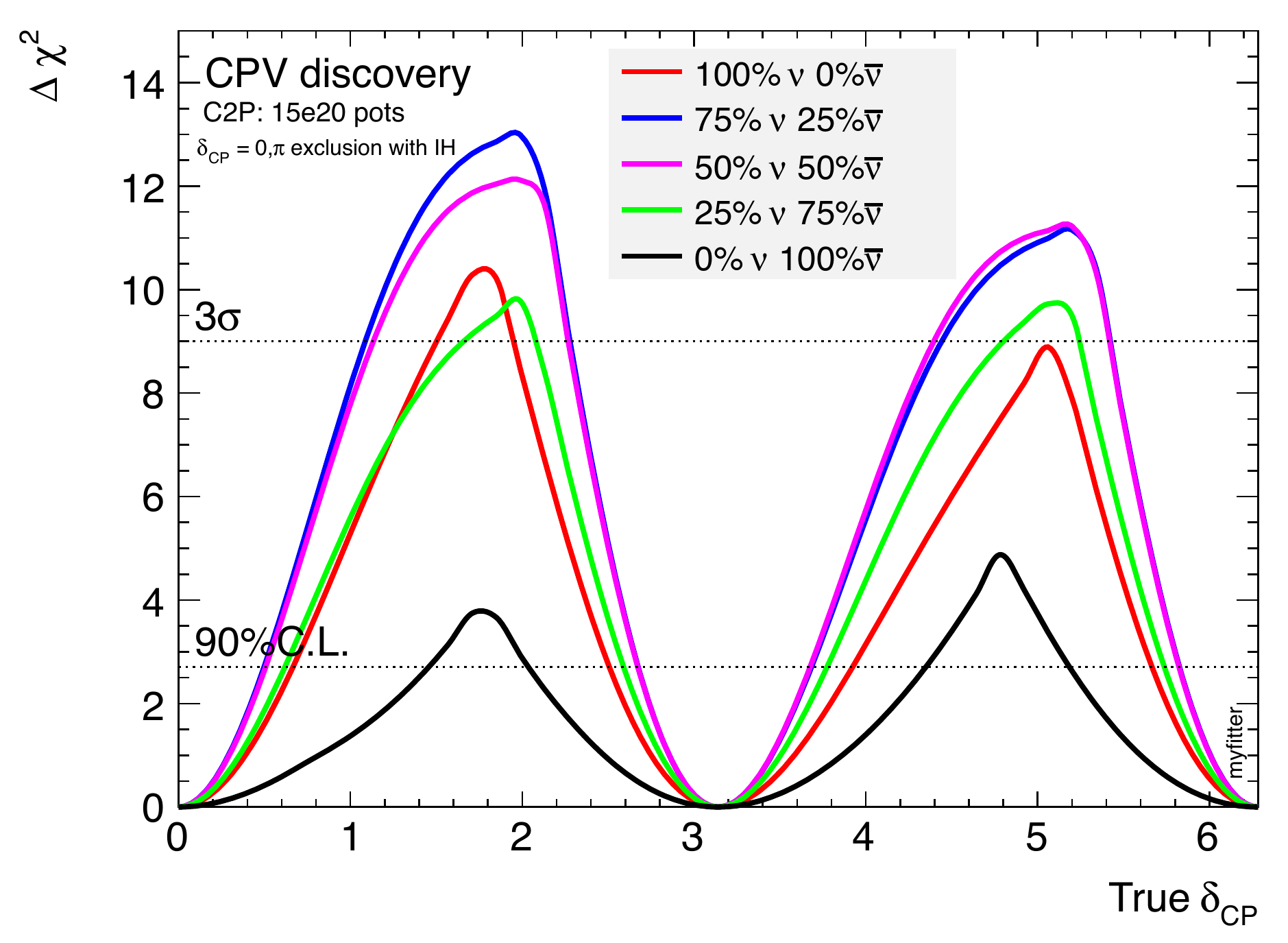}
}
\caption{\small{CPV sensitivity for different sharing between $\nu : \bar{\nu}$ running modes,
for $1.5$ $10^{21}$ protons on target and LBNO 20~kton detector. The upper plot
is for NH and the lower plot for IH.} }
\label{fig:sharing}
\end{center}
\end{figure}

For the CP phase measurement, the beam normalisation is set to $1.5$ $10^{21}$ protons on target (POT) (or
approximately 12~years of nominal running at the SPS),
and the optimisation of the beam sharing between $ \nu$ and $ \bar{ \nu}$ has been studied in detail. 
Figure~\ref{fig:sharing} shows the sensitivities for a non vanishing $ \delta_{CP}$ for the two mass hierarchies assuming different percentage of sharing assuming all the parameters in Table~\ref{tab:cuts} and~\ref{tab:cuts2}. 
Our simulations show a maximum of coverage in the case of 75 \% $ \nu$ - 25\% $ \bar{ \nu}$. 
This sharing will be assumed for all the studies presented in the next paragraphs.

\subsection{Significance of a first and second maxima analysis method}
\label{subsec:shapesignif}

\begin{figure}[h]
\centering
\includegraphics[width=0.85\textwidth]{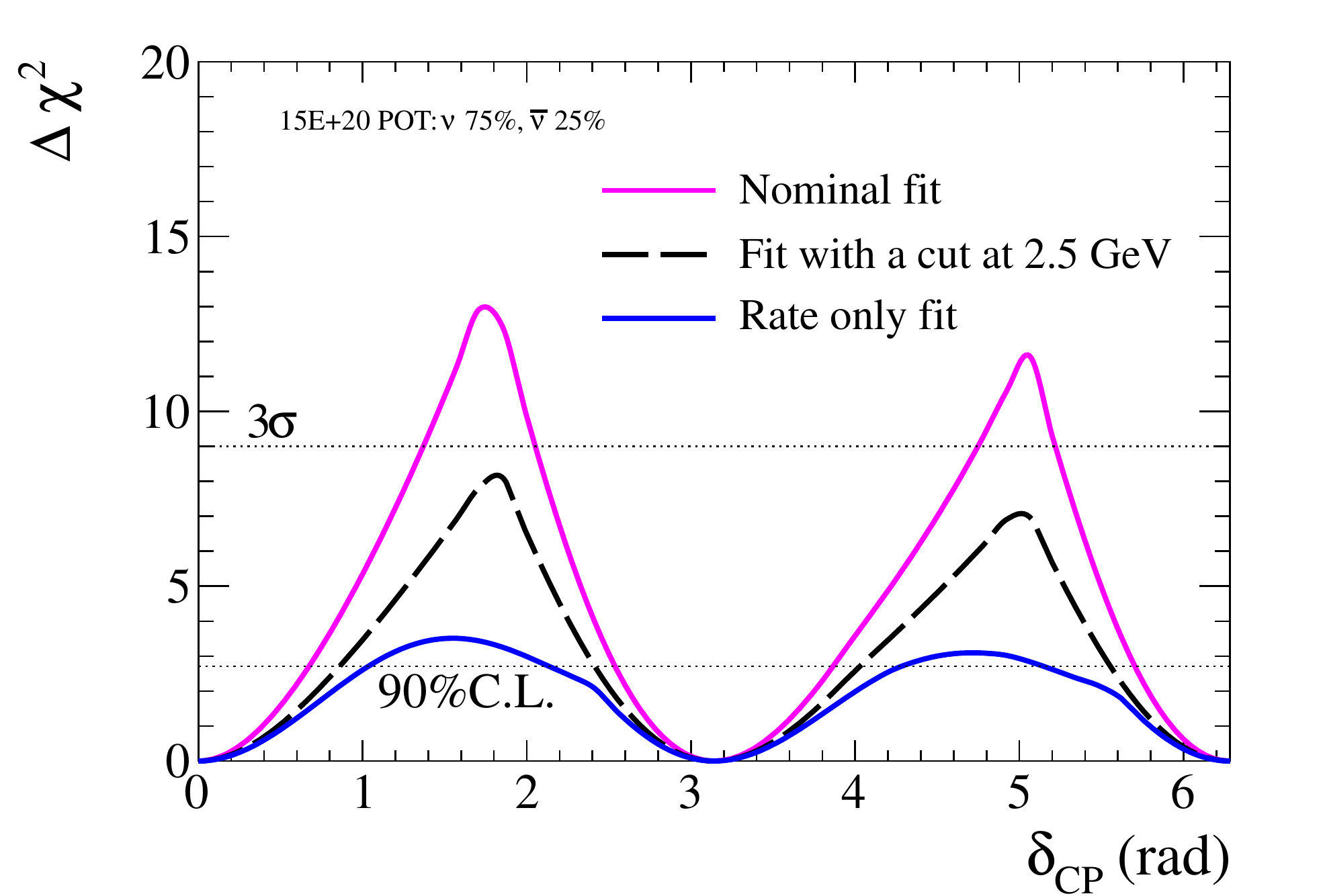}
\caption{Comparison of the CPV sensitivities of a rate only analysis, an analysis with a cut on a reconstructed energy at 2.5 GeV (excluding the $2^{nd}$ maximum), and the nominal case where the full event spectrum is used.}
\label{fig:oscshapestudy}
\end{figure}

The analysis method takes into account the information contained in the whole shape of the e-like event distributions in both the ranges of the  1\textsuperscript{st} and the 2\textsuperscript{nd}  oscillation maximum.
To consider both the oscillation maxima as well as the spectral shape is a very powerful method to extract $\delta_{CP}$ and to confirm the oscillatory behaviour predicted in the three neutrino oscillation schema together with matter effects. This approach is the only one proposed to put in evidence the theoretical framework of oscillations as a whole.
In fact, the spectrum shape as well as the number of events strongly depend on the value of $\delta_{CP}$ in particular in the energy region corresponding to the 2\textsuperscript{nd} maximum. 
We have compared the significance of our standard method to a first maximum only and a rate only analysis. 
The study of the significance of the events around the 2\textsuperscript{nd} oscillation maximum was done by evaluating the CPV sensitivity with a cut on the reconstructed energy of the e-like events placed at 2.5 GeV. This effectively removed all information about the 2\textsuperscript{nd} maximum from the e-like sample. 
In addition we have tested the importance of performing an analysis based on the e-like event distributions by a rate only analysis evaluation. 
The rate only measurement leads to a drastic loss of sensitivity of the experiment to the CPV. 
These studies are shown in Figure~\ref{fig:oscshapestudy}. 
The important quantity in this plot is the width of the interval
below the curve for a given confidence level, which tells us the fraction of
unknown parameter space for which we would be able to discover CP violation.
As can be seen in this plot, the rate only measurement leads to a drastic loss of sensitivity of the experiment to the CPV. The power of measuring events over an energy range that covers the 1\textsuperscript{st} and the 2\textsuperscript{nd} oscillation maxima is also evident.

\subsection{Impact of prior uncertainties on the  $\delta_{CP}$ discovery potential}
\label{sec:priors}

\begin{figure}[!h]
\begin{center}
\mbox{
\includegraphics[angle=0,width=0.85\textwidth]{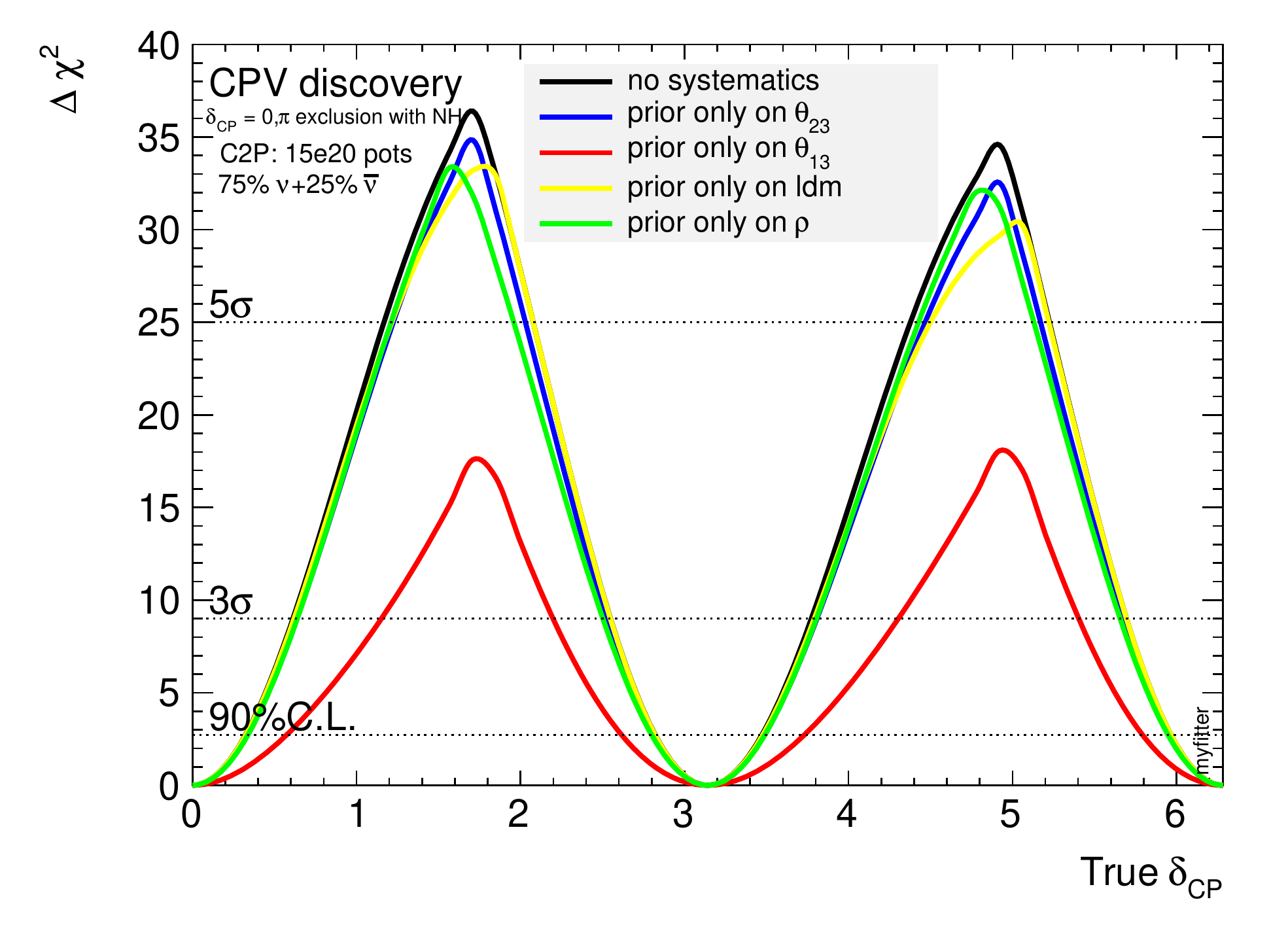}
}
\caption{\small{Impact of systematic errors: CPV sensitivity of LBNO phase I as a function
of $\delta_{CP}$, with only statistical  and no systematic errors (black), and 
effect of each prior on the oscillation parameters (blue, red, yellow, green). }}
\label{fig:priors}
\end{center}
\end{figure}

The effects of the prior uncertainties on the oscillation parameters have been studied in detail. 
The CP phase space coverage has been evaluated setting one prior at time for each oscillation parameter according to Table~\ref{tab:cuts}.
This is shown in Figure~\ref{fig:priors}  where it is evident that the  priors with the largest impact is that on $\theta_{13}$.% and $\theta_{23}$.  

\begin{figure}[ht!]
\begin{center}
\mbox{
\includegraphics[angle=0,width=0.5\textwidth]{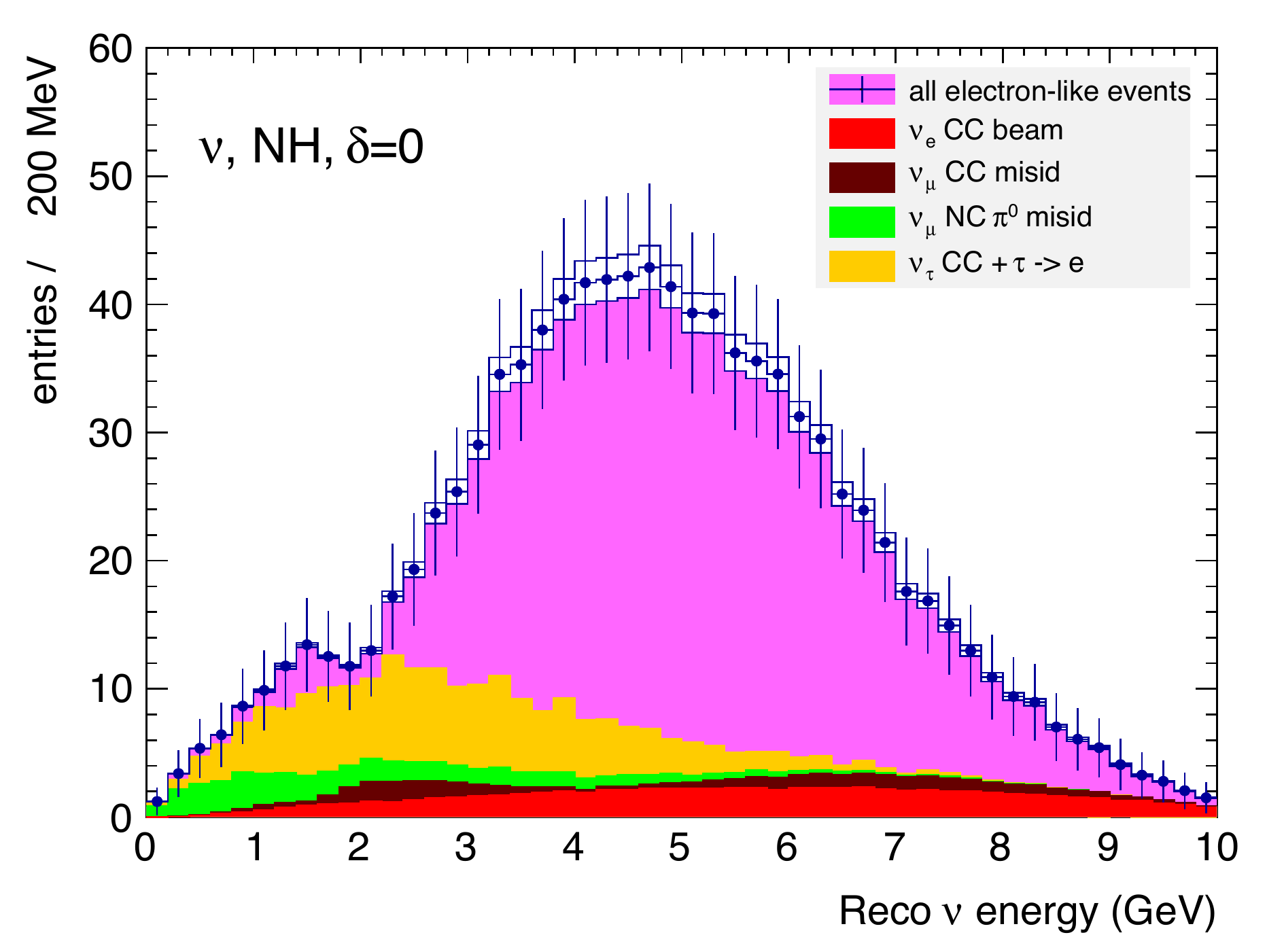}
\includegraphics[angle=0,width=0.5\textwidth]{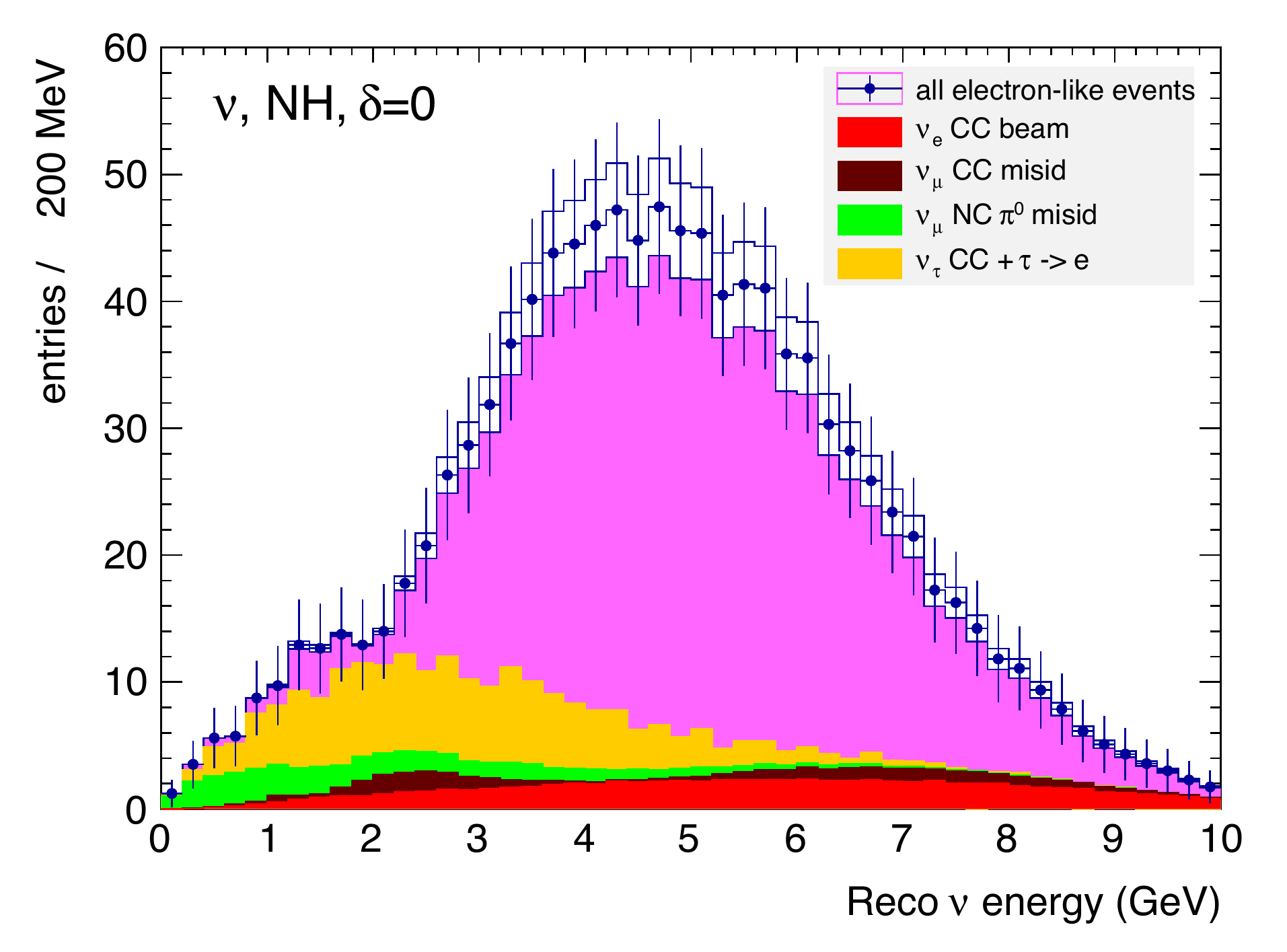}
}
\caption{\small{Measured e-like spectrum. Left: Maximum and minimum band if $\sin^2\theta_{23}$ is varied by $\pm1\sigma$. Right: Maximum and minimum band if $\sin^22\theta_{13}$ is varied by $\pm1\sigma$ in the appearance channel.} }
\label{fig:reco_priors}
\end{center}
\end{figure}

In Figure~\ref{fig:reco_priors} we show the effects on the expected electron neutrinos energy spectrum when values of  $\theta_{13}$ and $\theta_{23}$  are varied by $\pm 1\sigma$ for both the appearance and the disappearance channel. This effect is represented by a white band whose binned histograms limits correspond to the reconstructed energy spectrum assuming the minimum and the maximum value of the two mixing angles inside their $1\sigma$ range of variability. The statistical error for each bin is also shown. 
We would like to stress the $1\sigma$ effect represented by the white band is a fully correlated effect whereas the statistical error is bin-to-bin uncorrelated.
Each one of these effects (uncertainty on  $\theta_{13}$ and  $\theta_{23}$) is thus 
more important than the statistical error in each bin which fluctuates in both directions around the average
bin value represented in the histogram.

\subsubsection{Influence of $\theta_{13}$ on the  $\delta_{CP}$ discovery potential}
\label{sec:theta13}

\begin{figure}[htb]
\begin{center}
\mbox{
\includegraphics[angle=0,width=0.75\textwidth]{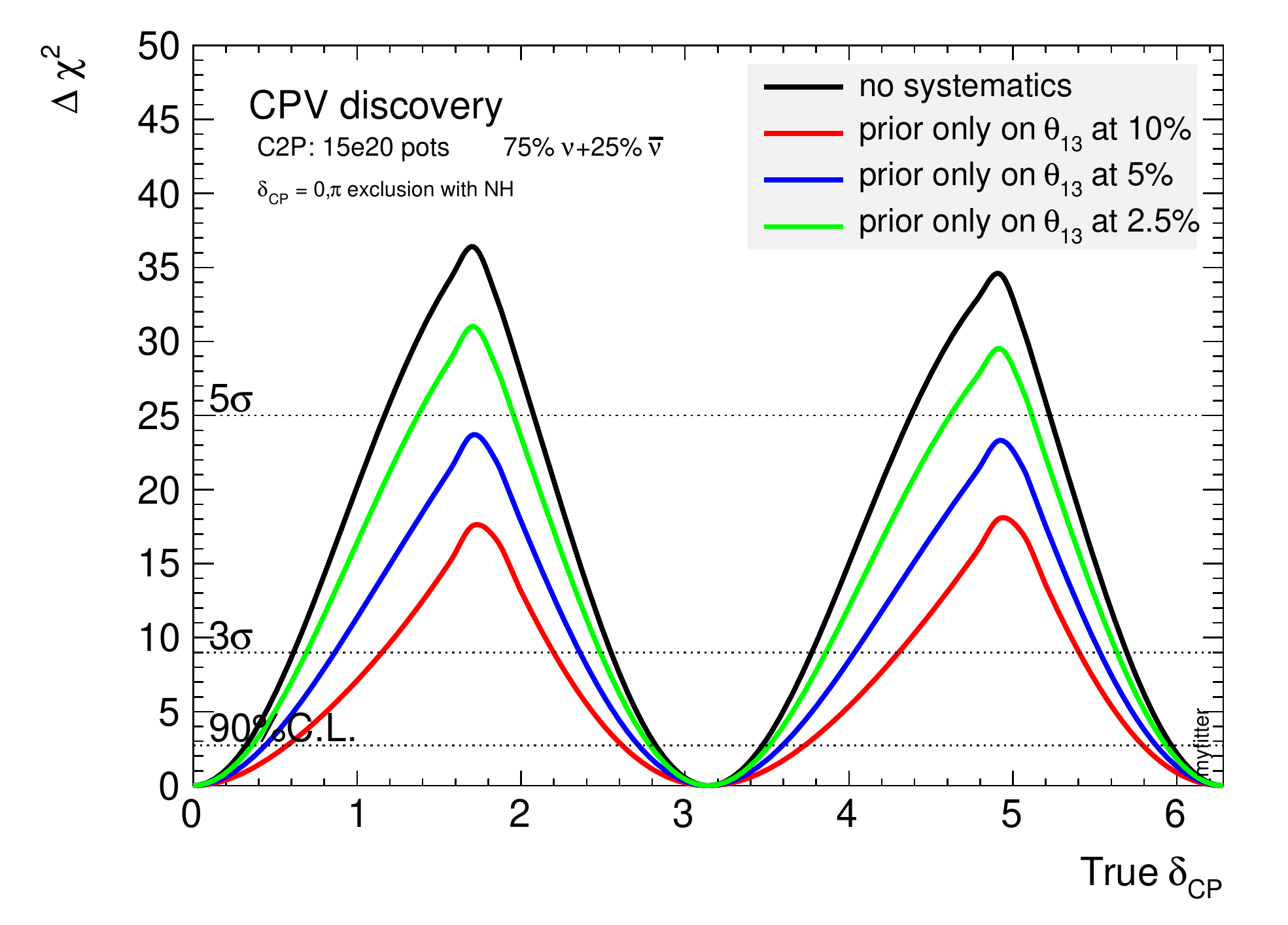}
}
\caption{\small{Impact of systematic errors: CPV sensitivity of LBNO phase I as a function
of $\delta_{CP}$, with only statistical  and no systematic errors (black), and 
effect of the error on the $\sin^22\theta_{13}$ parameter prior of $\pm10\%$ (green),
$\pm5\%$ (blue), $\pm2.5\%$ (red). }}
\label{fig:theta13a}
\end{center}
\end{figure}

\begin{figure}[!h]
\begin{center}
\mbox{
\includegraphics[angle=0,width=0.75\textwidth]{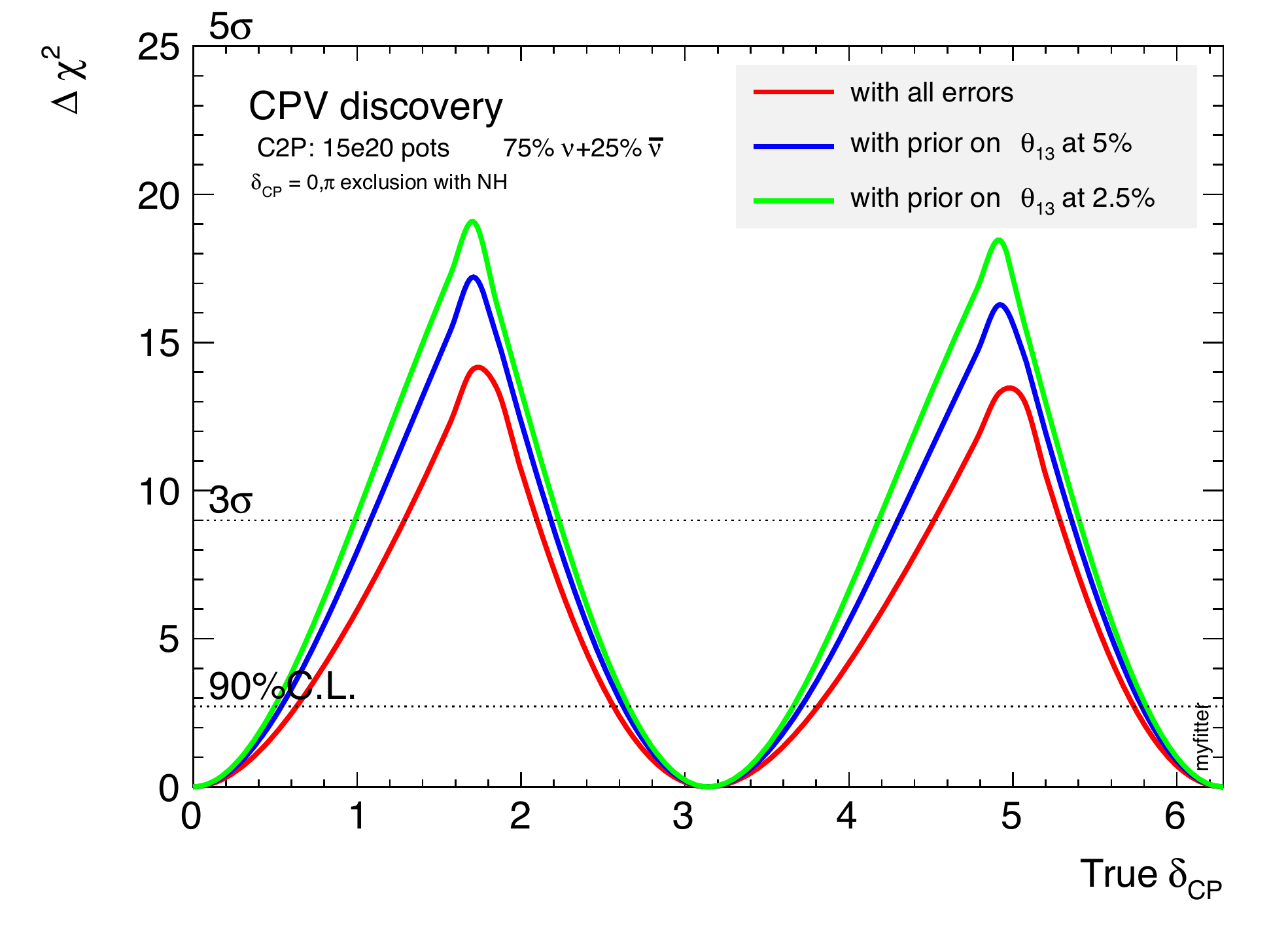}

}
\caption{\small{Same as Figure~\ref{fig:theta13a} but with all other systematic errors included.}}
\label{fig:theta13b}
\end{center}
\end{figure}

The effect of the knowledge of  $\theta_{13}$ has been studied in detail. As stated in the previous paragraphs the knowledge of  $\theta_{13}$ has a very important role on
the $\delta_{CP}$ discovery potential. 
It is of great value to reach a more precise measurement of this angle 
in order to increase the $\delta_{CP}$ sensitivity. 
In Figure~\ref{fig:theta13a} we show the effect of varying the prior on $\theta_{13}$ between 0\% and 10\% when all the other systematic errors on the oscillation parameters are set to 0\%.  
In Figure~\ref{fig:theta13b} we show the effect on the $\delta_{CP}$ sensitivity with all the systematic errors included. 
We find that the prior knowledge of $\theta_{13}$ is important to constrain $\delta_{CP}$, even
in presence of all other systematic errors.

\subsubsection{Influence of $\theta_{23}$ on the  $\delta_{CP}$ discovery potential}

\begin{figure}[!ht]
\begin{center}
%\begin{tabular}{c c c}
%\subfloat[C2P]{
%%\includegraphics[angle=0, width = 0.7\textwidth]{figures/theta23_study.pdf}
\includegraphics[angle=0, width = 0.8\textwidth]{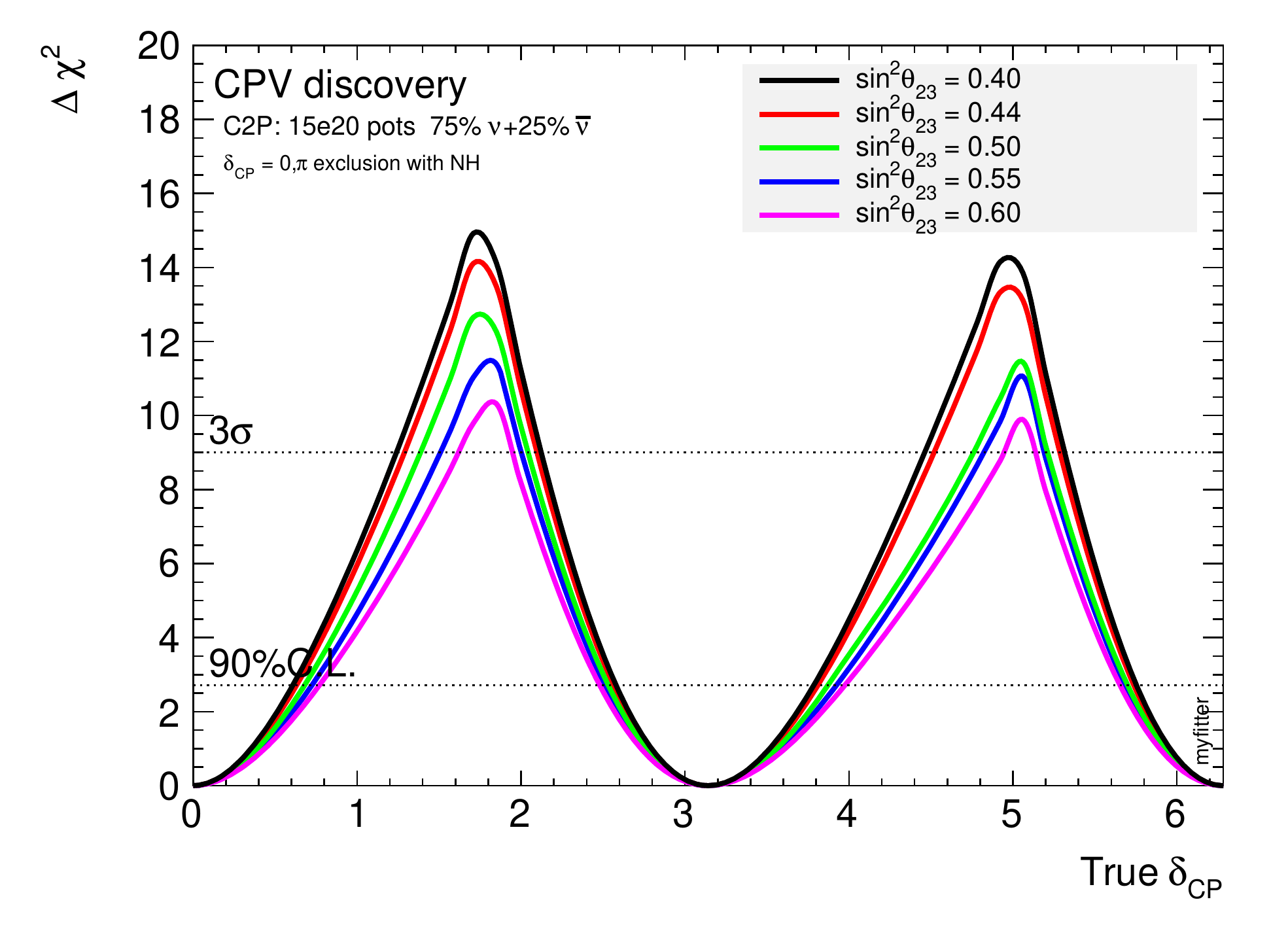}
%} &%
%~~ &% 
%\subfloat[P2P]{\includegraphics[clip, trim=15 10 50 0, width = 0.45\textwidth]{figures/theta_23_P2P.pdf}}
%\end{tabular}
\caption{\small Impact of the prior value of $\theta_{23}$: CPV sensitivity of LBNO phase I as a function
of $\delta_{CP}$ for a range of values of $\theta_{23}$.}
\label{fig:CPVtheta23}
\end{center}
\end{figure}

Now that the value of $\theta_{13}$ mixing angle has been measured, the knowledge of the mixing angles
which describe the PMNS matrix has changed significantly.  
Whilst previously $\theta_{13}$ was not known, the uncertainty on it had a dominant influence on the possible
%the uncertainty in $\theta_{13}$ had a dominant influence on the possible
discovery reach of long-baseline facilities, now it makes sense to investigate also the influence of $\theta_{23}$ (excluding $\delta_{CP}$)
 whose uncertainty has as well a large impact. 
 Its true value influences the sensitivity of CPV as shown 
in figure~\ref{fig:CPVtheta23}.
We see that variations in $\theta_{23}$ induce only slight
changes in the sensitivity to $\delta_{CP}$ at the $90\%$~CL. At higher significances,
the true value of $\theta_{23}$ plays a more important role. At the $3\sigma$
confidence level, the change in $\theta_{23}$ by 1$\sigma$ can make the
difference between having no ability to measure CP violation and being able to
exclude CP violation for around $30\%$ of the parameter space.
 
The dependence of the discovery reach on $\theta_{23}$ can be understood
analytically by following the method introduced in Ref.~\cite{Coloma:2012wq}.
This procedure can be modified to account for more complicated scenarios by
combining neutrino and antineutrino running, including matter effects and
non-trivial flux profiles; however, for our purposes of extracting the
dependence of $\Delta\delta_{CP}$ on $\theta_{23}$, this will not change the
functional behaviour and has been omitted for clarity.
To find the dependence on $\theta_{23}$, we use the approximate form for
the probabilities of the form of Eq.~\ref{eq:oscprob}.
For the known value of $\theta_{13}$, 
we can find approximate forms for the $\theta_{23}$ dependence of the
probability and its derivative by retaining their leading order behaviour
\begin{align*} P(\nu_\mu\to\nu_e) &\propto \sin^2\theta_{23},\quad \mathrm{and}\quad \frac{\partial
P}{\partial \delta_{CP}} \propto \sin(2\theta_{23}).  \end{align*}
Using these two expressions, we can compute the leading order dependence of
$\Delta\delta_{CP}$ on $\theta_{23}$ to be 
\begin{equation}\Delta\delta_{CP} = 
\frac{\sqrt{P(\nu_\mu\to\nu_e)}}{\frac{\partial P}{\partial \delta_{CP}}}=
C \frac{\sin\theta_{23}}{\sin(2\theta_{23})} \propto \sec\theta_{23},\label{eq:Deltadelta} \end{equation}
where $C$ is a constant factor. Using Eq.~\ref{eq:Deltadelta}, we
can see that the precision to $\delta_{CP}$ decreases with increasing $\theta_{23}$ 
within the currently allowed region. In
figure~\ref{fig:CPVtheta23}, $\Delta\delta_{CP}$ is indicated by the width of the region
of good fit around $\delta_{CP}=0$ and we can see that this interval grows with
increasing $\theta_{23}$. Exact numerical computations confirm the 
validity of the analytical expression.

\subsection{Impact of event normalization systematics on  the $\delta_{CP}$ discovery potential }
\label{sec:norm}

\begin{figure}[!ht]
\begin{center}
\includegraphics[angle=0,width=0.85\textwidth]{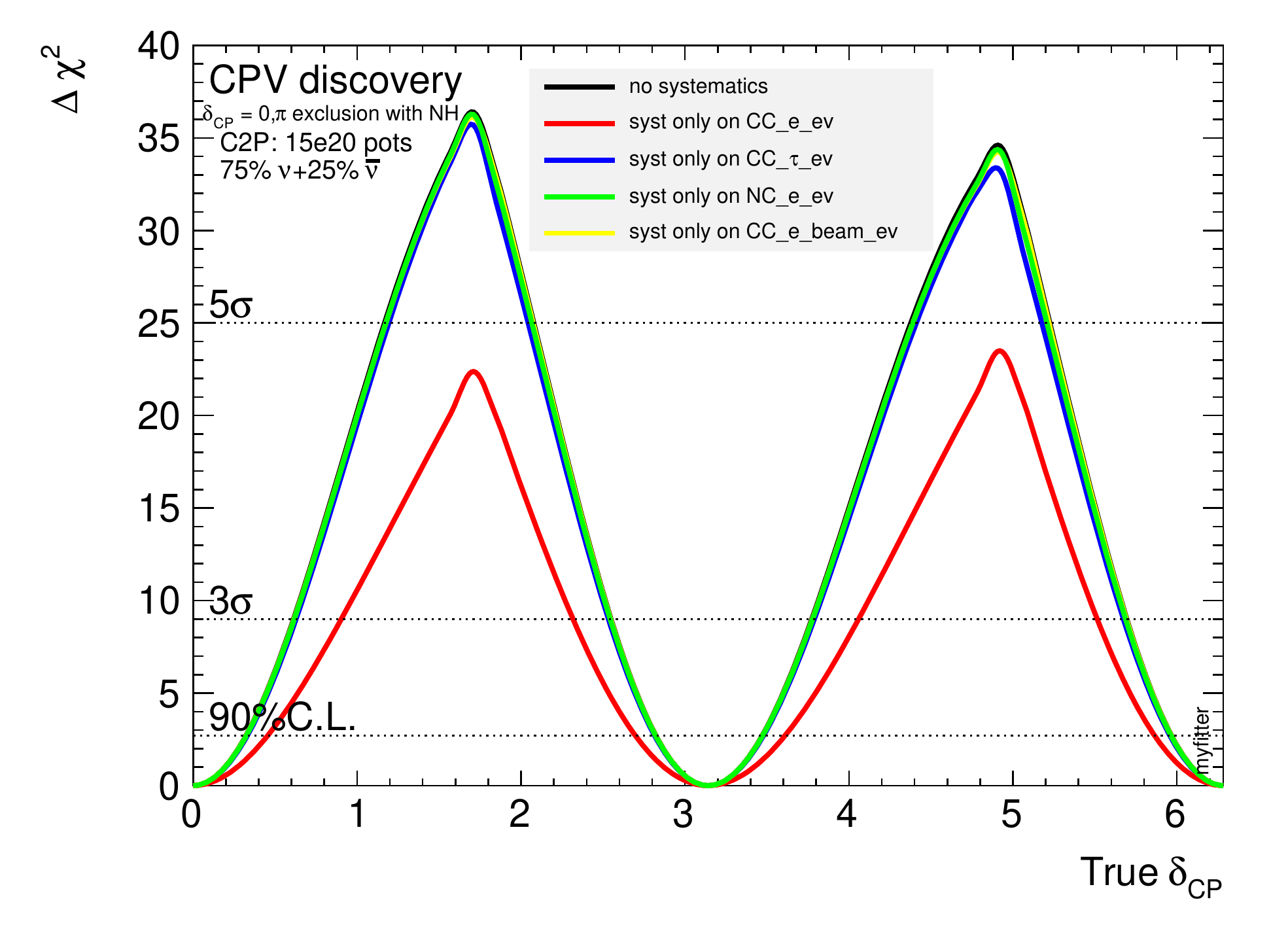}
\caption{\small{Impact of systematic errors: CPV sensitivity of LBNO phase I as a function
of $\delta_{CP}$, with only statistical  and no systematic errors (black), and 
effect of the error on the normalisation of the signal and backgrounds. }}
\label{fig:reco_norm}
\end{center}
\end{figure}

\begin{figure}[!ht]
\begin{center}
\mbox{
\includegraphics[angle=0,width=0.5\textwidth]{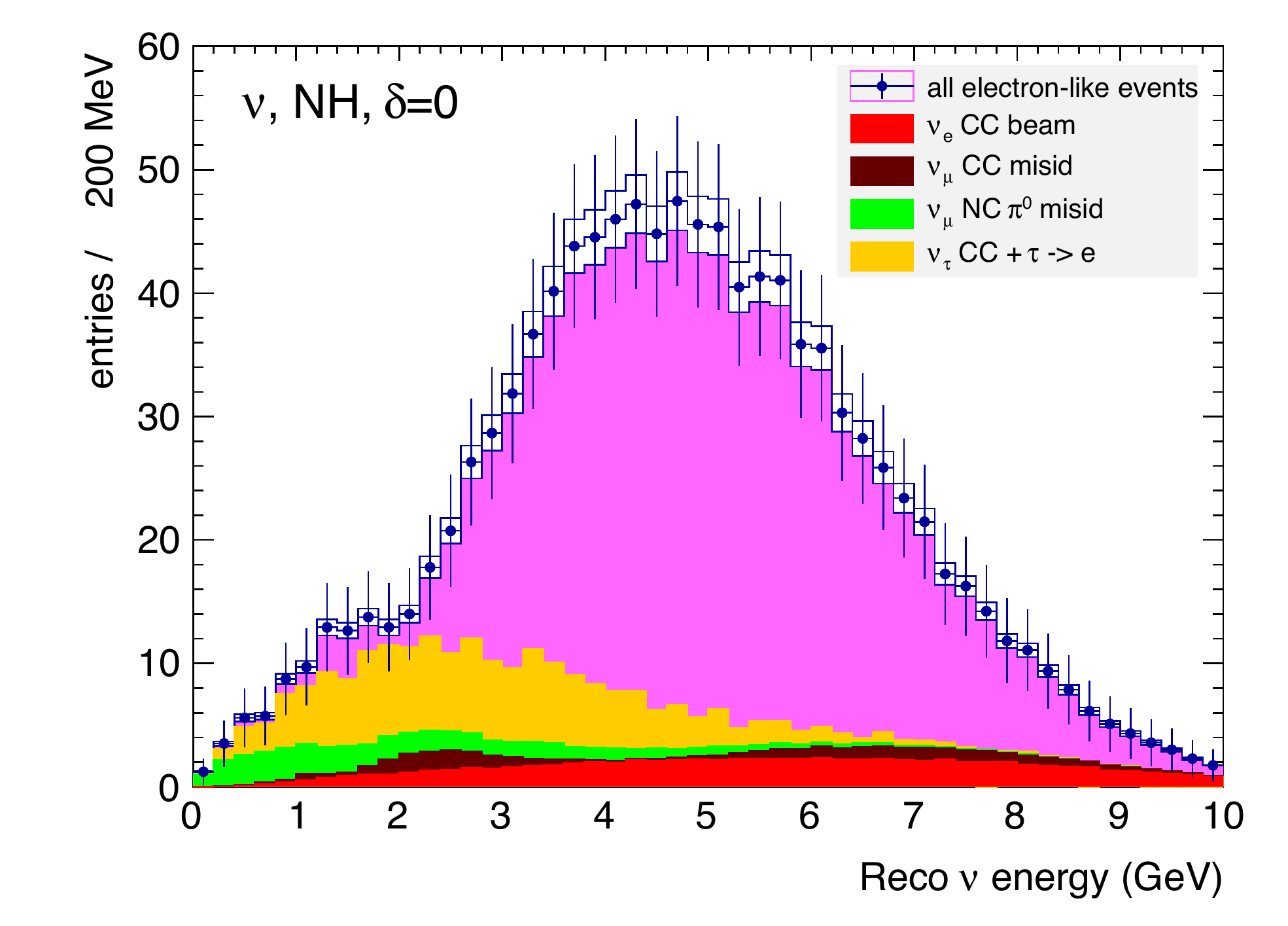}
\includegraphics[angle=0,width=0.5\textwidth]{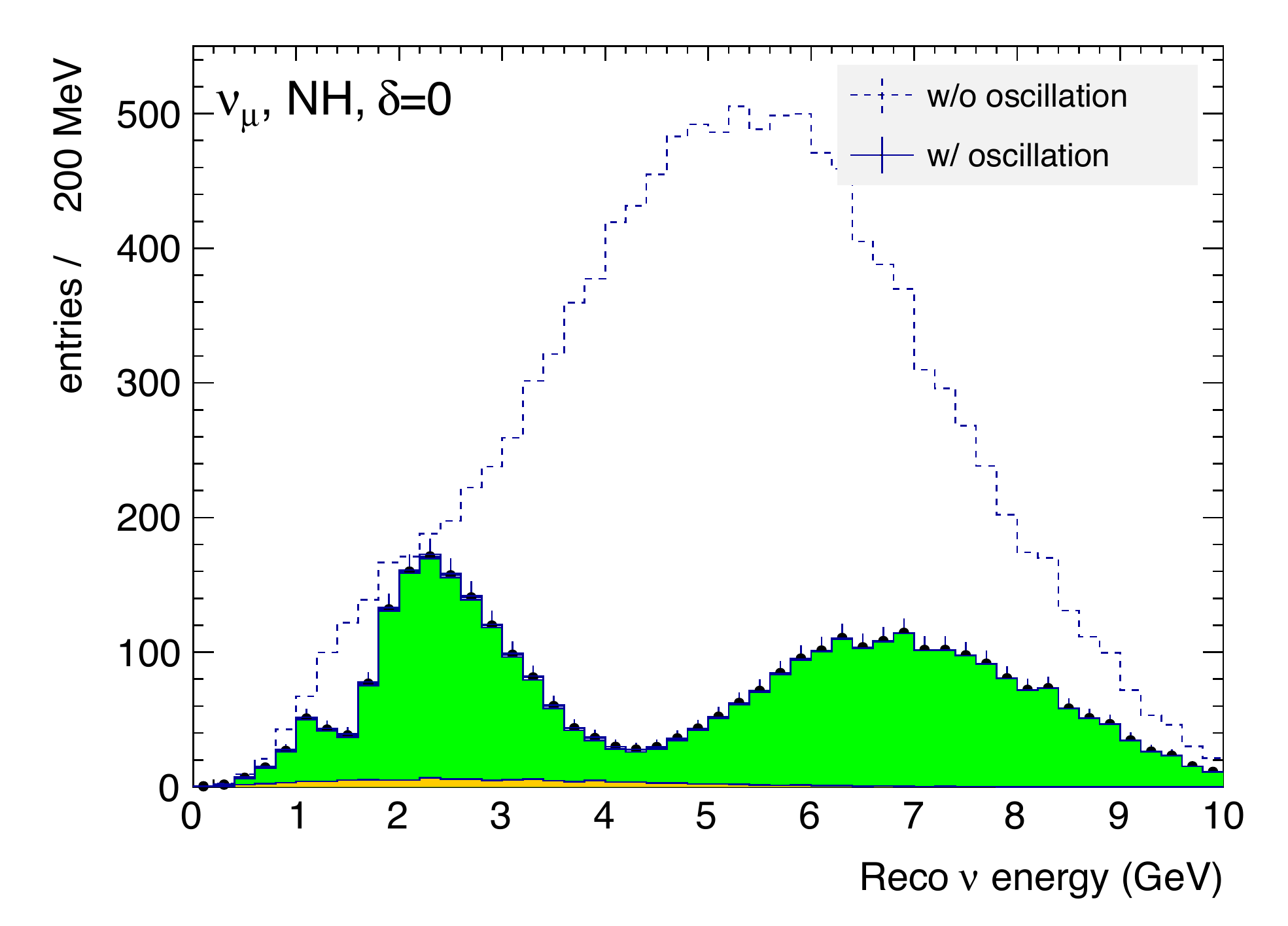}
}
\caption{\small{Measured neutrino spectra for (left) e-like appearance (right) muon-like disappearance channels, when all the normalisation errors listed in Table~\ref{tab:cuts2} are varied by $\pm1\sigma$  in a fully correlated way.
Statistical error are also shown.}}
\label{fig:reco_norm2}
\end{center}
\end{figure}

The impact of systematics due to the knowledge of signal and background normalization has also been studied.
Results are shown in Figure~\ref{fig:reco_norm} and~\ref{fig:reco_norm2}.
In Figure~\ref{fig:reco_norm} the impact of each systematic effect on the $\delta_{CP}$ sensitivity is shown: it is evident that the most important systematic error is the one on the signal channel normalization, as could be expected.
In Figure~\ref{fig:reco_norm2}, the variability band due to the effects of systematics is compared to the statistical error for the appearance and disappearance channels.
For the disappearance channel the effect is negligible.
Errors on normalizations have been considered, very conservatively,  to be fully correlated according to Table~\ref{tab:cuts2}. Their effect is smaller than the statistical uncertainty.
This study shows also the importance to have a near detector in order to reduce the effect of these uncertainties.

\subsection{Statistical power as a function of exposure}

\begin{figure}[!ht]
\begin{center}
\mbox{
\includegraphics[angle=0,width=0.8\textwidth]{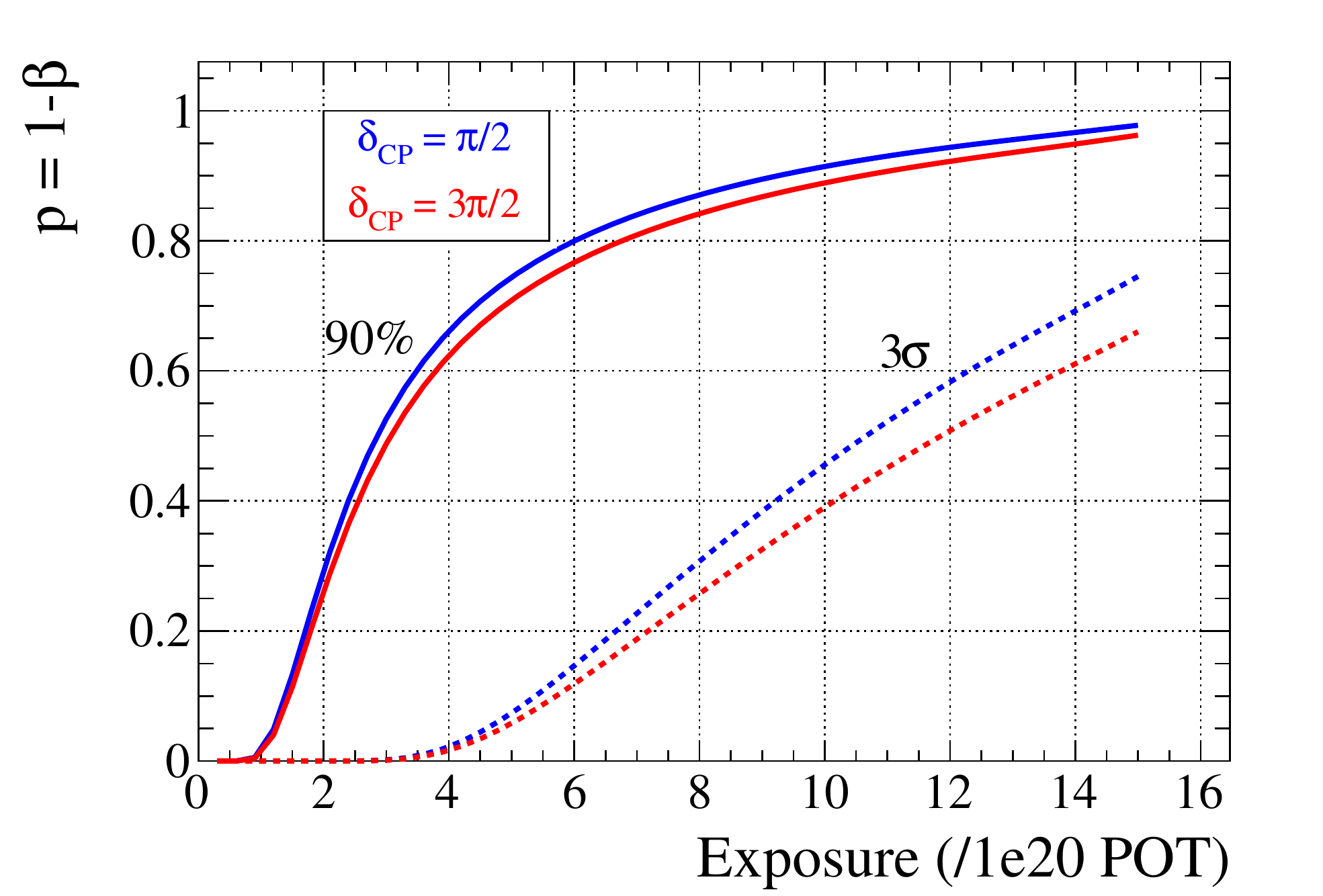} 
}
\caption{\small{Statistical power for CPV discovery as a function of exposure for $90\%$ and $3\sigma$ CL assuming NH.
The far detector of 20~kton LAr and 750~kW SPS neutrino beam are assumed. We use the nominal conservative
systematic errors (see text).}}
\label{fig:cppower}
\end{center}
\end{figure}

The statistical power of LBNO for CPV determination as a function of exposure is shown in Figure~\ref{fig:cppower}, for the two different $CL$s of $90\%$ and $3\sigma$. The two most favourable cases, $\delta_{CP}=\pi/2$ or $3\pi/2$, are considered.

\section{Ultimate CPV sensitivity}

We have seen that the LBNO Phase I has significant physics goals, in particular it is guaranteed to
be fully conclusive for MH discovery with an expected $5\sigma$~C.L. over the full range of
$\delta_{CP}$. On the other hand, the  CPV sensitivity reach is more difficult to predict,
since  ultimately is dependent on the achievable systematic errors and  on the true $\delta_{CP}$.
We have chosen to use presently realistic errors
on oscillation parameters and on the normalisation of the signal and backgrounds. With the
series of expected new measurements and possibly the addition of dedicated measurements
from experiments on hadro-production and neutrino cross-sections, it is conceivable
to think that the overall balance of 
errors could be reduced in the future, thereby improving further the expected CPV sensitivity 
of LBNO Phase I. The situation will keep to be monitored, as new experimental results
are published.

\begin{figure}[!ht]
\begin{center}
\mbox{\includegraphics[angle=0,width=0.75\textwidth]{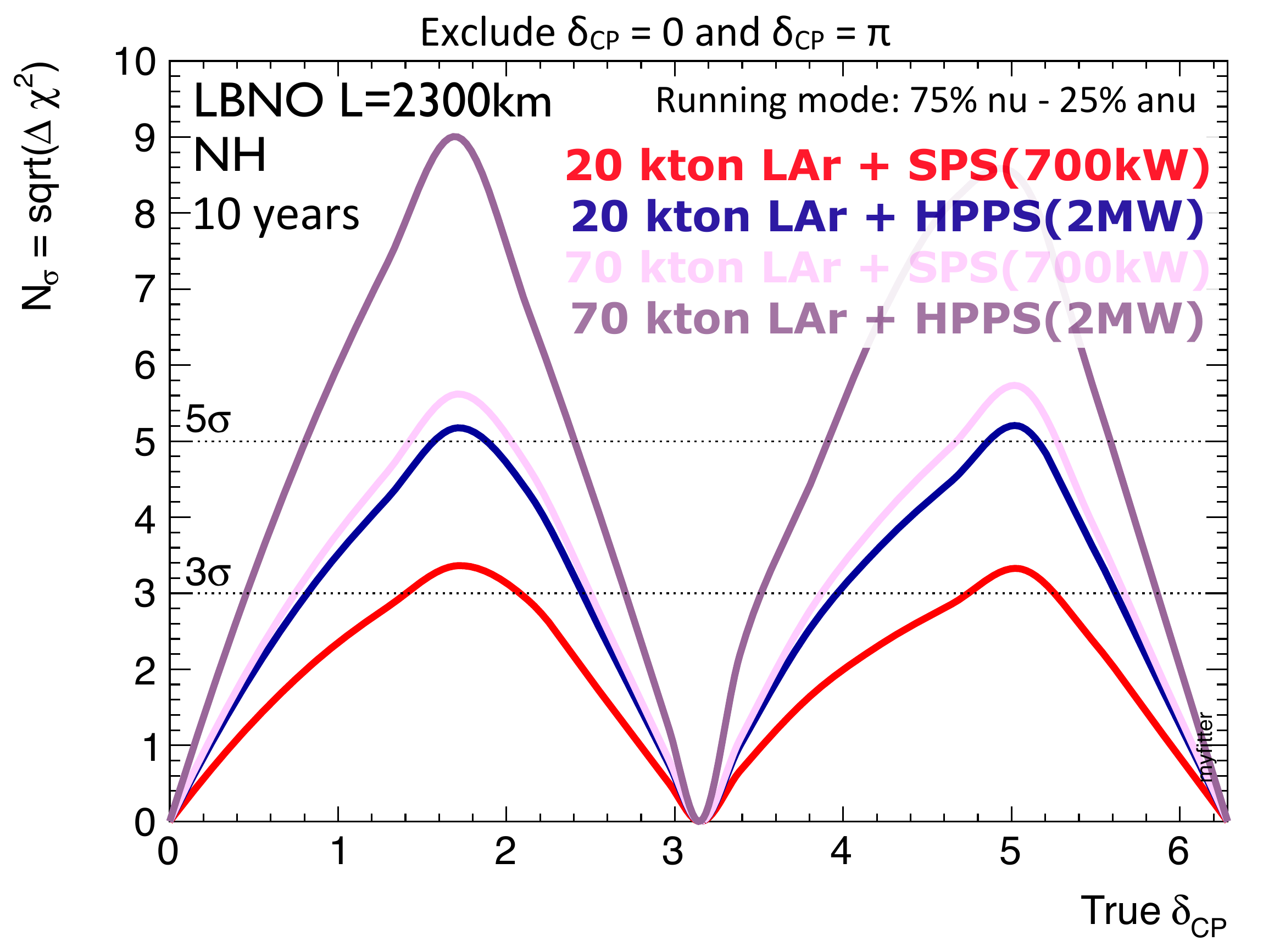}}
\caption{\small{CPV sensitivity as a function
of $\delta_{CP}$ for various upgrades of beam power with the HP-PS, and of the far detector mass, with 20~kton and 70~kton.}}
\label{fig:ultimateCPV}
\end{center}
\end{figure}

On the other hand, an alternative method is to increase the detector mass and the neutrino beam
power in order to decrease the statistical error around the 2nd oscillation maximum. Because of
the natural cut-off of the muon-neutrino flux spectrum at low energy,
and the linear increase of the total neutrino cross-section with energy,
the 2nd maximum is more difficult to study than the 1st maximum. However, this is still possible at
the LBNO baseline of 2300~km since the 2nd maximum is at an accessible energy of  $\sim$1.5~GeV.
Since the CP-asymmetry at the 2nd maximum is more sensitive to $\delta_{CP}$ than at the 
first maximum, a significant gain is obtained by populating this region with oscillation
events. This is one
of the main goals of the LBNO Phase II. The expected CPV 
sensitivity as a function
of $\delta_{CP}$ is shown in Figure~\ref{fig:ultimateCPV}
for various upgrades of beam power with the HP-PS, and of 
the far detector mass, from 20~kton to 70~kton. With a new powerful proton driver such as 
the conceptual HP-PS and a 70~kton detector mass, the coverage 
at $>5\sigma$'s C.L. will be $\sim$54\% after 10~years.

\section{Summary and Conclusions}
\label{sec:sumconcl}
The  LBNO experiment is the outcome of intense and comprehensive design studies 
supported by the European Commission since 2008. In an incremental approach, we propose LBNO with a 
20~kton underground detector as the first stage of a new neutrino observatory able to address 
long-baseline neutrino physics as well as neutrino astrophysics. The programme 
has a clear long-term vision for future stages of the experiment, including the 
Neutrino Factory~\cite{Edgecock:2013lga}, for which the baseline of 2300 km is 
well adapted.

Unlike the attempts to infer MH with atmospheric neutrinos in multi-megaton low-threshold
detectors~\cite{Ribordy:2013xea,Franco:2013in}, such as the one proposed with 
PINGU~\cite{IceCube:2013aaa} or ORCA (for a discussion on the physics potential
see e.g.~\cite{ORCA:2014}), or with medium-baseline 
reactor experiments~\cite{Kettell:2013eos},
such as JUNO (see e.g.~\cite{Li:2013zyd}),
the accelerator-based approach of LBNO addresses both fundamental problems
of CPV and MH in clean and straightforward conditions, profiting from the ability
to reverse the focusing horns polarity and from the well-known and monitored 
fluxes,  which characterise accelerator-based neutrino beams.

In this paper, we have presented our state-of-the-art studies of the expected sensitivity to CPV and MH. 
We have addressed  the impact of the knowledge of the oscillation 
parameters and of the systematics errors of the experiment. 
We employed a  Monte-Carlo technique simulating a very large number
of toy experiments to estimate the confidence level of the MH and CPV measurements.
We find that, with the capability of reversing the horn focusing polarity, and
even under pessimistic assumptions on systematic errors,
LBNO alone provides a direct and
guaranteed discovery of MH with $\geq 3\sigma$($\geq5\sigma$) confidence level, 
independently of the value of the CP phase and the octant of $\theta_{23}$, within $\sim$2.5(5) years of 
CERN SPS running.
The first stage of LBNO will therefore discover the mass hierarchy with certainty.
 
LBNO  has also a unique sensitivity to CPV  through the exploration of the 
first and second oscillation maxima, making possible to study the $L/E$
modulation which should match that expected by $\delta_{CP}$ terms in
the oscillation probability. With conservative expectations on
the systematic errors and after
10~years of CERN SPS running,
a significance for CPV above $> 3\sigma$'s C.L. will be reached for $\sim25(40)$\% of the $\delta_{CP}$ 
values, under the assumption that $\sin^22\theta_{13}$ will be known from reactor experiments with a precision
of $\pm10(2.5)$\%. 

The ultimate CPV reach is sensitive to the knowledge of the oscillation parameters
and to the assumed flux, cross-section and detector-related 
systematic errors. 
The CPV reach is larger if  sources of systematic
errors can be controlled below the values conservatively assumed in our present study. 
In particular, improvements in
the present knowledge of differential neutrino interaction cross-sections would increase 
the expected CPV discovery potential of LBNO. 
Alternatively, with an increased exposure aimed at increasing the number of oscillated
around the 2nd maximum, a CPV discovery level $> 5\sigma$'s C.L. is reachable over
a wide range of $\delta_{CP}$ values.
With a new powerful proton driver such as 
the conceptual HP-PS and a 70~kton detector mass, the coverage 
at $>5\sigma$'s C.L. will be $\sim$54\% after 10~years.

We conclude that the control of systematic errors will be the critical challenge for all next generation long-baseline projects such as LBNE, LBNO and Hyper-Kamiokande. This study has presented a comprehensive overview of how experimental uncertainties, together with our limited knowledge of the oscillation parameter space, effects the physics reach of LBNO. 
Realistic assumptions regarding systematic uncertainties and analysis priors are mandatory in order to develop any new project of this scale.

\acknowledgments
We are grateful to the European Commission for the financial support of the project through the FP7 Design Studies LAGUNA (Project Number 212343) and LAGUNA-LBNO (Project Number 284518). 
We acknowledge the financial support of the UnivEarthS LabEx programme at Sorbonne Paris Cit\'e (ANR-10-LABX-0023 and ANR-11-IDEX-0005-02).
In addition, participation of individual researchers
and institutions has been further supported by funds from ERC (FP7).

\end{document}